\documentclass[11pt]{iopart}

\expandafter\let\csname equation*\endcsname\relax
\expandafter\let\csname endequation*\endcsname\relax 
\usepackage[english]{babel}
\usepackage{amsmath}
\usepackage{amssymb}
\usepackage{graphicx}
\usepackage{color}
\usepackage[utf8]{inputenc}
\usepackage{amsfonts}
\usepackage{graphicx} 
\usepackage{amsthm}
\usepackage{placeins}
\usepackage{verbatim}
\usepackage{booktabs}
\usepackage{color}
\usepackage{bbold}
\usepackage{comment}
\usepackage{url}
\usepackage[all]{xy}
\usepackage[width=17.5cm, vscale=0.8]{geometry}
\newcommand{\beq}{\begin{equation}}
\newcommand{\eeq}{\end{equation}}
\newcommand{\bea}{\begin{eqnarray}}
\newcommand{\eea}{\end{eqnarray}}

\newcommand{\order}{{\mathcal{O}}}
\usepackage{epsfig}
\newcommand{\bxi}{{\mbox{\boldmath $\xi$}}}
\newcommand{\bsigma}{{\mbox{\boldmath $\sigma$}}}
\newcommand{\bpsi}{{\mbox{\boldmath $\psi$}}}
\newcommand{\bbeta}{{\mbox{\boldmath $\eta$}}}

\newcommand{\rd}{{\rm d}}
\newcommand{\bm}{{\bf m}}
\newcommand{\bM}{{\bf M}}
\newcommand{\bv}{{\bf v}}
\newcommand{\bra}{{\langle}}
\newcommand{\ket}{{\rangle}}
\newcommand{\bA}{{\bf A}}
\newcommand{\bq}{{\bf q}}
\newcommand{\bD}{{\bf D}}
\newcommand{\bP}{{\bf P}}

\begin{document}
\title[A dynamical model of the adaptive immune system]{A dynamical model of the adaptive immune system: effects of cells promiscuity, antigens and B-B interactions}
\author{Silvia Bartolucci$^{1}$ and Alessia Annibale $^{1,2}$}
\address{$^1$ Department of Mathematics, King's College London, The Strand,
London WC2R 2LS, UK}
\address{$^2$ Institute for Mathematical and Molecular Biomedicine, King's College London, Hodgkin Building, London SE1 1UL, UK}\begin{abstract}
We analyse a minimal model for the primary response in the adaptive immune system comprising three different players: antigens, T and B cells. We assume B-T interactions to be diluted and sampled locally from heterogeneous degree distributions, which mimic B cells receptors' promiscuity.
We derive dynamical equations for the order parameters quantifying the B cells activation and study the nature and stability of the stationary solutions using linear stability analysis and Monte Carlo simulations.
The system's behaviour is studied in different scaling regimes of the 
number of B cells, dilution in the interactions and number of antigens.
Our analysis shows that: (i)
B cells activation depends on the number of receptors in such a way that 
cells with an insufficient number of triggered receptors cannot be activated; 
(ii) idiotypic (i.e. B-B) interactions enhance parallel activation of multiple clones, improving the system's ability to fight different pathogens in parallel; 
(iii) the higher the fraction of antigens within the host the harder is for the system to sustain parallel signalling to B cells, crucial for the homeostatic control of cell numbers.
\end{abstract}
\newpage
\tableofcontents
\maketitle
\section{Introduction}
\begin{figure}
\centering
\includegraphics[width=0.4\textwidth]{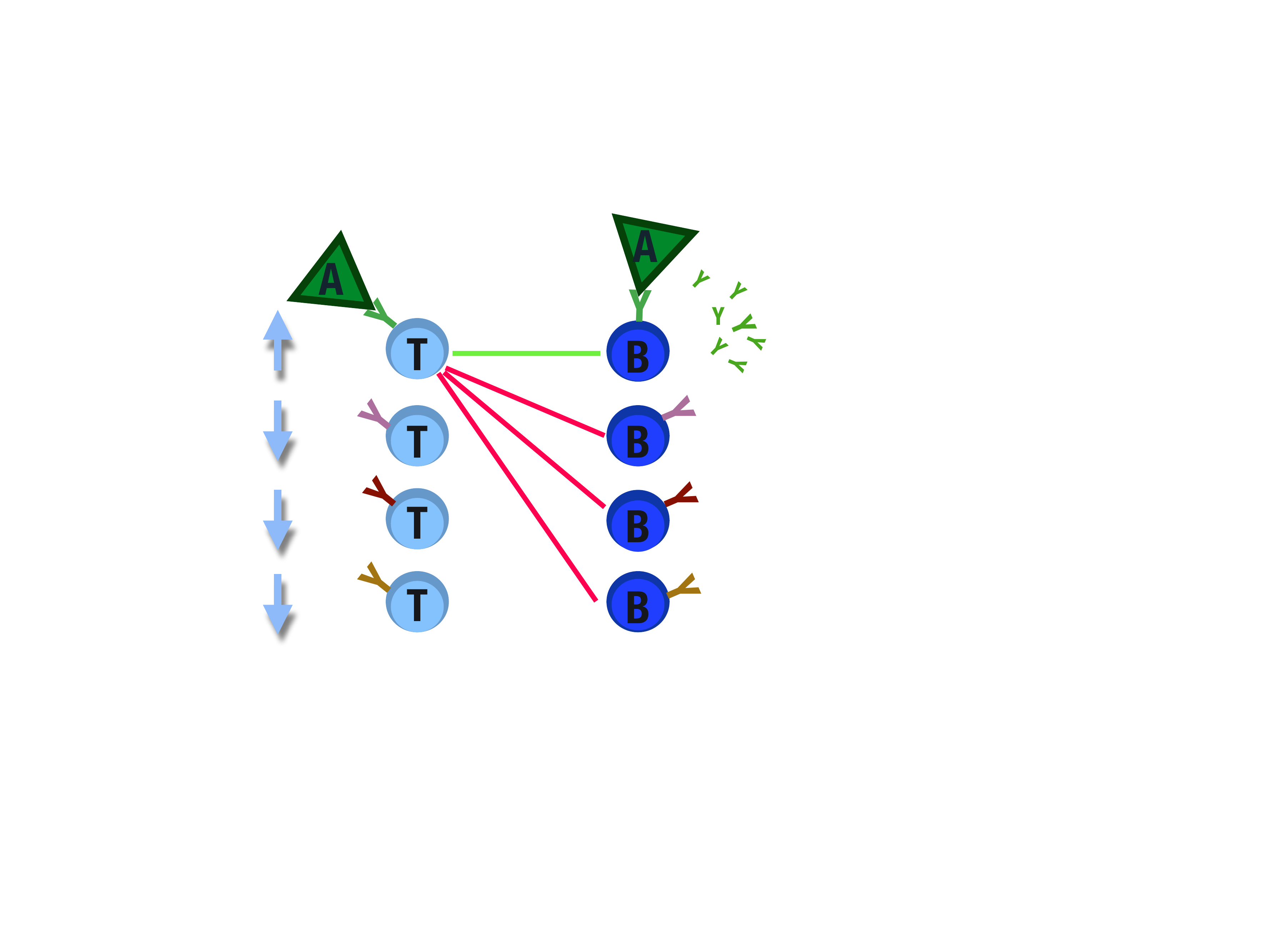}
\caption{Schematic representation of the antigen recognition process and immune response activation by T and B cells. The best-matching B and T cells independently detect the antigen: the active T clone sends excitatory cytokines (green links) to the B clone to start the antibodies production and suppress the non-complementary B clones with inhibitory signals (red). The other T cells are in a quiescent state.}
\label{fig:schemenet}
\end{figure}

The immune system is a complex collection of organs, tissues and cells which is present in all vertebrates and protects the organism from external pathogens \cite{Abbas}. In this work we introduce a minimal model to describe the primary response of the adaptive immune system, a network of highly specialised cells that produces a targeted reaction to specific antigens, {\em i.e.} viruses. The main players are B and T cells, two different types of lymphocytes. They independently recognise the antigen binding it with their receptors (fig. \ref{fig:schemenet}). Each group of T and B cells sharing the same receptors' structure (clone) is able to recognise and fight {\em only} a particular virus with complementary epitope.

The immune response by B cells is activated or suppressed according to a confirmation signal sent by T cells in the form of excitatory or inhibitory proteins, the cytokines. Such response (when activated) consists in the secretion of antibodies, proteins able to chemically bind and neutralise the antigen,  hence (possibly) avoiding the propagation of the infection. 
This two- signals mechanism prevents erroneous B cells activations against mismatching antigens or other cells of the organism.

In recent years, this system composed by a large number of interacting agents - T, B cells and antigens -  has been looked at through the prism of statistical mechanics to understand its global features and functionalities \cite{Parisi,WeisBB,mora,PRE}.
Following this promising line of research we propose here a model for the dynamics of T and B cells and antigens, extending preliminary proposals (see \cite{PRE,jphysaas} and references therein) to incorporate important biological features of  real immune systems. In particular, we study the effect of having  B cells with a variable number of receptors on their surfaces, one of the most important mechanisms preventing autoimmune reactions and diseases:
a {\em receptor editing process} is indeed very commonly observed during the B cell maturation, where self-reactive cells, which may be responsible for the onset of autoimmune responses, are suppressed at an early stage of the development by changing the number of receptors \cite{Kous}. 
We also introduce interactions between B cells, the so-called {\em idiotypic network}, first hypothesised by Jerne \cite{Jerne}: B cells receptors can not only recognise antigens but also other B cells with complementary receptors. In the presence of an antigen antibodies with complementary receptors, called {\em idiotypes} and denoted with Ab1, are produced and recognised by complementary antibodies (Ab2) which share structural features with the original antigen.
According to recent experimental studies \cite{autoimmuneBB1,autoimmuneBB2,autoimmuneBB3} the idiotypic network seems to play a central role in autoimmune diseases, supporting a cascade of autoantibodies production, which recognise each other and modulate the immune response. 
Finally, we incorporate the effect of an external antigenic field to investigate the {\em immunological memory} \cite{memory}, {\em i.e.} the ability of the immune system to produce a more effective and faster response at a second encounter with an antigen. 

From a statistical mechanics perspective this system is modelled as a bipartite network, where links between B and T-cells are sparse as biologically required. We introduce a phenomenological Hamiltonian description of the system and carry out a dynamical analysis of the network, evolving via a Glauber sequential update. 
Via non equilibrium statistical mechanical techniques initially developed for Hopfield neural networks and spin glasses \cite{hopfield,amit,MPV,amitbook,sollichcoolen} we derive the equations for the time evolution of a set of parameters, quantifying the immune response strength. We analyse the system's behaviour in different regions of the parameter's space ({\em i.e.} number of clones and triggered receptors, noise level, {\em etc.}) through linear stability analysis and Monte Carlo 
simulations.
The paper is structured as follows: in Sec. \ref{sec:model} we define the model, in Sec. \ref{sec:over} we give an overview of the results that we derive in 
Sec. \ref{sec:q}, \ref{sec:BB} and \ref{sec:ant}. We summarise our conclusions in \ref{sec:conclusions}.
\section{The model}\label{sec:model}
We model T-cells as binary variables or ``spins'' $\sigma_i=\pm 1$, $i=1,\dots,N$, which can be active {\em i.e.} secreting 
cytokines ($+1$) or quiescent ($-1$). B cells and antigens are characterised via their concentration, 
respectively $b_{\mu}\in\mathbb{R}$,  $\mu =1,\dots, P$ and $\psi_a\in\mathbb{R}, a=1,\dots,A$ with respect to a reference level.
Both the number of B clones and the number of antigens are sub-linear in the 
system size $N$, 
with $P= N^\delta, \ \delta\in[0,1)$ and 
$A=N^{a},\ a\in[0,1)$.
In the absence of antigens and interactions with T cells, the B clone sizes $b_\mu$ can be regarded as 
Gaussian variables \cite{memory}. Without loss of generality we take their mean to be 
zero and we denote by ${\bf A}^{-1}$ their covariance matrix.

\begin{figure}[htb!] 
\centering
\resizebox{0.5\columnwidth}{!}{\includegraphics{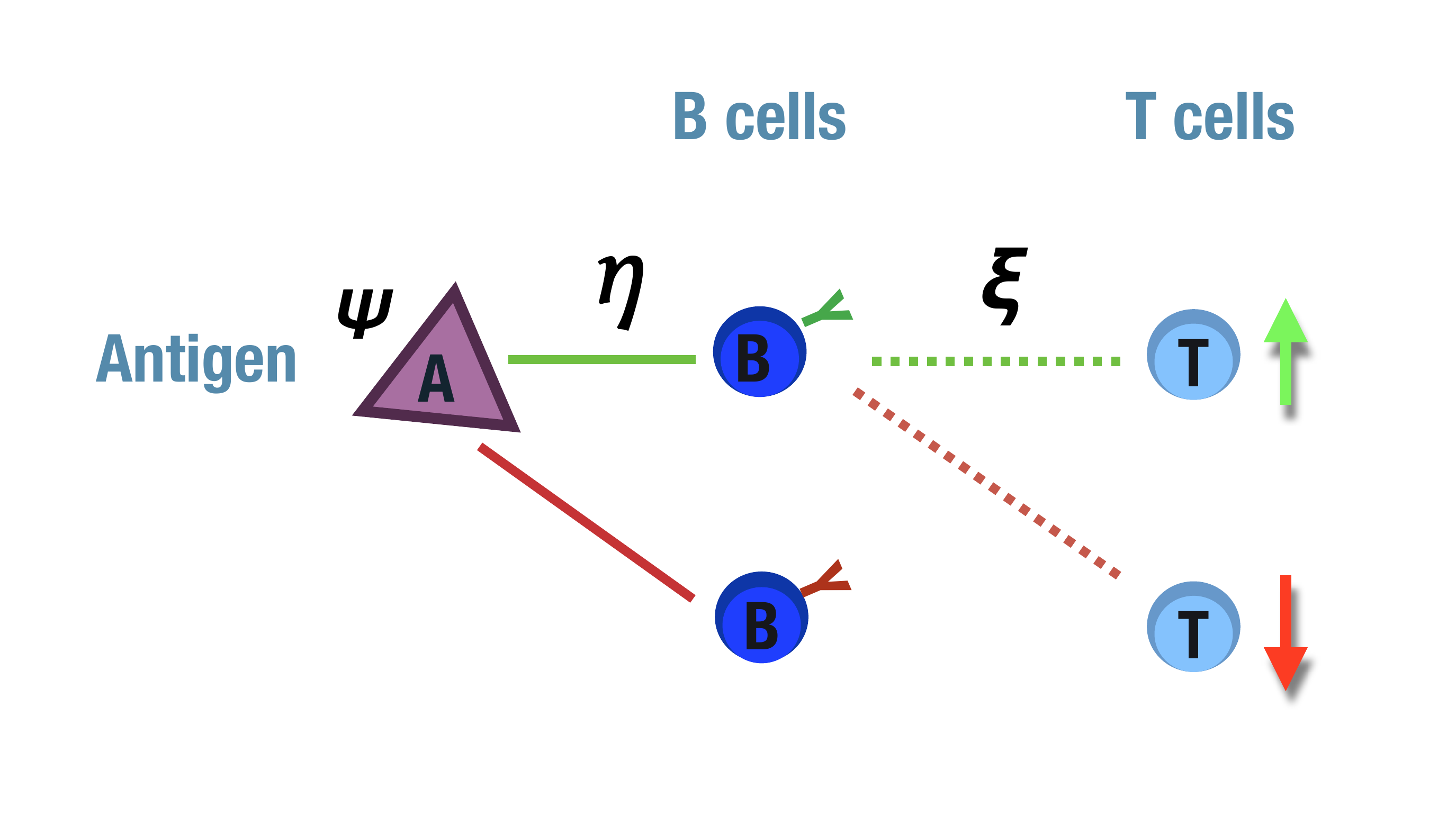}}
\caption{Schematic interactions between B, T cells and the antigen A. In the presence of an antigen with concentration $\psi$,  the complementary B cell will detect 
it (B-A interactions mediated via the matrix $\bbeta$) and will receive a confirmatory signal from the active T cells (represented by up arrows)
via the cytokines $\xi_i^{\mu}$ . 
}\label{fig:scheme}
\end{figure}

\begin{figure}[htb!]
\centering
\includegraphics[width=0.33\textwidth]{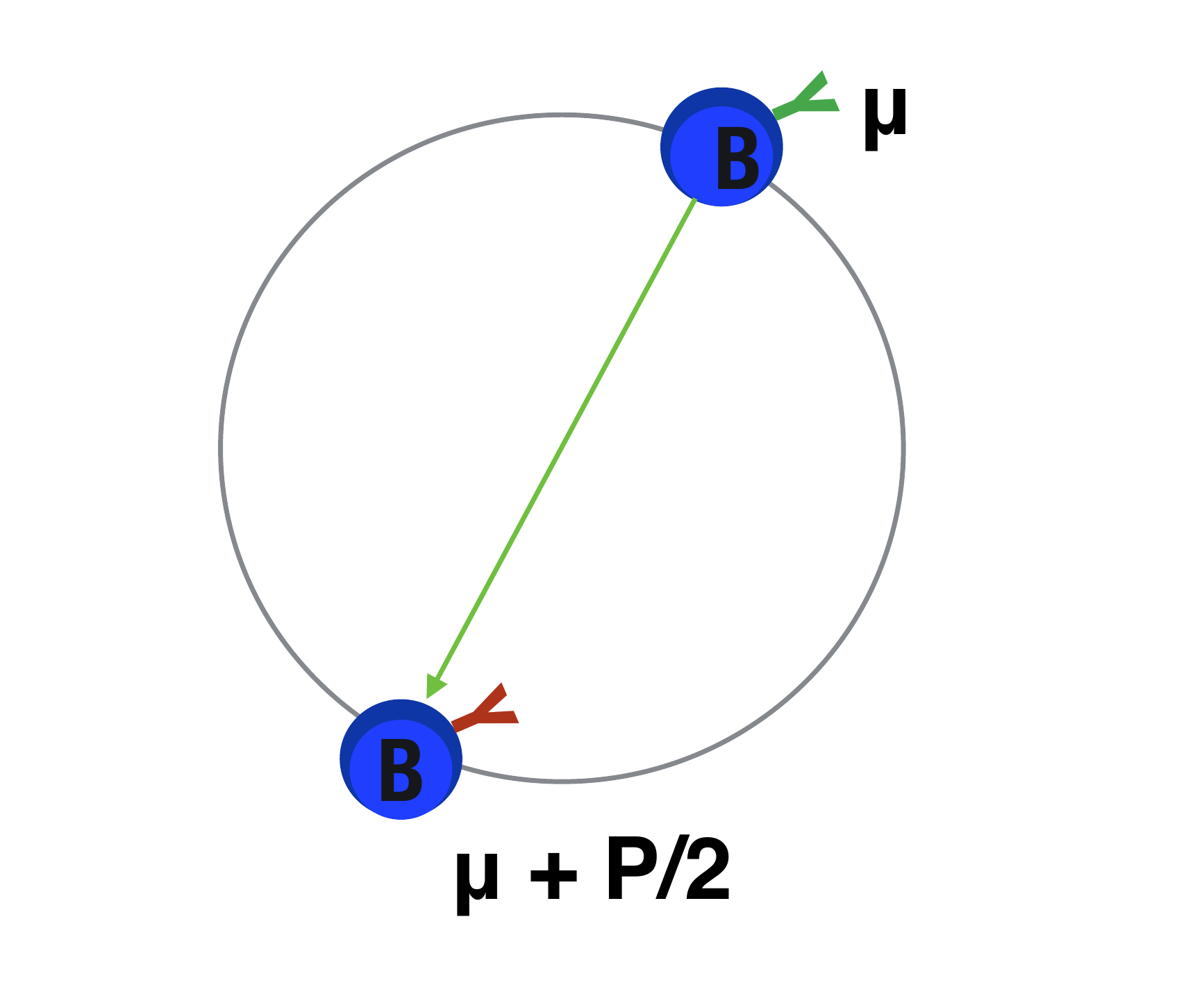}
\includegraphics[width=0.4\textwidth]{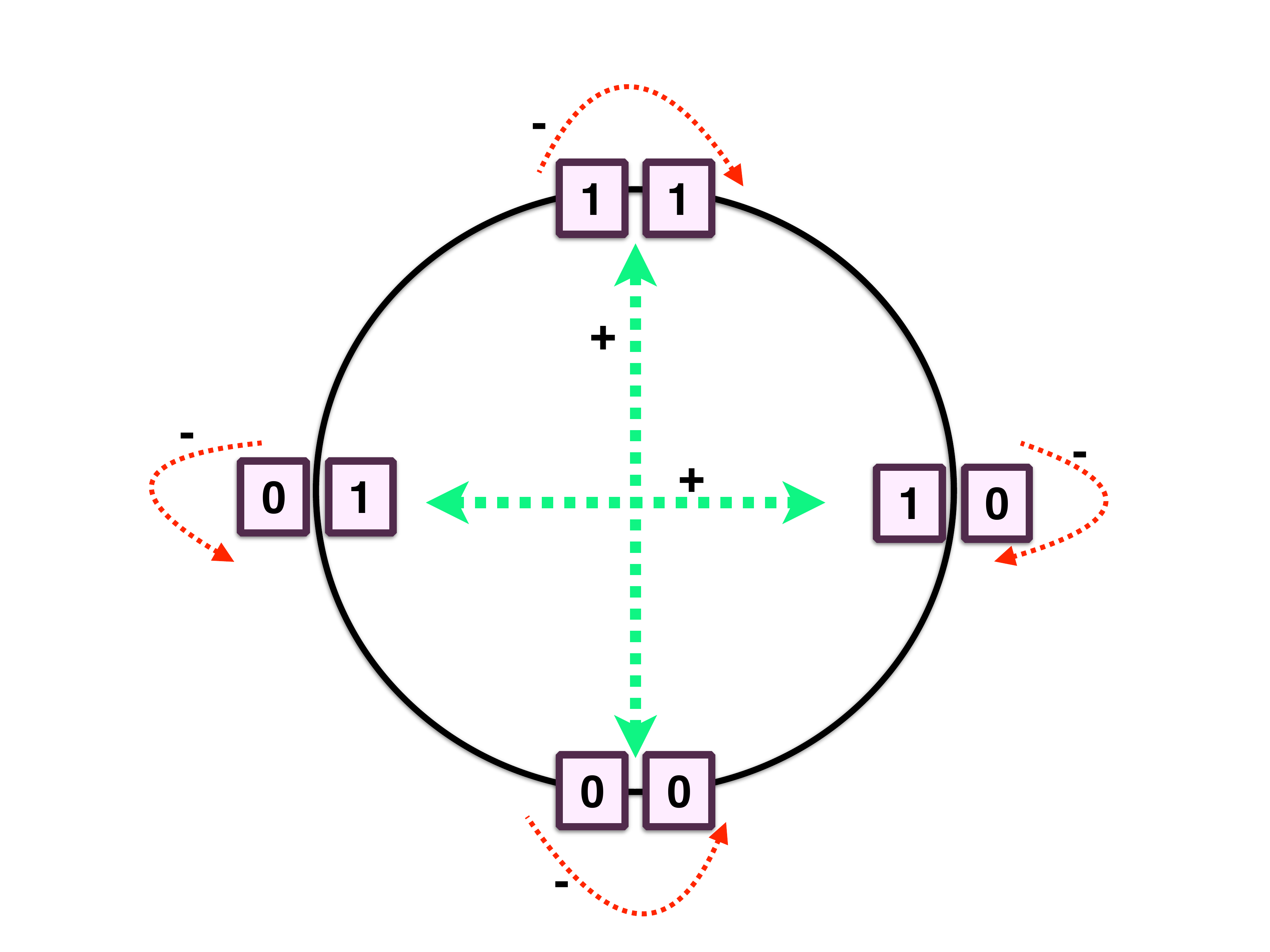}
\caption{Left: B-B interaction via $A_{\mu\nu}$. The expansion of the $\mu$-th clone is triggered by its complementary clone. 
Right: We represent the epitopes as binary strings, organising them on a ring and assuming that the complementary strings interact.}
\label{fig:BBTOP}
\end{figure}

\begin{figure}[htb!]
\centering
\includegraphics[width=0.4\textwidth]{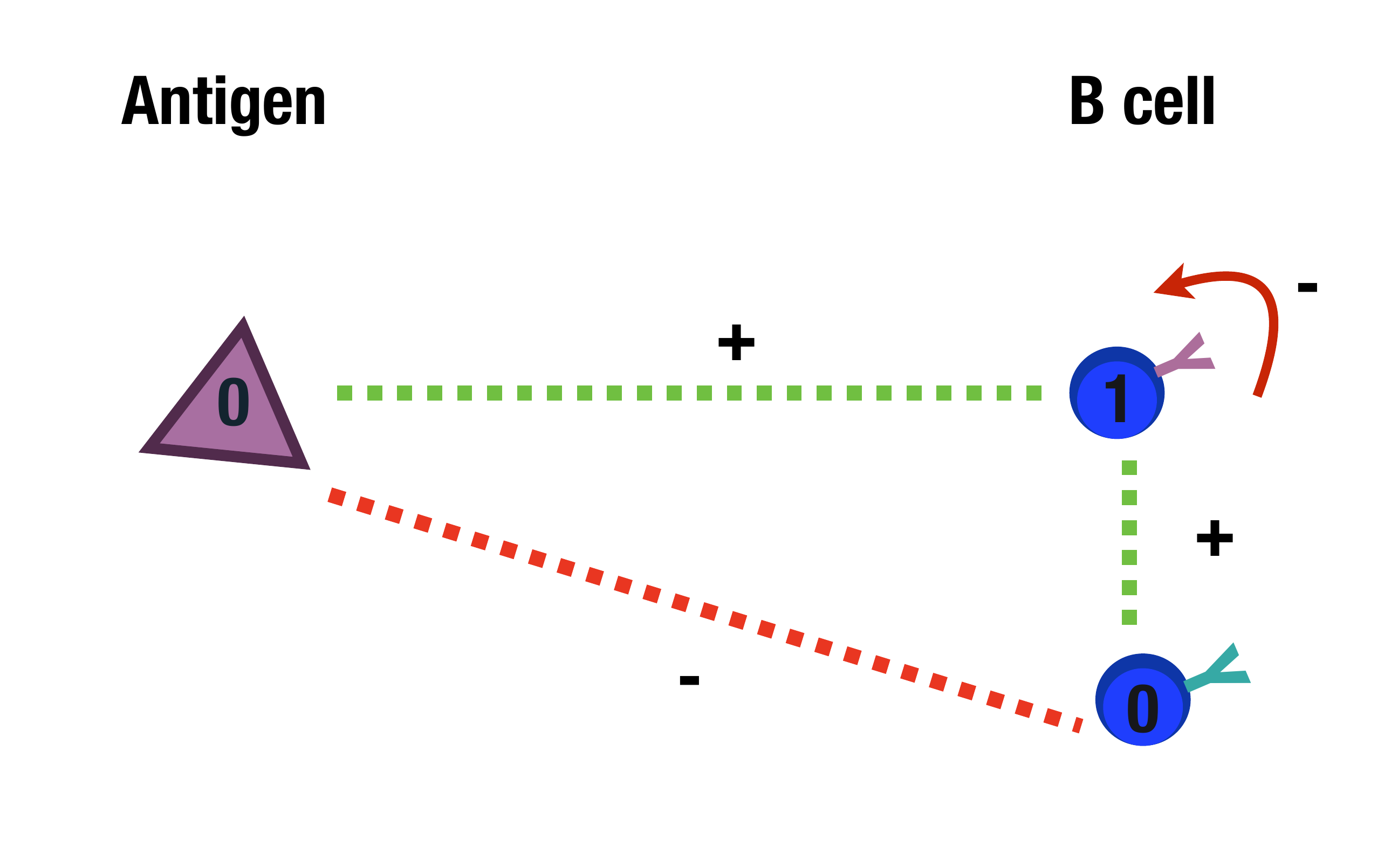}
\caption{Scheme of B-B and B-A interactions. Antigens and B cells are 
denoted by 
variables $0,1$ representing the shape of their receptors: $0-1$ variables 
represent 
complementary receptors (key-lock mechanism). Different B cells excite each 
other and each 
of them represses itself, while the antigen will excite complementary B 
repressing the 
identical one (0-0).}
\label{fig:BBA}
\end{figure}

The interactions between the $ i$-th T clone and the $\mu$-th B clone, mediated via the cytokines, are represented by the variables $\xi_i^{\mu}\in \{+1,-1,0\}$, respectively corresponding to excitatory, inhibitory or absent signals.
We will regard those as random variables drawn from
\bea
P(\bxi)=\prod_{i\mu}\left[\frac{q_{\mu}}{2N^{\gamma}}\big(\delta_{\xi_i^{\mu},1} + \delta_{\xi_i^{\mu},-1}\big) +\bigg(1-\frac{q_{\mu}}{N^{\gamma}}\bigg)\delta_{\xi_i^{\mu},0}\right] , \quad \quad q^{\mu}=\mathcal{O}(N^0) \quad\forall \mu,
\label{eq:xdistrib}
\eea
where the $q_\mu$'s are drawn from a distribution $\mathcal{P}(q)=P^{-1}\sum_\mu\delta_{q,q_\mu}$ and control the degree of B cells promiscuity, 
{\em i.e.} their ability to communicate with different T cells, via different receptors. 
The fraction of non-zero B-T links determines the degree of dilution of the system: for $\gamma=0$ the system 
is finitely diluted, whereas for $\gamma>0$ the system is extremely diluted.
The combined interacting system of B-T cells and the antigens can be phenomenologically described by the following Hamiltonian:
\bea
\mathcal{H}(\bsigma,{\bf b})= -\sum_{\mu=1}^P b_{\mu}\left(\sum_{a=1}^A\psi_{a}\eta_{a}^{\mu} +N^{1-\gamma}
\sum_{i=1}^N\xi^{\mu}_i\sigma_i\right)+\frac{1}{2\sqrt{\beta}} \sum_{\nu,\mu=1}^P
b_{\mu}A_{\mu\nu}b_{\nu}.
\label{eq:Hamil}
\eea
The first term takes into account the interactions between antigens and B cells via the matrix $\eta_{a}^{\mu}$, the second term is related to B - T 
interactions mediated by cytokines, the third one takes into account the effect of the idiotypic network (B-B interactions) (fig. \ref{fig:scheme}).
B-B interactions are mediated via ${\bf A}=\{A_{\mu\nu}\} $: 
according to the theory of idiotypic interactions \cite{Jerne,BBRaff}, 
B clones can 
recognise not only antigens but also antibodies with complementary epitopes creating a network of imitative interactions between B cells.
We represent epitopes as binary strings and assume that complementary strings, 
like e.g. $010\ldots$ and $101\ldots$, excite each other. Also, we assume that we can order the strings on a ring 
in such a way that each string sits close to affine strings and opposite to complementary ones (fig. \ref{fig:BBTOP}, right).
Hence, we suppose that the $\mu$-th B clone expansion is triggered by the ($\mu+P/2$)-th B clone, which is precisely complementary to that B 
clone (fig. \ref{fig:BBTOP}, left). While complementary B cells excite each other, we assume that each B cell suppresses itself, 
to prevent uncontrolled production of a single cell type. 
Therefore, we are led to use for the B-B interactions the Toeplitz matrix
\bea
A_{\mu\nu}=\delta_{\mu\nu}-k\delta_{\mu,(\nu + P/2)\ \mathrm{mod}\ P},
\eea
with $k\in[0,1)$ representing the strength of idiotypic interactions. 
Its inverse is 
\bea
(A^{-1})_{\mu\nu}=\frac{1}{1-k^2}\delta_{\mu\nu} + \frac{k}{1-k^2}\delta_{\mu,(\nu + P/2)\ \mathrm{mod}\ P}.
\label{eq:inverse}
\eea
We note that ${\bf A}$ is positive definite and symmetric and the following relations hold: $A_{\mu\nu}= A(\mu-\nu)$ with $A(n+P)=A(n)$ and $A(n)=A(-n)$. The eigenvalue spectrum $\rho(a)$ of ${\bf A}$ has a finite limit as $N\to\infty$, ensuring the correct scaling of the Hamiltonian.
B-antigen interactions are taken into account via the matrix $\eta_{a}^{\mu}$: antigens will excite complementary B and they will repress the identical one, in a similar way as B clones detect and excite each other as shown in fig. \ref{fig:BBA}.
We can further investigate the B-Antigen interactions analysing the stochastic process that governs the dynamics of B cells concentration.
Assuming that the B cells concentration evolves according to a gradient descent on the Hamiltonian \eqref{eq:Hamil}, we have
\bea 
\hspace*{-1.5cm}\frac{{\rm d}b_{\mu}}{\rm d t} = -\frac{\partial \mathcal{H}}{\partial  b_{\mu}}+noise = \sum_{a=1}^A\psi_{a}\eta_{a}^{\mu} +N^{\gamma-1}\sum_{i=1}^N\xi^{\mu}_i\sigma_i-\frac{1}{2\sqrt{\beta}}\sum_{\nu=1}^PA _{\mu\nu}b_{\nu} +noise.
\label{eq:bconc}
\eea 
Denoting with $\psi_\mu$ the concentration of the antigen complementary to the $b_\mu$ clone, we need the $b_\mu$ concentration 
to increase when $\psi_\mu\neq 0$. Also, we assume that clone 
$b_{\mu+p/2}$, complementary to $b_\mu$ and thus carrying the same epitope as antigen $\psi_\mu$, is inhibited by the presence of 
antigen $\psi_\mu$, and we denote $k_1$ the strength of this inhibition.
This leads to a matrix for the B-Antigen interaction in the form 
\bea
\eta_{a}^{\mu}=\delta_{\mu a}-k_1\delta_{\mu,(a+P/2)\ \mathrm{mod}\ P}
\label{eq:eta}
\eea
According to \eqref{eq:bconc}, the concentration of the $\mu$-th clone also increases in the 
presence of excitatory signals received by T cells (second term) while the 
third suppressive term is related to B-B interactions acting as a threshold to be overcome to start the immune response. 
Assuming that the total B clones concentration is conserved on average
leads to a relation between $k$ and $k_1$
\bea
\frac{\rm d}{\rm d t}\sum_\mu b_\mu=0\rightarrow\frac{1-k_1}{1-k}=\frac{1}{2\sqrt{\beta}}\frac{\sum_\nu b_\nu}{\sum_a\psi_a}
\eea
which depends on the steady state concentrations of B cells and antigens and on the inverse noise level $\sqrt\beta$. 
For simplicity we will set $k=k_1$,  
which leads to $\mbox{\boldmath$\eta$}={\bf A}$ and to an equilibrium ratio between B cells and antigens concentration only controlled by the noise level in 
the system.
\subsection{Dynamical equations}
At equilibrium at inverse noise level $\sqrt{\beta}$, 
(consistent with our assumption ${\bf b}\sim {\mathcal N}(0,{\bf A}^{-1})$) 
we expect the joint distribution $P(\bsigma,{\bf b})$
to be given by the Boltzmann distribution
\bea
P(\bsigma, {\bf b})=\frac{{\rm e}^{-\sqrt{\beta}\mathcal{H}(\bsigma, {\bf b})}}{Z}
\eea
The equilibrium marginal distribution for the $\bsigma$ is found, by 
integrating out the variables $b_\mu$, 
\bea
P(\bsigma)&=&\frac{1}{Z}\int {\rm d} {\bf b} {\rm e}^{
\sqrt{\beta}\left[
\sum_{\mu}b_{\mu}(\sum_{a}\psi_{a}\eta_{a}^{\mu} +N^{\gamma-1}\sum_i\xi^{\mu}_i\sigma_i)-\frac{1}{2} \sum_{\nu\mu}b_{\mu}A_{\mu\nu}b_{\nu}\right]}
\nonumber\\
&=&\frac{1}{Z'}{\rm e}^{\beta \left[
\frac{N^{2(\gamma-1)}}{2}\sum_{ij}\sigma_i\sigma_j\sum_{\mu,\nu}\xi^{\mu}_i(A^{-1})_{\mu\nu}\xi^{\nu}_j
+ N^{\gamma-1}\sum_{i a} \sigma_i \psi^{a}\sum_{\mu,\nu}\eta_{a}^{\mu}(A^{-1})_{\mu\nu}\xi^{\nu}_i\right]},
\label{eq:marginal_eq}
\eea 
in the Boltzmann form with effective Hamiltonian $\mathcal{H}(\bsigma)$, involving only 
interactions between T cells,
\beq
\mathcal{H}(\bsigma)=-\frac{1}{2}N^{2(\gamma-1)}\sum_{ij}\sigma_i\sigma_j\sum_{\mu,\nu}\xi^{\mu}_i(A^{-1})_{\mu\nu}\xi^{\nu}_j
+ 
N^{\gamma-1}\sum_{i a} \sigma_i \psi^{a}\sum_{\mu,\nu}\eta_{a}^{\mu}(A^{-1})_{\mu\nu}\xi^{\nu}_i
\label{eq:Hsigma}
\eeq
with the separable form $J_{ij}=N^{2(\gamma-1)}\bxi_i {\bf A}^{-1} \bxi_j$, where $\bxi_i=(\xi_i^1,\ldots,\xi_i^P)$, 
and thus describing an associative network with diluted patterns $\{\bxi^\mu\}$, 
encoding fighting strategies against different antigens \cite{jphysaas}. 
In the regime we consider here $\delta<1$, the associative network is away from saturation. 
Analysis near saturation have been carried out in statics for $\bA={\bf 1}$, mostly for
${\mathcal P}(q)=\delta(q-c)$ \cite{saturation,Peter}.
We can rewrite the Hamiltonian (\ref{eq:Hsigma}) as
\bea 
\mathcal{H}(\bsigma)=-\frac{1}{2}{\bf M}^T(\bsigma){\bf A}^{-1} {\bf M}(\bsigma)+\bpsi^T \bbeta {\bf A}^{-1} {\bf M}(\bsigma)
\eea 
in terms of the order parameters ${\bf M}=(M_1,\ldots,M_P)$, where 
\bea M_{\mu}(\bsigma)=\frac{1}{N^{1-\gamma}}\sum_{i=1}^N\sigma_i\xi^{\mu}_i
\label{eq:over}
\eea 
quantifies the strength of the excitatory signal on B clone 
$\mu$ and thus its activation. Here $\bpsi=(\psi_1,\ldots,\psi_P)$ and we denotes
$\bv^T$ the transpose of $\bv$.
Next, we assume a Glauber sequential dynamics for the variables $\bsigma$, 
converging to the equilibrium measure $P(\bsigma)$, so that the instantaneous 
probability of finding the system in state 
$\bsigma=(\sigma_1, \dots,\sigma_N)$ at time $t$ is governed by the master equation
\begin{eqnarray}
\frac{\partial{P_t(\mbox{\boldmath$\sigma$}})}{\partial{t}}= \sum_{i=1}^N\big[P_t(F_i {\bsigma})w_i(F_i {\bsigma})-P_t(\mbox{\boldmath$\sigma$})w_i(\mbox{\boldmath$\sigma$})\big]\ ,
\label{eq:master}
\end{eqnarray}
where $F_i$ is the $i$-th spin-flip operator $F_i(\sigma_1,..,\sigma_i,...,\sigma_N)= (\sigma_1,...,-\sigma_i,...\sigma_N)$
and transition rates between $\bsigma$ and $F_i\bsigma$ have the Glauber form
\begin{eqnarray}
w_i(\bsigma)= \frac{1}{2}[1-\sigma_i\tanh{(\beta h_i^{{\rm eff}}(\bsigma))}]\ ,
\label{eq:trans}
\end{eqnarray}
with effective field 
\bea 
h_i^{{\rm eff}}=N^{\gamma-1}\left( \bM^T {\bf A}^{-1}\bxi_i
+\bpsi^T \bbeta {\bf A}^{-1} \bxi_i\right)
\eea 
ensuring convergence to \eqref{eq:marginal_eq}. Here  
$T=1/\beta$ represents the rate of spontaneous spin-flip, hence 
the effective noise of the $T$ cell dynamics.
Using \eqref{eq:master}, we write the dynamics for the B cells activation, described by the probability of finding the system in a macroscopic state ${\bf M}$ at 
time $t$, namely  
\begin{eqnarray}
P_t({\bf M})= \sum_{\bsigma} P_t(\mbox{\boldmath$\sigma$})\delta({\bf M}-{\bf M}(\bsigma)).
\end{eqnarray}
Via a Kramers-Moyal expansion for large system size and away from the saturation regime it is possible to show that $P_t({\bf M})$ 
evolves according to a Liouville equation \cite{jphysaas}. It follows that ${\bf M}$ evolve deterministically with a dynamics described by
\bea 
\frac{{\rm d }{\bf M}}{{\rm d} t}=\bigg\langle\bxi^T\tanh\left[\beta\left(\bM^T {\bf A}^{-1} \bxi+
\bpsi^T {\bf C}\bxi\right)\right]\bigg\rangle_{\bxi} - {\bf M}
\label{eq:m}
\eea
where 
${\bf C}=\bbeta {\bf A}^{-1}$ and $\bra \cdot\ket_\bxi$ denotes the average over the distribution 
(\ref{eq:xdistrib}).
For $\mbox{\boldmath$\eta$}$ defined in \eqref{eq:eta} and $k=k_1$, we have ${\bf C}={\bf 1}$, and each order parameter $M_\mu$ evolves accordin
\bea 
\frac{{\rm d }M_{\mu}}{{\rm d} t}=\bigg\langle\xi^{\mu}\tanh\bigg[\beta(\bM^T {\bf A}^{-1} \bxi + 
\bpsi^T \bxi)\bigg]\bigg\rangle_{\bxi} - M_{\mu}\ .
\label{eq:mm}
\eea
\begin{figure}[htb!]
\centering
\includegraphics[width=0.4\textwidth]{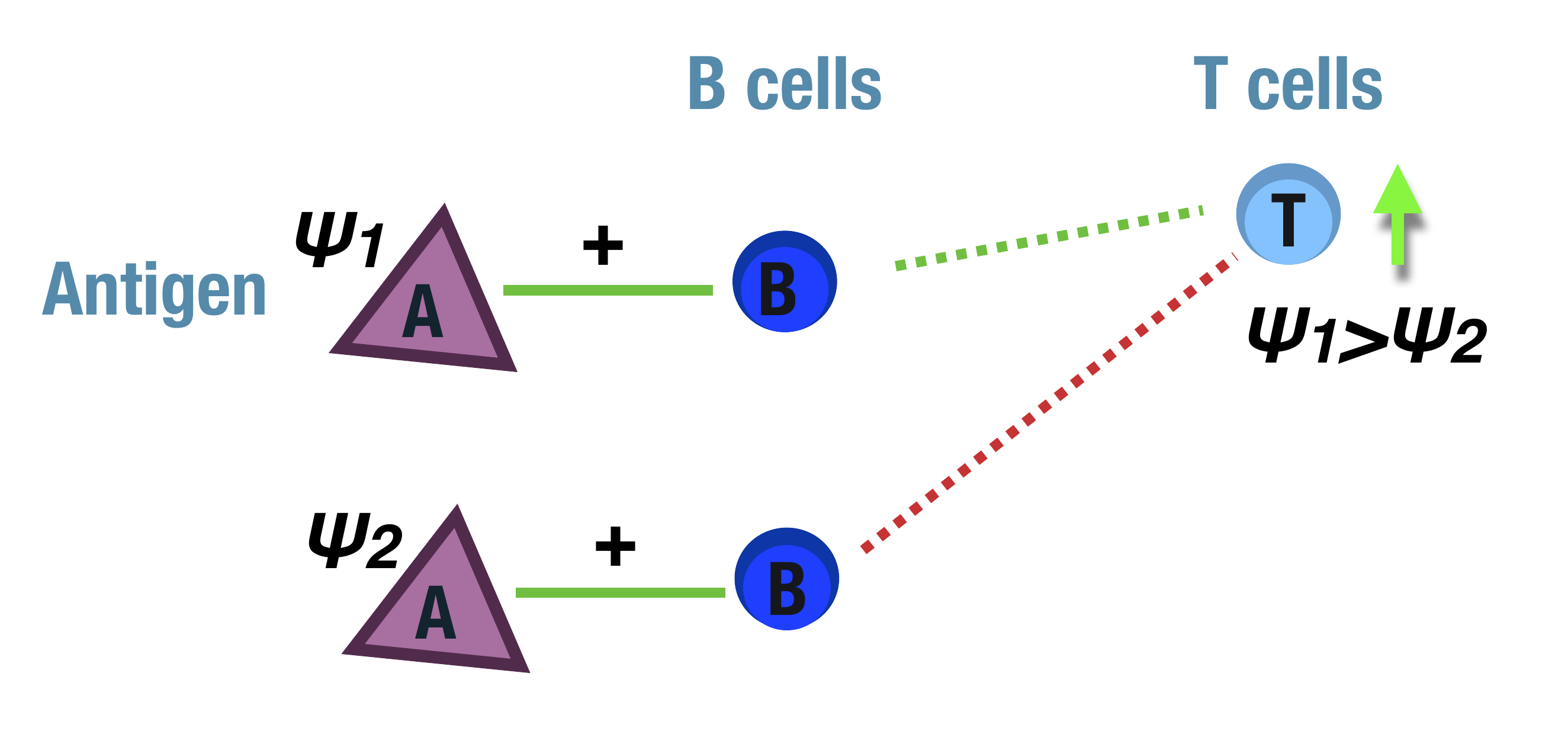}
\includegraphics[width=0.4\textwidth]{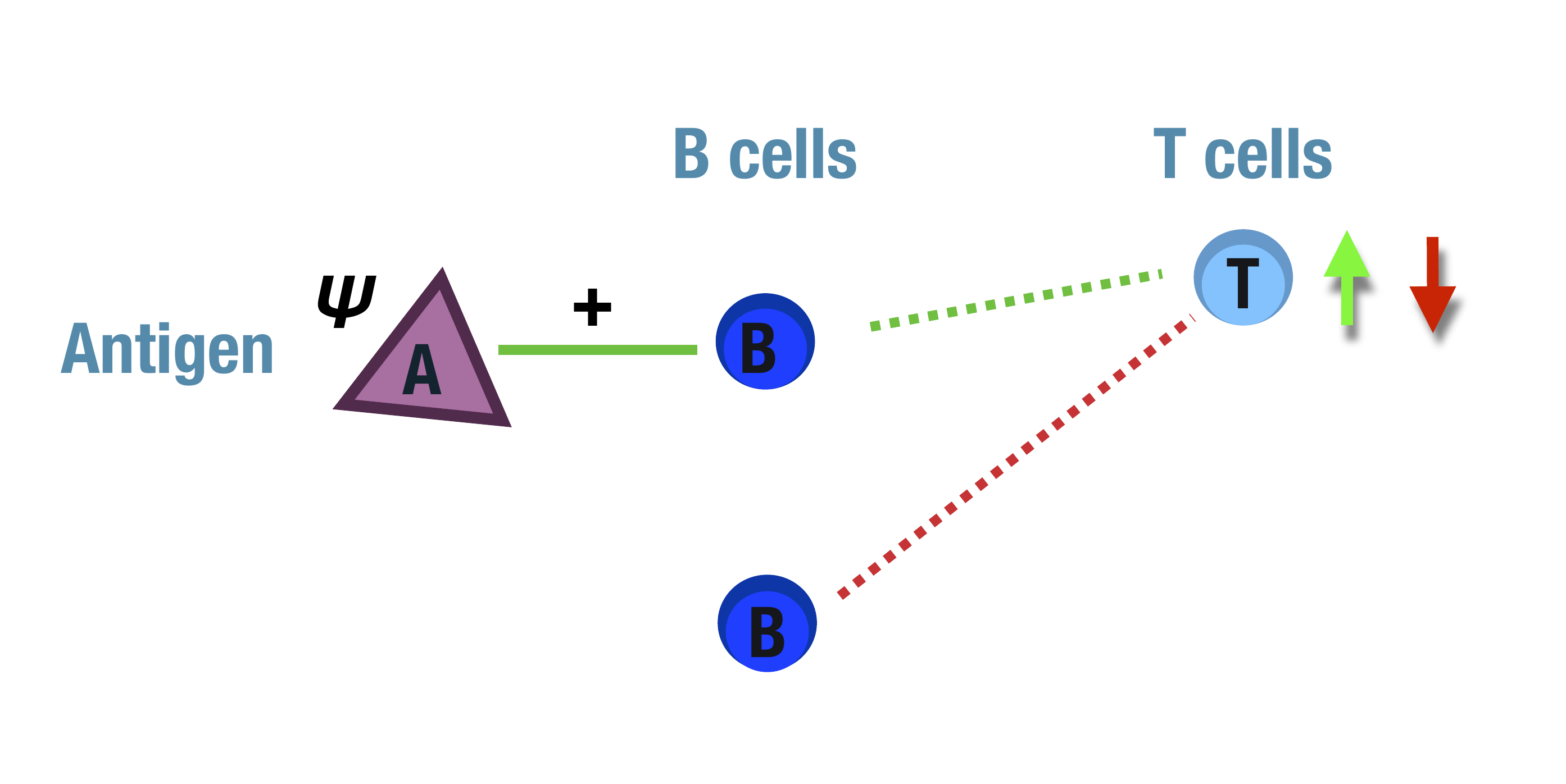}
\caption{Left: effect of having ${\bf C}= {\bf 1}$, in the presence of two antigens whose complementary B cells are signalled by different cytokines from the same 
T cell:
the field acting on the T cell is $\psi_1-\psi_2$ and T gets activated only if $\psi_1>\psi_2$. 
Right: an illustration of the effect of having ${\bf C} = {\bf A}^{-1}$, which may result in an inhibitory signal on T cells from inactive  B cells.}
\label{fig:2antig}
\end{figure}
Different choices of the matrix $\bbeta$ or of the constants $k, k_1$ will lead to different forms of the 
matrix ${\bf C}$, however the choice ${\bf C} = {\bf 1}$ seems to ensure the strongest 
excitatory signal to antigen-activated B clones. As an illustration, let us consider the simple case where we have just one 
antigen {\em i.e.} $\psi_1$: for the choice ${\bf C}={\bf 1}$ the field acting on the 
$i$-th T cell exciting the complementary clone $b_1$ is $\psi_1$, 
hence $i$ receives an activation signal only from the 
clone B activated by the antigen. 
With two antigens present,  
whose complementary B cells are signalled by different cytokines from the same 
T cell, the field acting on the latter is $\psi_1-\psi_2$ and the T cell 
will get activated only if $\psi_1>\psi_2$, as biologically desired (fig. \ref{fig:2antig}, left).
For ${\bf C}\neq {\bf 1}$ there might be a negative interference on the desired signal.
For example, for the alternative choice $\bbeta={\bf 1}$, leading to ${\bf C} = {\bf A}^{-1}$,
the field acting on T in the presence of a single virus $\psi_1\neq 0$
would be $\frac{\psi_1}{1-k^2}[\xi^1_i + k\xi^{1+\frac{P}{2}}]$, 
meaning that for $\xi^1=+1$, $\xi^{1+\frac{P}{2}}=-1$ the 
T cell would receive, 
besides the excitatory signal from the B cell activated by the antigen, an inhibitory signal from a 
non-activated B-cell (fig. \ref{fig:2antig}, right). As a result the overall field and immune 
response will decrease.
Similar arguments can be given for $\bbeta$ defined in (\ref{eq:eta}) with $k_1\neq k$.

\section{Overview of results}\label{sec:over}
The sparsity of the B-T connections makes the system able to activate multiple B clones in parallel. This multitasking capability is one of the core 
features of the immune system that, in normal conditions, can control and block several simultaneous antigenic invasions. We find that the 
parallel activation of B clones may occur in \emph{symmetric} fashion, 
where all infections are fought with the same strength, or in  
a \emph{hierarchical} fashion, where the system prioritises immune responses against specific pathogens. We are able to identify the system's parameters, 
such as noise, 
number of links, number of different infections, which induce the switch from the symmetric to the hierarchical operational mode. This might be potentially 
useful to 
investigate the causes of dramatic failures in the functioning of the immune system. The switch to a hierarchical immune response is in fact mostly related to the 
presence of strong infections, such as hepatitis and autoimmune disorders: while the immune system invests the highest amount of resources tackling the main disease, 
the progression of minor infections may become lethal \cite{inf,inf2}.

An important feature of the model is the 
dependence of the B cells activation on the number of receptors on their surface.
Results in Sec. \ref{sec:q} show that cells with few receptors may be transiently activated but fail to sustain the signal even if strongly 
excitatory (fig. \ref{fig:qmu}). The number of triggered receptors affects both the immune response strength and the critical temperature at which they 
get activated: both decrease with the number of triggered receptors. 
In addition, competition emerges between B clones to be activated. As the number of activated B 
clones increases, signalling pathways to 
inactive clones get more noisy due to the interference of active clones.
The critical temperature for the activation of inactive clones is a decreasing function of the fraction of active clones \eqref{eq:critican+1}. 

Another result of biological interest is that idiotypic {\em i.e.} 
B-B interactions contribute to the overall stability of the immune system, 
preventing unwanted activation and increasing the region where all clones are equally activated and 
ready to start an immune response upon arrival of new infections (Sec. \ref{sec:BB}). In particular, including B-B interactions in the model affects the critical temperature, 
in this case widening the region where a symmetric immune response is stable, as the interactions strength $k$ is increased (fig. \ref{fig:BBmedphase}). 

Finally, the role of antigens in B cells activation is understood by our model as that of external fields on coupled ferromagnetic systems \ref{sec:ant}. 
One of the interesting consequences is then the 
presence of hysteresis phenomena \cite{CWfield}, which could explain short-term memory effect in the immune response \cite{memory,Martin}, even in the absence of memory cells.
The effect of antigens on the immune system {\em basal activity} 
and surveillance is also of interest: the response of non-infected B clones decreases with the fraction of infected clones, due to the interference of strongly activated B 
cells, and is activated at a lower $T$, making the whole system unresponsive to new incoming viruses (fig. \ref{fig:m2noninf}).

\section{Effects of receptors' promiscuity}\label{sec:q}
We first consider the case where there is no antigen $\psi_a=0 ~\forall~a $, and no B-B interactions {\em i.e.} ${\bf A}={\bf 1}$ and focus on the 
effect of having a variable number of triggered receptors on different B clones, {\em i.e.} heterogeneous $q_\mu$ in \eqref{eq:xdistrib}. 
In order to compare activation of B clones with different numbers of 
receptors,  
it is convenient to look at the activation per receptor, given by the 
normalised order parameters 
$m_\mu=M_\mu/q_\mu$, $\mu=1,\ldots,P$, which take values in the range $[-1,1]$ for all clones $\mu$. 
The dynamical equations \eqref{eq:mm} then read
\bea 
\frac{{\rm d }m_{\mu}}{{\rm d} t}=\frac{N^{\gamma}}{q_\mu}\bigg\langle\xi_{\mu}\tanh\left(\beta \sum_{\nu}q_{\nu}\xi^{\nu}m_{\nu}\right)\bigg\rangle_{\xi}- m_{\mu}\ .
\label{eq:qm}
\eea 
One can show that at the critical temperature
$T_c=q_{{\rm max}}$, where $q_{{\rm max}}=\max_\mu [q_{\mu}]$, the system undergoes a phase transition, with  
the equilibrium phase at $T>T_c$ characterised by ${\bf m}={\bf 0}$, and B clones activation occurring at 
$T<T_c$, where ${\bf m}\neq {\bf 0}$. 
At the steady state (${\rm d}{\bf m}/{\rm d}t =0$), we have 
\bea
M_{\mu}= N^{\gamma}\langle\xi^{\mu}\tanh(\beta\sum_{\nu}\xi^{\nu} M_{\nu})\rangle_{\bxi}\ .
\eea
Taking the scalar product with ${\bf M}$ and using 
the inequality $|\tanh x|\leq |x|$, yields
\bea
{\bf M}^2 \leq N^{\gamma} \beta \sum_{\mu}M_{\mu}\sum_{\nu} M_{\nu}\langle\xi^{\nu}\xi^{\mu}\rangle_{\bxi}
=\beta \sum_{\mu}q_{\mu}M_{\mu}^2 \leq \beta q_{max}{\bf M}^2\ .
\eea
This implies ${\bf M}^2=0$, hence ${\bf m}=0$, for $\beta q_{{\rm max}}<1$, 
meaning that none of the B cells can get 
activated for noise levels above the critical value $T>q_{{\rm max}}$. 
Although ${\bf m}=0$ is a steady state solution of 
(\ref{eq:qm}) for any value of $T$, a linear stability analysis 
shows that it becomes 
unstable for $T<q_{\rm max}$. To this purpose, we compute the Jacobian of the 
linearised dynamics about the steady state ${\bf m}^\star={\bf 0}$ 
\bea
J_{\mu \nu}=\frac{\partial F_{\mu}^{(1)}({\bf m})}{\partial m_{\nu}}\bigg|_{{\bf m}={\bf m}^\star},
\quad\quad\quad F_{\mu}^{(1)}({\bf m})=\frac{N^\gamma}{q^{\mu}}\langle \xi^{\mu}\tanh(\beta
\sum_{\nu=1}^P\xi^{\nu} q^{\nu} m_{\nu})\rangle_{\bxi} - m_{\mu}
\label{eq:Jacobian}
\eea 
which gives
\bea
J_{\mu \nu}= \frac{N^{\gamma}\beta q_{\nu}}{q_{\mu}}\langle \xi^{\nu}\xi^{\mu}[1-\tanh^2(\beta\sum_{\nu} \xi^\nu q_{\nu}m^\star_{\nu})]\rangle_{\bxi}-\delta_{\mu\nu}.
\label{eq:JacobianM}
\eea
Substituting ${\bf m}^{\star}={\bf 0}$ we get 
$J_{\mu \nu}= (\beta q_\mu-1)\delta_{\mu\nu}$. 
This is a diagonal matrix, hence the largest eigenvalue, which gives the 
stability of ${\bf m}^\star=0$, 
is $\lambda_{\rm max}=\beta q_{\rm max}-1$. This gets positive for $\beta q_{\rm max}>1$, showing that non-zero solutions ${\bf m}\neq 0$ will bifurcate away 
from ${\bf m}=0$ at $T=q_{\rm max}$. We inspect the structure and the stability of the bifurcating solutions first for a toy model with just two B-clones 
and then for the general case with $P$ B-clones.

\subsection{A toy model with two B-clones}
\begin{figure}[htb!]
\centering
\includegraphics[width=0.4\textwidth]{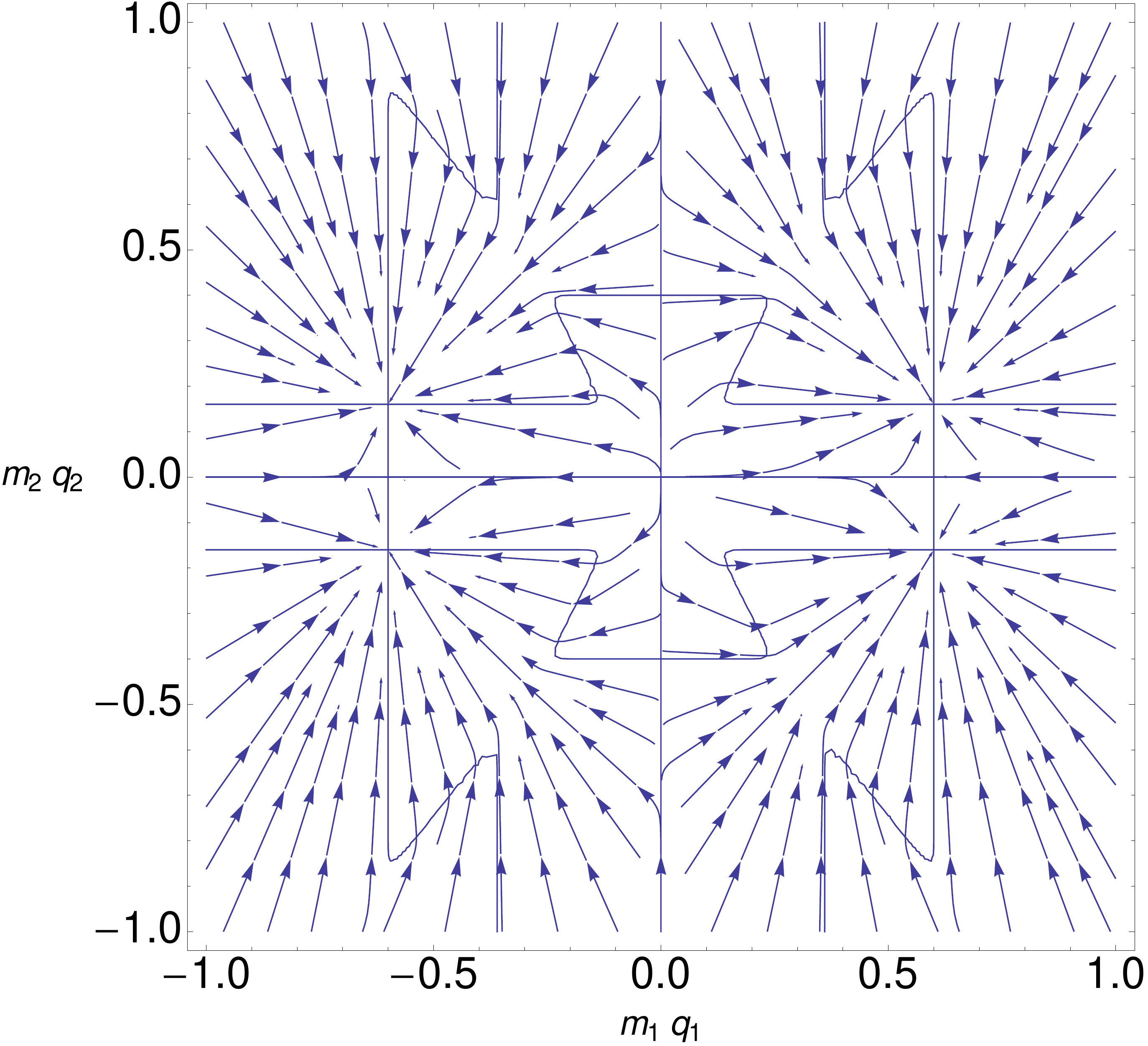}
\includegraphics[width=0.4\textwidth]{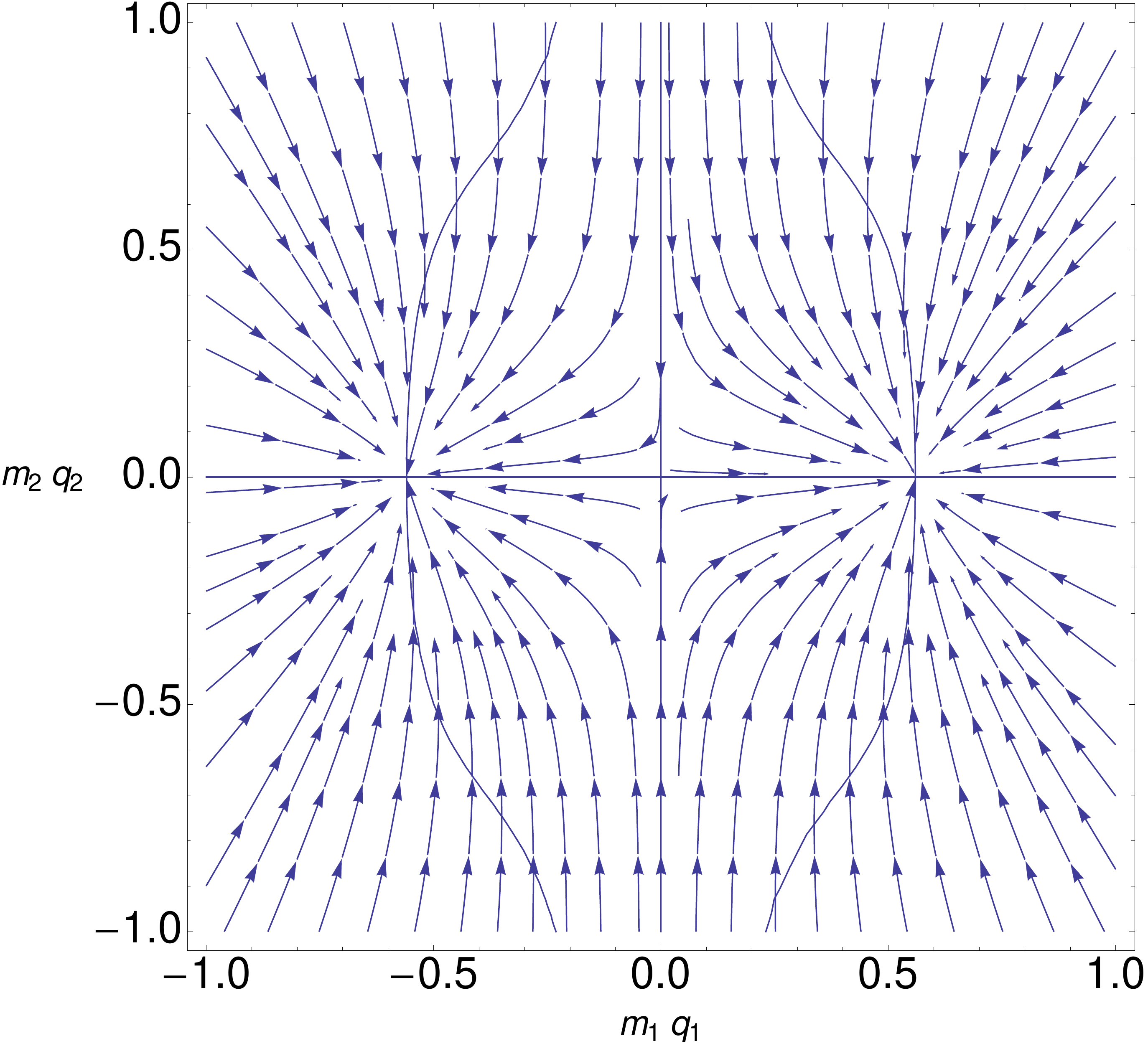}
\caption{Phase portrait of the dynamical system (\ref{eq:dm1}, \ref{eq:dm2}) 
for $q_1=0.6$, $q_2=0.4$ at low temperature $T=0.01$ (left) and high temperature $T=0.2$ (right). 
Red lines represent null-clines and stationary states are at the intersection of null-clines.}
\label{fig:qmu}
\end{figure}

\begin{figure}[htb!]
\centering
\includegraphics[width=0.4\textwidth]{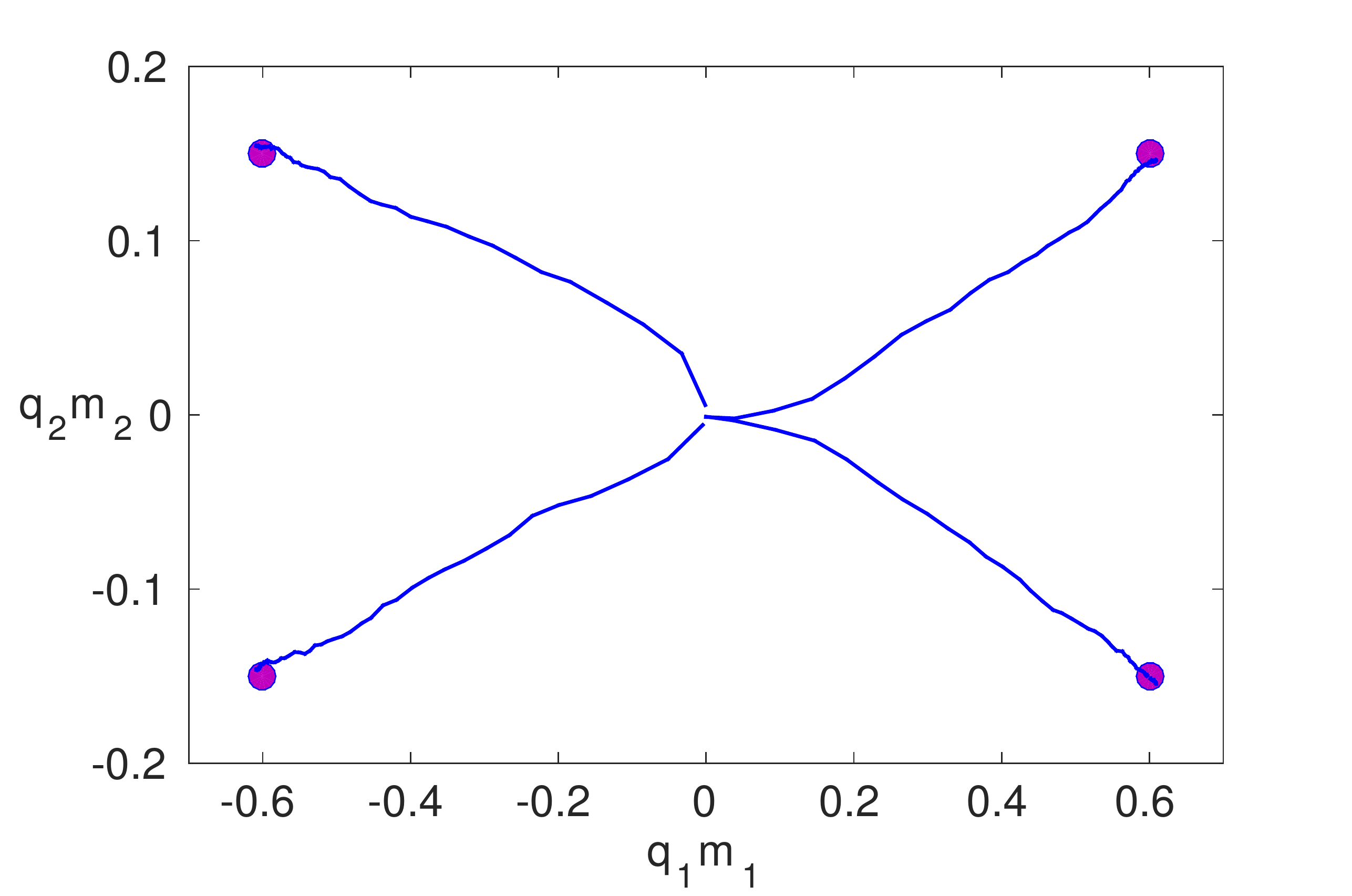} 
\includegraphics[width=0.52\textwidth]{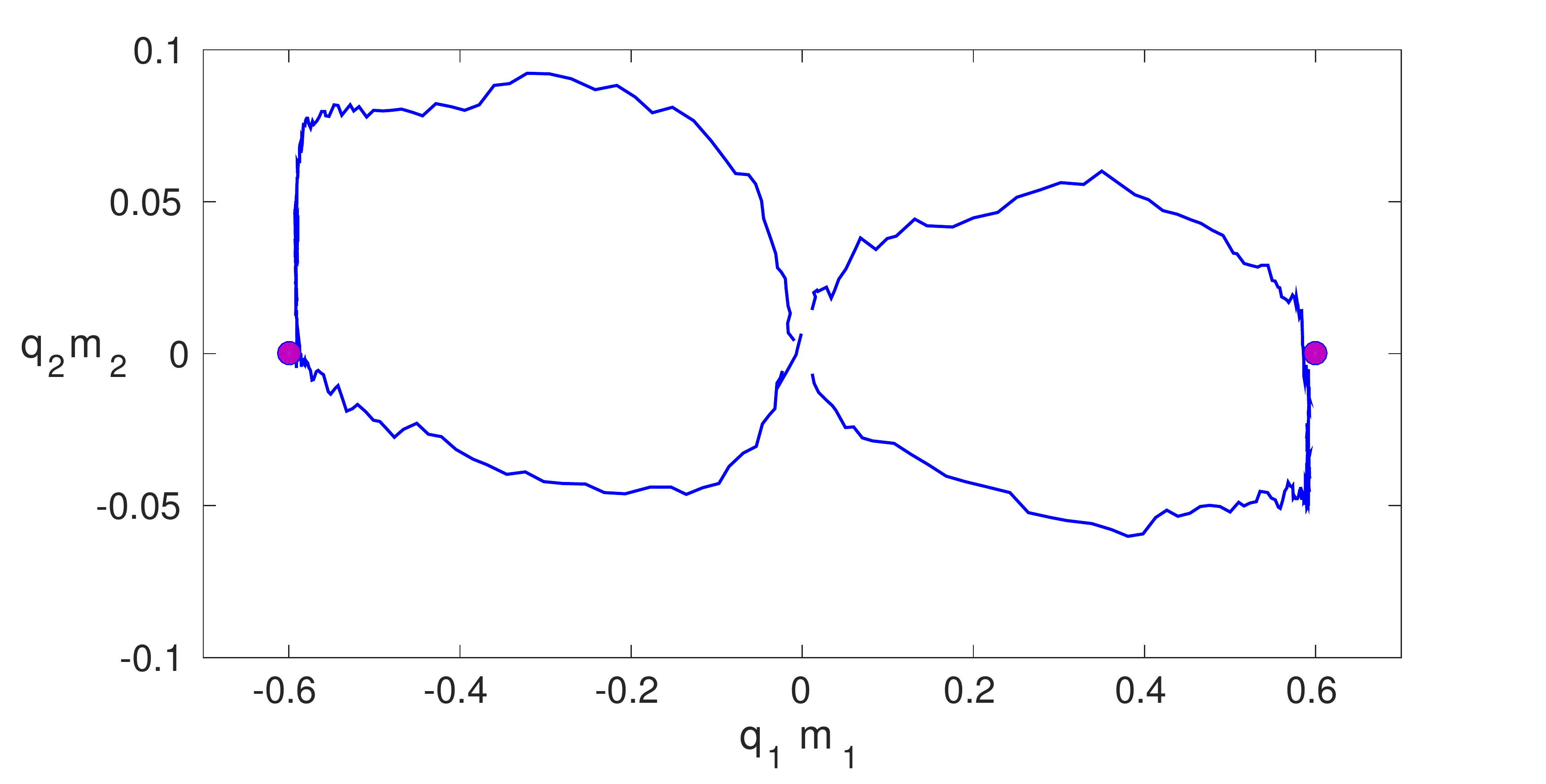} 
\caption{Monte Carlo simulations with $10^4$ spins for $q_1=0.6$, $q_2=0.4$ at low temperature $T=0.01$ (left) and at high temperature $T=0.2$ (right). 
Markers represent the numerical solutions of (\ref{eq:dm1}, \ref{eq:dm2}) .}
\label{fig:sim}
\end{figure}

\begin{figure}[htb!]
\centering
\includegraphics[width=0.5\textwidth]{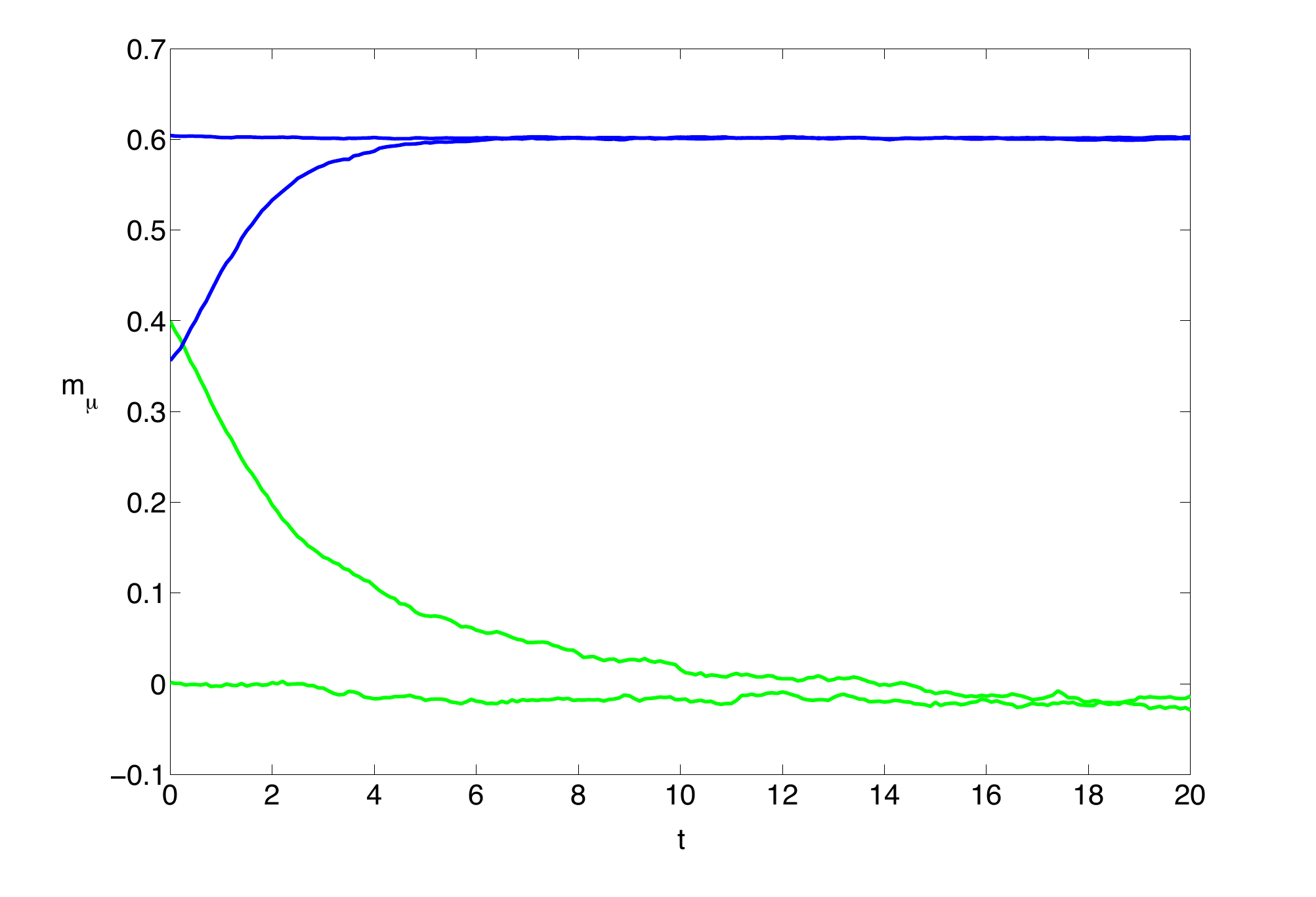}
\caption{Monte Carlo simulations with $N=10^4$ spins, with $q_1=0.6$, $q_2=0.4$ at $T=0.2$.  Clone activation $m_1$, $m_2$ as a function of time for different initial conditions. Cells with few receptors, $q_2$ (green) fail to be activated even if triggered with a strong signal.}
\label{fig:different}
\end{figure}

\begin{figure}[ht!]
\centering
\includegraphics[width=0.4\textwidth]{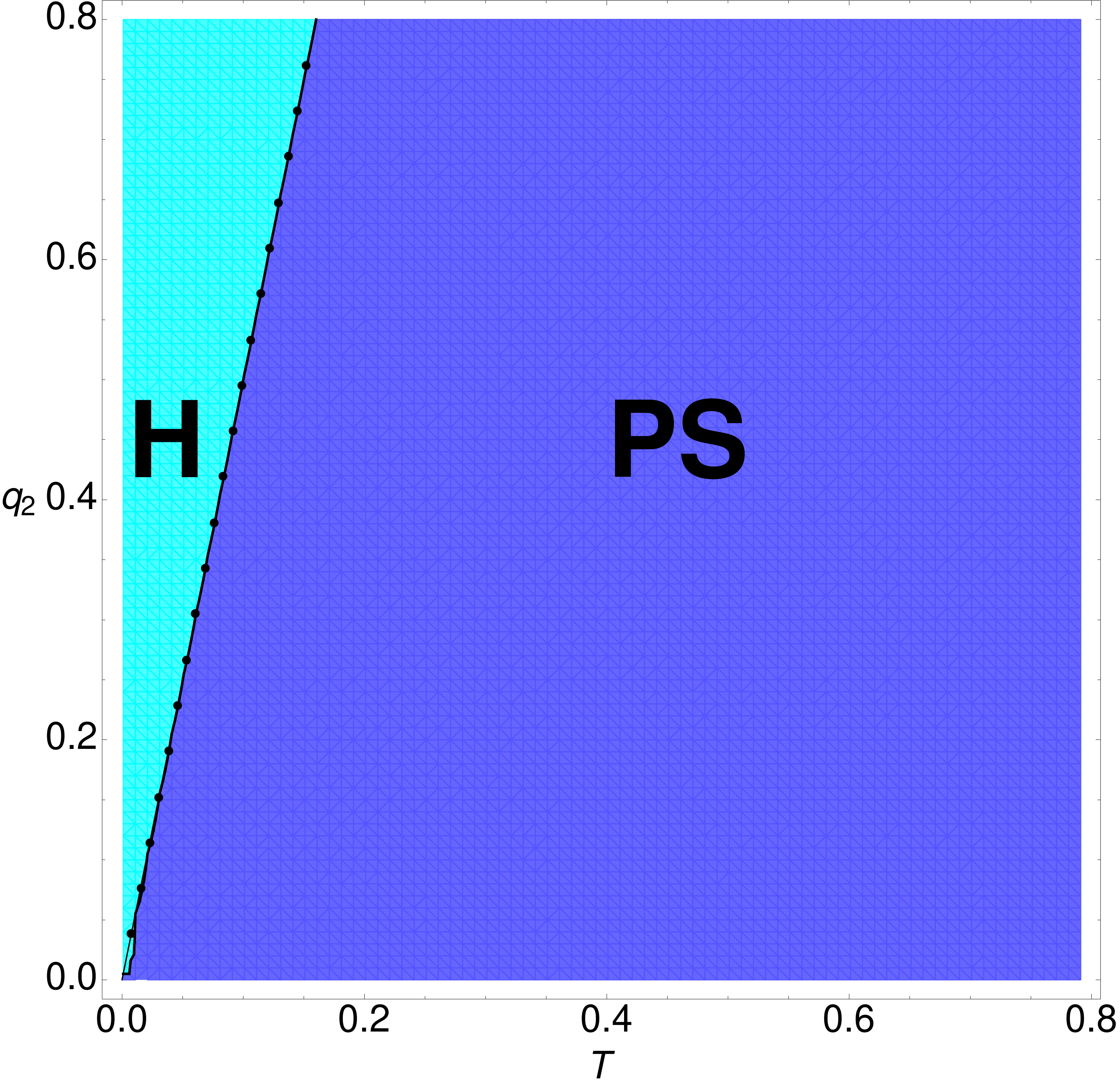}
\caption{Phase diagram in the space $(T,q_2)$ fixing $q_1=0.8$ obtained from the condition $\lambda_2<0$ \eqref{eq:eigenpure}. The dotted line represents the theoretical critical temperature \eqref{eq:tclow}.
In the ($\bf PS$) region the pure state is stable ($m_1\neq0, m_2=0$). At low $T$, B clones are hierarchically activated ({\bf H}).}
\label{fig:phasepure}
\end{figure}

\begin{figure}
\centering
\includegraphics[width=0.48\textwidth]{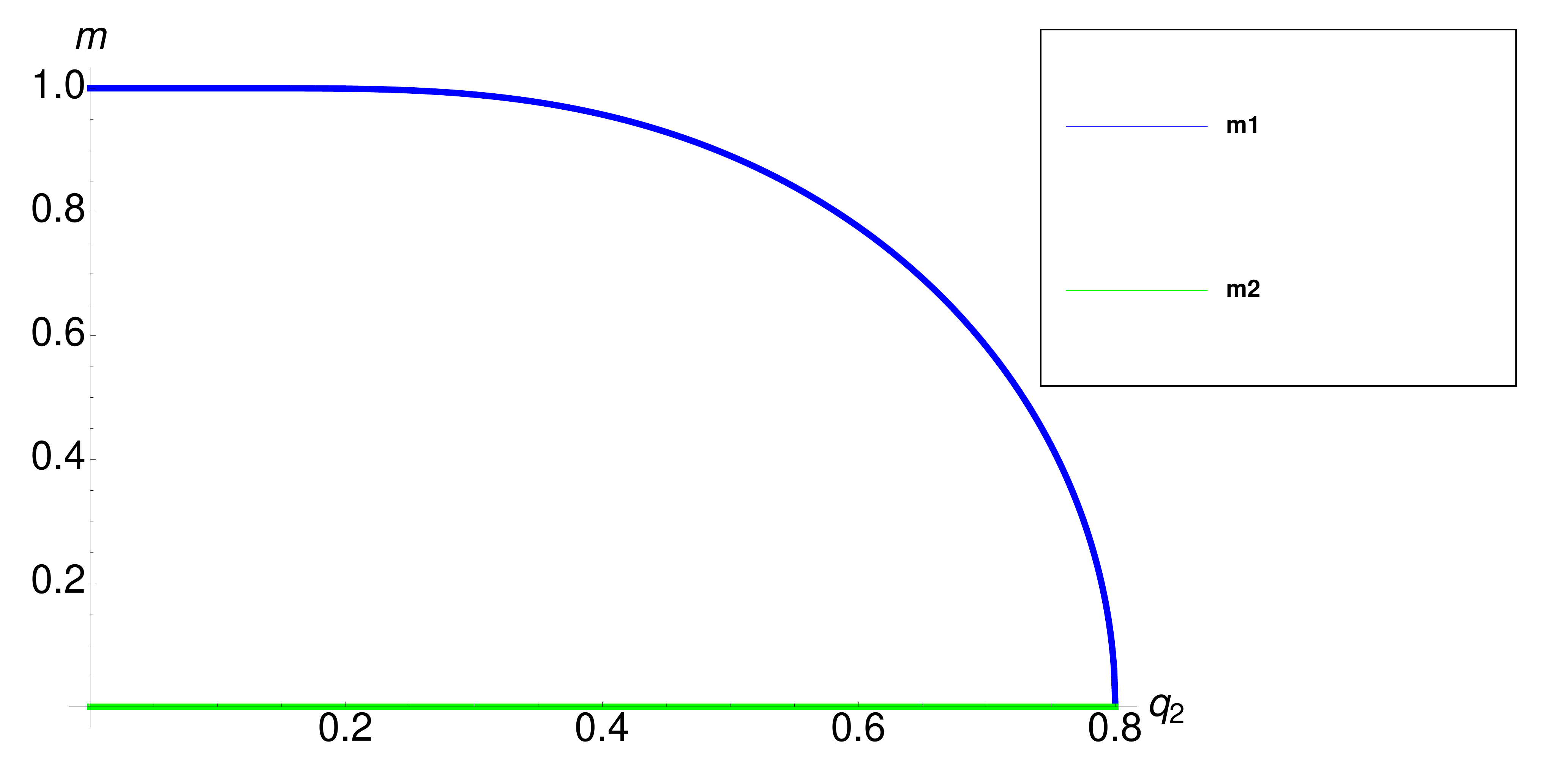}
\includegraphics[width=0.48\textwidth]{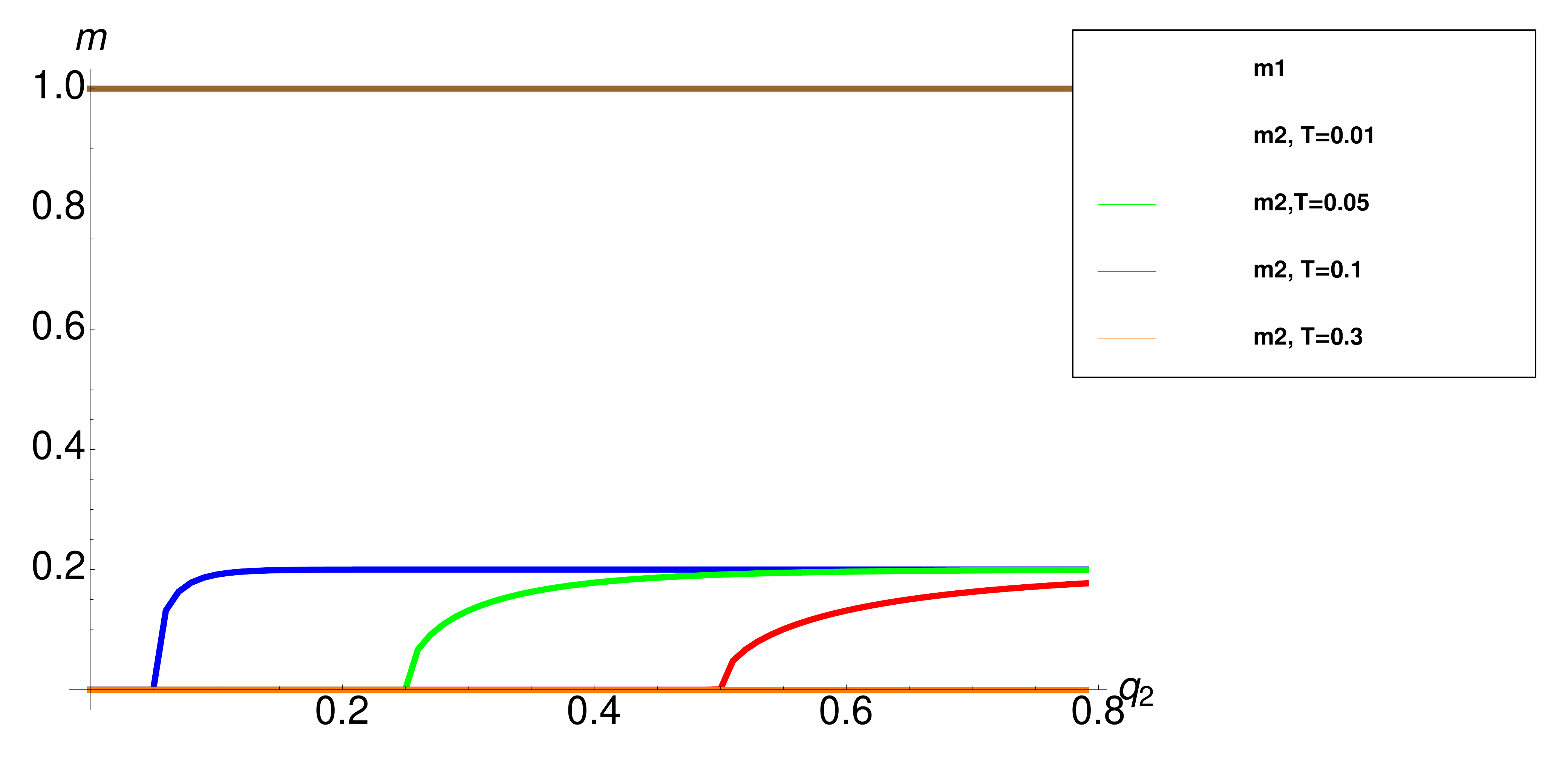}
\caption{Plot of $m_1, m_2$ as a function of $q_2$.  Left : $q_1=0.8$ at $T=q_2$,  Right: $q_1=0.8$, $T=0.01, 0.05, 0.1, 0.3$. }
\label{fig:m1critica}
\end{figure}

Here we study a toy model with $P=2$ B clones, and assume $q_1>q_2$. For $\gamma>0$ this model
reduces to two independent Curie-Weiss ferromagnets, with critical temperatures $q_1$ and $q_2$ respectively (see discussion 
in Sec. \ref{sec:qmed}). Hence, the most interesting case is obtained for $\gamma=0$.
The state ${\bf m^{\star}}=(0,0)$ is the only steady state of the system for $T>q_1$, but it destabilises for $T<q_1$.  
We can understand the system's behaviour below criticality by solving numerically its dynamical equations
\begin{align}
\hspace*{-2cm}
\frac{{\rm d }m_1}{{\rm d} t
} &= (1-q_2)\tanh(\beta q_1m_1) +\frac{q_2}{2}[\tanh(\beta(q_1m_1+q_2m_2)) +\tanh(\beta(q_1m_1-q_2m_2)) ] -m_1\ ,
\label{eq:dm1}
\\
\frac{{\rm d }m_2}{{\rm d} t
}&= (1-q_1)\tanh(\beta m_2) +\frac{q_1}{2}[\tanh(\beta(q_1m_1+q_2m_2)) -\tanh(\beta(q_1m_1-q_2m_2) )] -m_2\ .
\label{eq:dm2}
\end{align}
In fig. \ref{fig:qmu} we show the flow diagram and the stable fixed points 
at different temperatures: first we notice that B 
cells with a higher promiscuity produce a higher immune response ($m_1$), 
whereas a lower promiscuity $q_{\mu}$ results in a lower or null activation 
($m_2$) depending on the
temperature. Hence, the number of receptors on B cells' surface 
affects the responsiveness of B clones. 
In particular, if $T$ is high, only clones with the highest number of 
receptors are activated and the system's fixed point corresponds to the 
pure state $m_1\neq 0, m_2=0$. Lowering $T$ induces the activation of cells with fewer receptors but with a lower intensity ($m_1>m_2$).
Theoretical results are consistent with Monte Carlo simulations, shown in fig. \ref{fig:sim}. Furthermore, Monte Carlo 
simulations match  
experimental results \cite{ScienceT} showing that cells with very few receptors are triggered transiently but fail to be activated in the long run, since the 
number of receptors is not sufficient to allow these cells to sustain the signal, even if strongly excitatory. 
Fig. \ref{fig:different} shows that our model can reproduce this effect: 
cells with few receptors (green) have a lower activation even if triggered by a strong signal (initial condition) and tends to be switched 
off after a short transient, conversely cells with a higher number of 
receptors (blue) produce a strong immune response, even if triggered by a 
low signal 
(initial condition).
Analytically, we can 
investigate the structure of the first states to bifurcate below $T_c$
by expanding the steady state equations, obtained setting 
${\rm d}m_{1,2}/{\rm d} t=0$
in (\ref{eq:dm1}, \ref{eq:dm2}), for small $m_1, m_2$, close to criticality, 
{\em i.e.} at $\beta q_1= 1+\epsilon$. We get as possible solutions 
\bea 
m_1^2=3\epsilon-3\frac{q_2^3}{q_1^2}m_2^2\label{eq:q1smallm1}\ ,\\
m_2^2=3\frac{q_1^2}{q_2^2}(1+\epsilon)-3\frac{q_1^3}{q_2^3}-3\frac{q_1^3}{q_2^2}m_1^2 \ ,
\label{eq:q1smallm2}
\eea
as well as $m_1=0, m_2=0$. Solution (\ref{eq:q1smallm2}) is unphysical as it stays $\order{(1)}$ at $\epsilon=0$ for $q_1\neq q_2$, hence 
$m_2=0$. Inserting $m_2=0$ in \eqref{eq:q1smallm1}, we get $m_1^2=3\epsilon$.
Hence, the first state bifurcating away from ${\bf m}=(0,0)$ is in the form ${\bf m}= (\sqrt{3\epsilon},0)$. 
At high temperature (below criticality) only the clone with the higher number 
of receptors is switched on. 
For $q_1=q_2$ we clearly retrieve the results in \cite{jphysaas} and both B clones are activated with the same intensity below criticality.
Next we derive the region in the phase diagram 
where cells with fewer receptors become responsive. 
Naively one might expect that $m_2$ becomes active at $T\simeq q_2$. 
In reality, heterogenities in the cell promiscuities deeply affects cell 
responsiveness and cells with fewer receptors will remain quiescent even at very low $T$. 
By Taylor expanding \eqref{eq:qm} for small $m_2$ in powers of 
$\epsilon=q_2\beta-1$, we obtain
\bea
m_1=\tanh(\beta q_1 m_1)\simeq\tanh\bigg(\frac{q_1}{q_2} m_1\bigg)\ ,\\
m_2 ^2 = 3\epsilon-3 \frac{q_1^3}{q_2^2} m_1^2\ .
\eea
Since $m_1$ is $\mathcal{O}(1)$ the non-zero solution for $m_2$ is impossible close to $T=q_2$. 
The pure state ${\bf m} =(m_1,0)$ will then have a wider stability region, 
that can be found 
by analysing the eigenvalues of the Jacobian \eqref{eq:Jacobian} at ${\bf m^\star} =(m_1,0)$,
\bea
\lambda_1 = \beta q_1 -  \beta q_1\tanh^2( \beta q_1m_1)-1\ ,\\
\lambda_2 =  \beta q_2-  \beta q_1 q_2 \tanh^2( \beta q_1m_1)-1\ .
\label{eq:eigenpure}
\eea
Analytically we can calculate $\lambda_{1,2}$ near $T\simeq 0$
\bea
\lambda_1\simeq-1\ ,\\
\lambda_2\simeq\beta q_2 -\beta q_1 q_2 -1\ .
\label{eq:limit0}
\eea
and near $T_c$, {\em i.e.} at $\beta q_1= 1+\epsilon$
\bea
\lambda_1\simeq -2\epsilon \ ,\\
\lambda_2\simeq \frac{q_2\epsilon +q_2-3\epsilon q_1q_2-q_1 }{q_1} +\mathcal{O}(\epsilon^2)\ .
\label{eq:tcexp}
\eea 
\begin{figure}[htb!]
\centering
\includegraphics[width=0.4\textwidth]{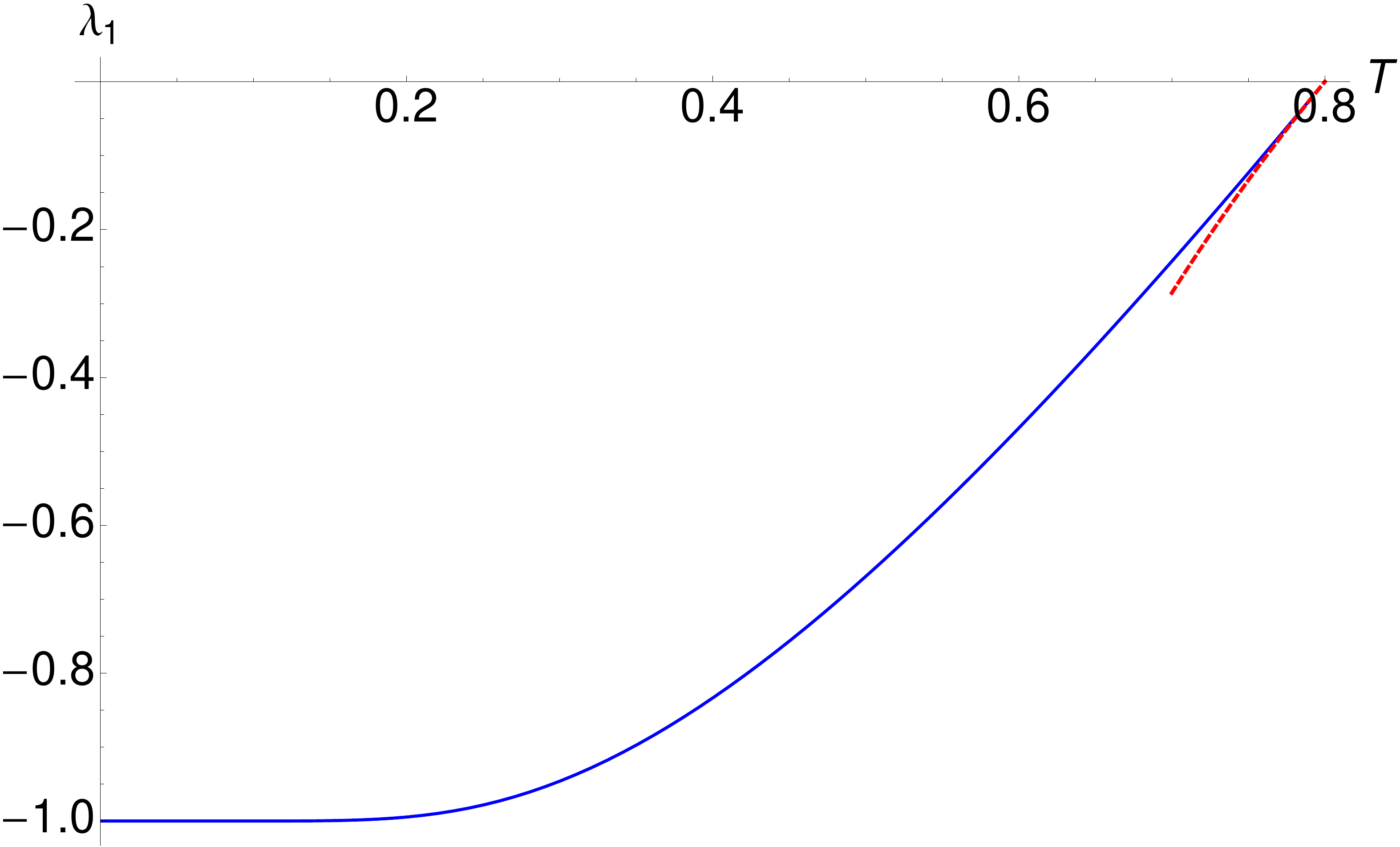}
\includegraphics[width=0.38\textwidth]{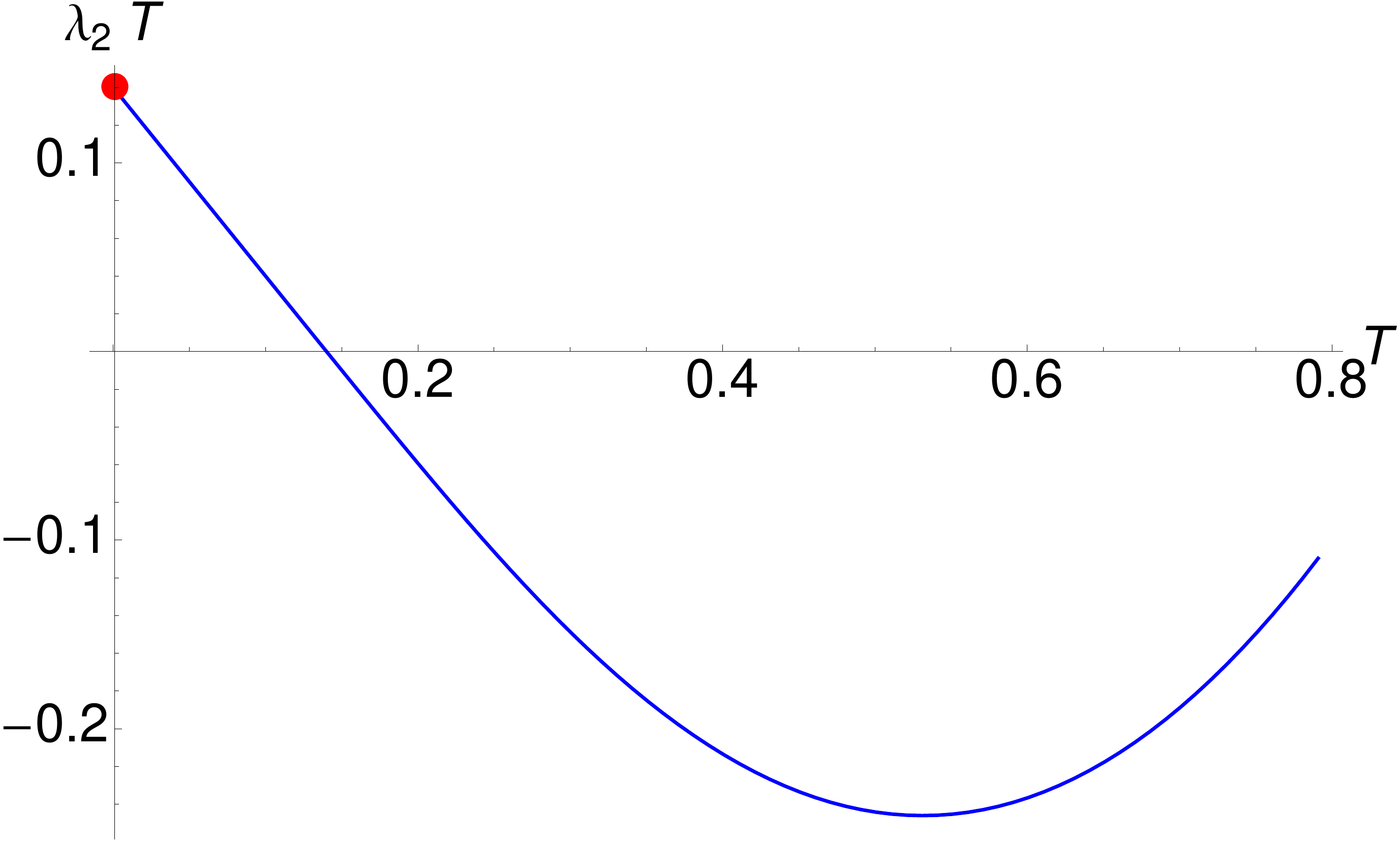}
\caption{Eigenvalues \eqref{eq:eigenpure} as a function of $T$  for $q_1=0.8$ and $q_2=0.7$. 
Left: $\lambda_1$, the red dashed line represents the behaviour near $T\simeq T_c$ \eqref{eq:tcexp}. 
Right: $\lambda_2 T$, the red marker represents the limit at $T\simeq 0$ \eqref{eq:limit0}.}
\label{fig:eigensec}
\end{figure}
For intermediate $T$ we compute $\lambda_{1,2}$ numerically. Plots of eigenvalues as a function of the temperature are shown in fig. \ref{fig:eigensec} where the theoretical predictions for $T\simeq0$ \eqref{eq:limit0} and $T\simeq T_c$ \eqref{eq:tcexp} are highlighted.
We note that  $\lambda_1<0, \forall~ T$, hence the stability of the pure state is determined by the sign of $\lambda_2$.
 In fig. \ref{fig:phasepure} we show a contour plot of $\lambda_2=0$ in the $T-q_2$ plane 
for different values of $q_1$.
The linear behaviour can be understood as follows. 
In the pure state region we have, using the steady state equation 
$m_1=\tanh (\beta q_1 m_1)$,
\bea
\lambda_1= \beta q_1 (1-m_1^2)-1\ ,\\
\lambda_2=\beta q_2 (1-q_1 m_1^2)-1\ .
\eea
As $T$ decreases (below $q_2$) $m_1$ increases, so that 
the eigenvalues stay negative and 
stability is 
ensured, until $m_1$ reaches its maximum value $m_1=1$. At this point,  
a further decrease of the temperature will make $\lambda_2$ positive,  
destabilising the pure state, and a new state $(m_1,m_2)$ will take over.
The temperature at which bifurcations from the pure state are expected 
is thus found from the condition $\lambda_2(T,m_1=1)= 0$, as 
\bea
 T= q_2(1-q_1)
 \label{eq:tclow}
 \eea
which is in agreement with the critical temperature computed numerically 
(fig. \ref{fig:phasepure}). Deviations from the linear behaviour are expected 
in the regime $q_1\simeq q_2$,
where a symmetric activation occurs for $q_1\leq 1/3$ \cite{jphysaas}.
In fig. \ref{fig:m1critica} we plot $m_1,m_2$ as a function of $q_2$
in the ({\bf PS}) region (left) and
crossing the critical line where $m_2$ becomes non-zero (right).
In the latter region, {\em i.e.} for $q_2>T/(1-q_1)$, the stable state is 
${\bf m}=(1,m_2)$ where $m_2$ is the $T$-dependent 
stationary solution of (\ref{eq:dm2}) at $m_1=1$.
In particular, for $T=0$ the stable state is 
 \bea
 m_1=1\ ,\\
 m_2=1-q_1\ ,
 \eea
in agreement with simulations and flow diagrams 
(fig. \ref{fig:qmu}, \ref{fig:sim}).
In conclusion, the system can activate 
clones with different numbers of receptors simultaneously for 
$T<q_2(1-q_1)$. The activation is hierarchical ({\bf H}), with clones 
with higher promiscuity being prioritised with respect to the others. 
In particular, clones with the highest number of receptors
are activated with the strongest possible 
signal in a wide region of the phase diagram.

\subsection{The case of $P$ B clones with a variable number of receptors}\label{sec:qmed}
\begin{figure}[htb!]
\centering
\includegraphics[width=0.48\textwidth]{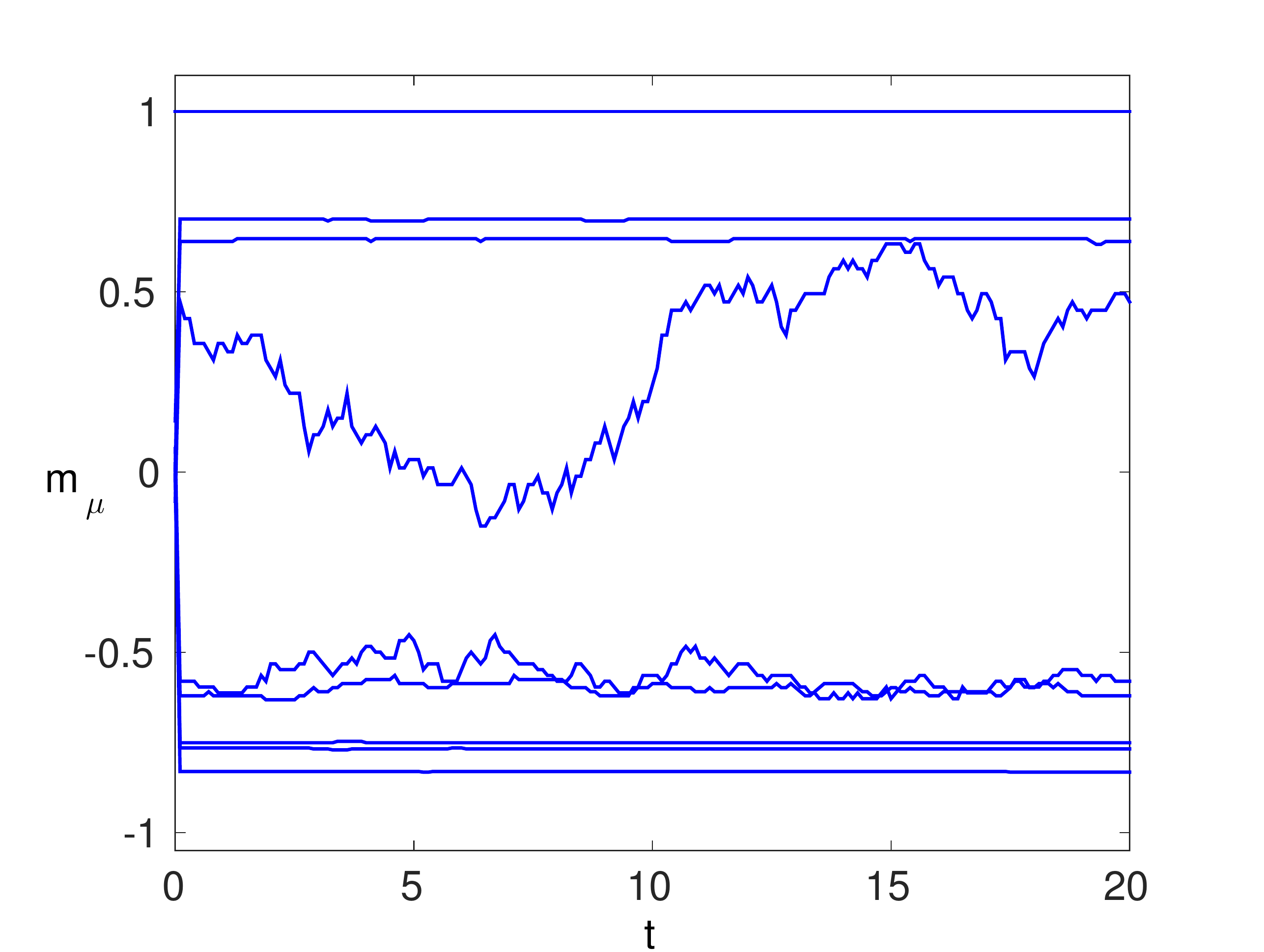}
\includegraphics[width=0.48\textwidth]{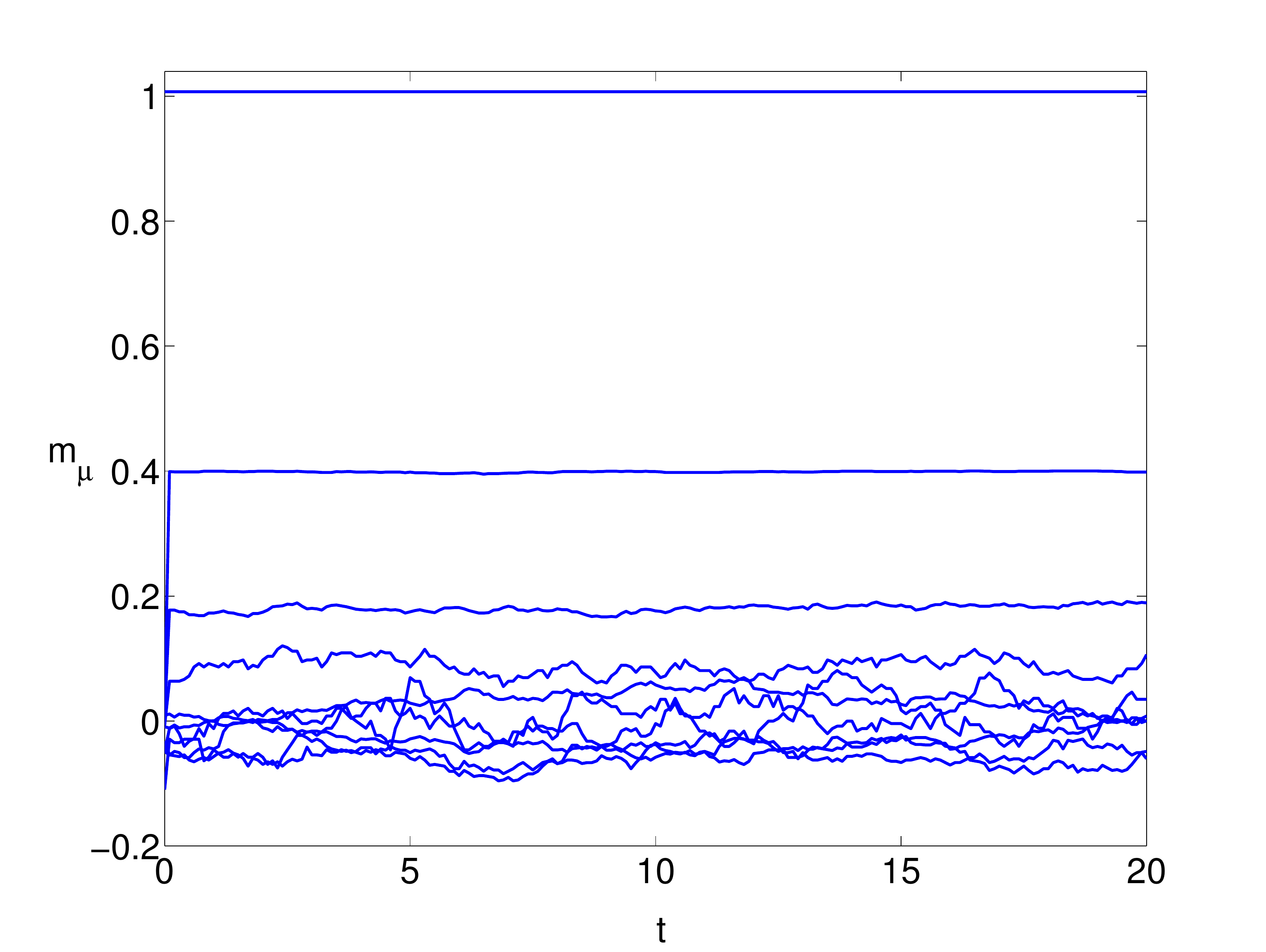}
\caption{Monte Carlo simulations with $N=10^4$ spins, $q_{\mu}=\epsilon^{\mu}q$. $T=0.01$, $\epsilon=0.7$, $q_1=6$. Left: $\delta=0.2, \gamma=0.25$. In the non-competing regime ($\delta<\gamma$) {\em all} clones are activated. 
Right: $\delta=\gamma=0.25$. In this case only few clones with the highest number of receptors are active.}
\label{fig:qmed}
\end{figure}

Next we study the case where the number of B clones is 
$P= N^\delta$ where $\delta\in[0,1)$, and $N$ is the number of T-clones.
The dynamical equations are
\bea 
\frac{{\rm d}  m_{\mu}}{{\rm d} t}=\frac{N^{\gamma}}{q_{\mu}}\bigg\langle\xi_{\mu}\tanh\left(\beta \sum_{\nu}q^{\nu}\xi^{\nu}m_{\nu}\right)\bigg\rangle_{\bxi}- m_{\mu}\ ,
\label{eq:mqmed}
\eea 
which can be rewritten as
\bea 
\frac{{\rm d}  m_{\mu}}{{\rm d} t}=\bigg\langle\tanh\left(\beta(q^{\mu} m_{\mu}+ \sum_{\nu\neq\mu}q^{\nu}\xi^{\nu}m_{\nu})\right)\bigg\rangle_{\xi_\mu \neq 0}- m_{\mu}\ .
\eea 
Upon introducing the noise distribution
$P_{\mu}(z|\{m_{\nu},q_\nu\})=\langle\delta(z- \sum_{\nu\neq\mu}\xi^{\nu}q_\nu m_{\nu})\rangle_{\bxi}$ on clone $\mu$, we have
\bea 
\frac{{\rm d}  m_{\mu}}{{\rm d} t}= \int {\rm d} z P_{\mu}(z|\{m_{\nu}\}) \tanh\left(\beta(q_\mu m_{\mu}+ z)\right) - m_{\mu}
\label{eq:coupled}
\eea 
where  $P_{\mu}(z|\{m_{\nu},q_\nu\})$ can be written, using the Fourier 
representation of the Dirac delta and carrying out the average 
over $\bxi$, as
\bea 
\hspace*{-1cm}P_{\mu}(z|\{m_{\nu},q_\nu\})&=&\int_{-\infty}^{+\infty}\frac{{\rm d} \omega}{2\pi} {\rm e}^{{\rm i} z \omega}\langle\prod_{\nu\neq\mu}^P {\rm e}^{- {\rm i} \omega 
\xi^{\nu} q_\nu m_{\nu}}\rangle_{\bxi}
=\int_{-\infty}^{+\infty} \frac{{\rm d} \omega}{2\pi} {\rm e}^{{\rm i} z \omega }{\rm e}^{\sum_{\nu\neq\mu} \frac{q^{\nu}}{N^{\gamma}}
[\cos(\omega q_\nu m_{\nu})-1]}\ .
\label{eq:noisedist}
\eea
If $\gamma>0$, extending the sum in the exponent to all patterns will add a negligible contribution in the thermodynamic limit, hence, as $N\to \infty$ 
all clones will have the same noise distribution
\bea P_{\mu}(z|\{m_{\nu},q_\nu\})\to P_P(z|\bm,\bq)=\langle\delta(z-\sum_{\nu=1}^P q^\nu\xi^\nu m_\nu)\rangle_{\bxi}
\label{eq:distriblim}
\eea
We note that the sum on the RHS of 
(\ref{eq:distriblim}) is at most $\order{(\sim N^{\delta-\gamma})}$, hence 
for $\delta<\gamma$ it is  
negligible in the thermodynamic limit. This yields
$P_P(z|\bm,\bq)\to \delta(z)$ as $N\to\infty$, so that  
equations (\ref{eq:coupled}) decouple and the system reduces, 
for $\delta<\gamma$, to a 
set of indipendent 
Curie-Weiss ferromagnets, each evolving according 
\bea
\frac{\rd m_{\mu}}{\rd t}=\tanh(\beta q_{\mu}m_{\mu}) -m_\mu.
\eea 
At the steady state, each B clone $\mu$ becomes active at its own critical temperature $T_c=q_\mu$, independently of the other clones (fig. \ref{fig:qmed}, left).
In contrast, for $\delta\geq \gamma$ the noise distribution 
$P_P(z|\bm,\bq)$ has a finite width, due to clonal interference, and the 
equations for the evolution of clonal 
activations are coupled. In this regime, clones compete to be activated 
and the ones with fewer triggered receptors will fail to be switched on 
(fig. \ref{fig:qmed}, right).
In the following section we will analyse the effect of receptors heterogeneity in the regime of competing clones. We will show that clonal
interference affects both critical temperature and intensity of B clones activations. 

\subsubsection{ Bifurcations near the critical temperature and stability region in the regime of competing clones.}\label{sec:qmedbif}
In this section we study the bifurcations away 
from ${\bf m}=(0,\dots,0)$ below the critical temperature $T_c=q_{{\rm max}}$. 
Without loss of generality, we assume $q_{\rm max}=q_1$. 
We Taylor expand the steady state equations obtained by setting 
$\rd m_\mu/\rd t=0$ in \eqref{eq:mqmed}, 
for small $m_{\mu}$ at $\beta q_{1}=1+\epsilon$
\bea
m_{\mu}\simeq\frac{N^{\gamma}}{q_{\mu}}\beta \sum^P_{\nu}q_{\nu} \langle\xi^{\mu}\xi^{\nu}\rangle m_{\nu}-\frac{N^{\gamma}}{q_{\mu}}\frac{\beta^3}{3} \sum^P_{\nu\rho\lambda}q_{\nu}q_{\rho}q_{\lambda} \langle\xi^{\mu}\xi^{\nu}\xi^{\rho}\xi^{\lambda}\rangle m_{\nu}m_{\rho}m_{\lambda}=\\
= \beta q_{\mu}m_{\mu} -\frac{\beta^3 q_{\mu}^3}{3}m_{\mu}^3- \beta  \frac{q_{\mu}}{N^{\gamma}}m_{\mu}\sum^P_{\rho\neq\mu}m_{\rho}^2 q_{\rho}^3 \ .
\eea
For any $\mu$, $m_{\mu}=0$ is always a solution.
Non-zero solutions are given, for $\mu=1$, by 
\bea 
\beta q_1-1-\frac{\beta^3 q_1^3}{3}m_1^2 +\frac{\beta q_1^4 }{N^{\gamma}}m_1^2 - \frac{\beta^3}{N^{\gamma}}q_1\sum^P_{\rho>1}m_{\rho}^2q_{\rho}^3=0
\eea
that at $\beta q_1= 1+\epsilon$ and for large $N$ yield
\bea 
m_1^2= 3\epsilon - \frac{3}{q_1^2N^{\gamma}}\sum^P_{\rho>1}m_{\rho}^2q_{\rho}^3\ .
\label{eq:mbif1}
\eea 
For $\mu\neq1$, we have 
\bea 
\beta q_{\mu}-1-\frac{\beta^3 q_{\mu}^3}{3}m_{\mu}^2 +\frac{\beta q_{\mu}^4 }{N^{\gamma}}m_\mu^2 - \frac{\beta^3}{N^{\gamma}}q_{\mu}\sum^P_{\rho>1}m_{\rho}^2q_{\rho}^3=0
\eea 
which gives, for $N\to\infty$ and to $\order{(\epsilon^0)}$,
\bea 
m_{\mu}^2= \frac{3}{q_{\mu}^3}(q_1^2q_{\mu}-q_1^3)-\frac{3}{N^{\gamma}q_{\mu}^2}\sum^P_{\rho>1}m_{\rho}^2q_{\rho}^3 +\mathcal{O}(N^{-\gamma})\ .
\eea
Summing over $\mu>1$ we get 
\bea 
\sum_{\mu>1}^{P}m_{\mu}^2 q_{\mu}^3
\bigg(1+\frac{3}{N^{\gamma}}\sum_{\rho>1}^{P}q_{\rho}\bigg)=
3\bigg(\sum_{\mu>1}^{P}(q_{\mu}-q_1)\bigg)
\eea
Since $q_{\mu}<q_1$, $\forall \mu>1$, this equality can never be satisfied, 
showing that $m_{\mu}=0, \forall \mu>1$.
Substituting this result into \eqref{eq:mbif1}, we find $m_1^2=3\epsilon$,
hence the first state to bifurcate away from 
${\bf m}=(0,\dots,0)$ is ${\bf m}= (\sqrt{3\epsilon},0,\dots,0)$. 
Next we calculate the region where the pure state $(m_1,0,\ldots,0)$ stays stable, by inspecting the sign of the eigenvalues of the Jacobian of the linearised equations of motion about the steady state. 

For a steady state with the general structure ${\bf m}= (m_1,m_2,\dots, m_n, 0,\dots,0)$, where $n$ 
is the fraction of activated clones, 
the Jacobian \eqref{eq:JacobianM} has a block structure,
where diagonal terms are, for $\mu\leq n$
\bea 
J_{\mu \mu}=\beta q_{\mu}( 1-\langle\tanh^2(\beta(q_{\mu}m_{\mu} +\sum_{\nu\neq\mu}^n\xi^{\nu} q_{\nu}m_{\nu}))\rangle_{\bxi})-1
\label{eq:lambda1}
\eea
and for $\mu> n$ 
\bea 
J_{\mu \mu}=\beta q_{\mu}( 1-\langle\tanh^2(\beta
\sum_{\nu}^n\xi^{\nu} q_{\nu}m_{\nu})\rangle_{\bxi})-1,
\label{eq:lambda2}
\eea
while off-diagonal elements are, for $\mu,\nu \leq n$
\bea 
J_{\mu \nu}=-\frac{\beta q_{\mu}}{N^{\gamma}}\langle\tanh^2(\beta\sum_{\nu}^n\xi^{\nu} q_{\nu}m_{\nu})\rangle_{\bxi}-1
\eea
and $J_{\mu\nu}=0$ otherwise.
For $N\to\infty$ the matrix becomes diagonal with 
eigenvalues $\lambda_\mu=J_{\mu\mu}$ 
given by (\ref{eq:lambda1}) for $\mu\leq n$ 
and (\ref{eq:lambda2}) for $\mu>n$.
In the pure state ${\bf m^{\star}}= (m_1,0,\dots,0)$, where only one clone is 
activated, we have 
\bea 
\lambda_1= \beta q_{1}( 1-\langle\tanh^2(\beta q_1m_{1})\rangle_{\bxi})-1 \ ,
\label{eq:l1}
\\
\lambda_\mu=\beta q_\mu( 1-\langle\tanh^2(\beta \xi^{1} q_{1}m_{1})\rangle_{\bxi})-1\quad\quad\mu>1 \ .
\label{eq:l2}
\eea 
Near the critical temperature $T_c=q_1$, setting $\beta q_1=1+\epsilon$
and using $m_1^2\simeq 3\epsilon$ gives
\bea 
\lambda_1= \beta q_{1} -1 - \beta^3 m_1^2 q_1^3 = -2\epsilon <0 
\\
\lambda_\mu=\beta q_\mu -1- \beta^3 q_1^2 q_\mu\langle \xi_{1}^2 m_{1}^2\rangle_{\bxi} = (1+\epsilon)\frac{q_\mu}{q_1} -1,\quad  \quad\mu>1
\eea 
showing that 
for $q_\mu< q_1$ the pure state is stable near criticality, 
as opposed to the case $q_\mu=q_1, \forall \mu$ 
studied in \cite{jphysaas}, where all clones are activated with 
the same intensity below $T_c$.
In the opposite limit $T\to 0$, we get from (\ref{eq:l1}), (\ref{eq:l2}) 
\bea
\lambda_1\simeq -1\ ,\\
\lambda_\mu \simeq \beta q_\mu\bigg(1-\frac{q_1}{N^\gamma}\bigg) -1\simeq \beta q_\mu -1\ . \eea
showing that the pure state is unstable at low temperature. 

\begin{figure}
\centering
\includegraphics[width=0.45\textwidth]{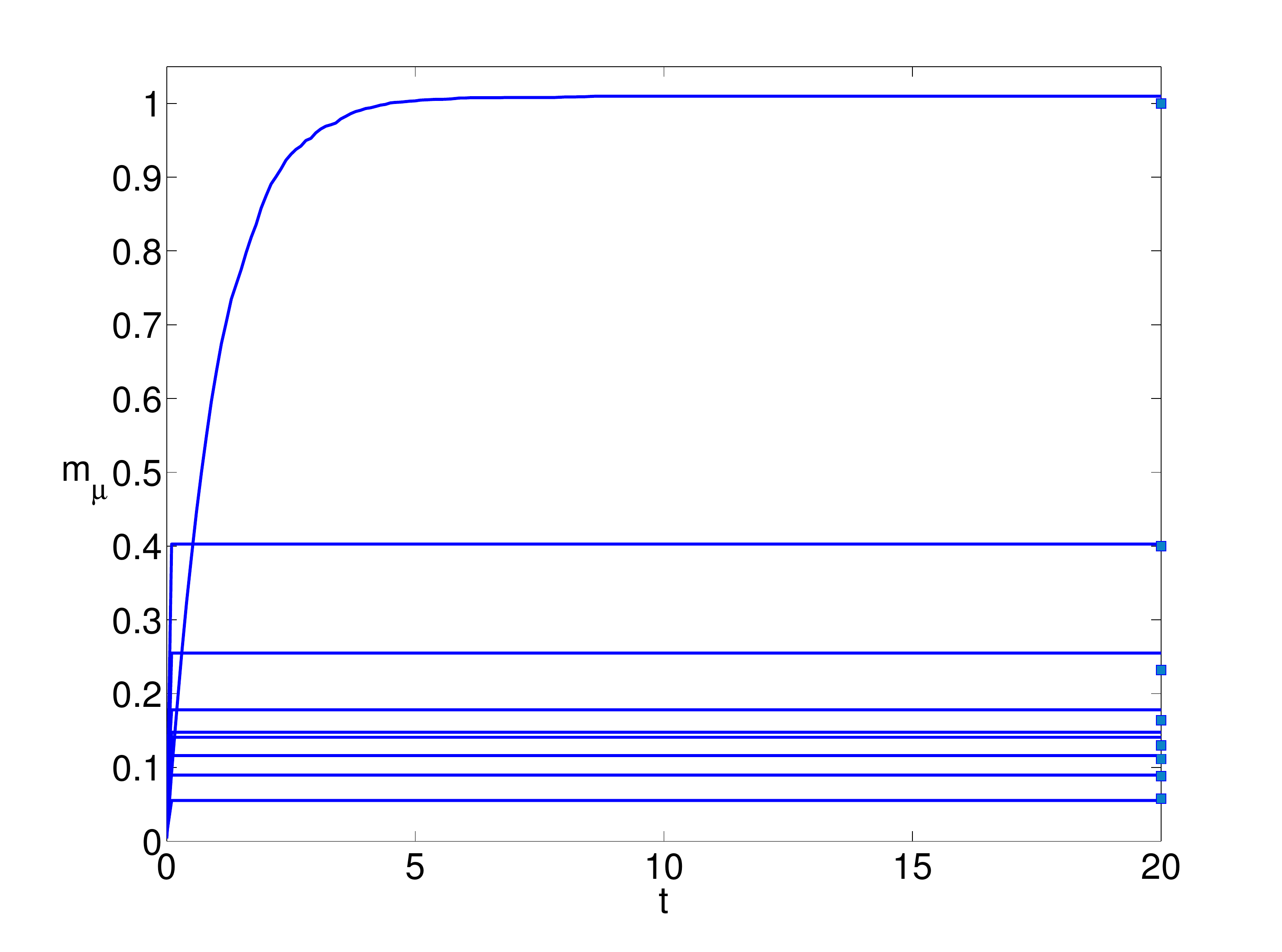}
\caption{Monte Carlo simulations with $N=10^4$ spins, $q_{\mu}=\epsilon^{\mu}q$. $T=0.001$, $\epsilon=0.7$, $q_1=6$, $\delta=\gamma=0.25$. 
The markers represents the theoretical predictions at $T=0$ \eqref{eq:qhier}.}
\label{fig:qhier}
\end{figure}

Indeed, as the temperature is lowered 
below $q_1$, we expect clones with fewer receptors to get 
active. In particular, at $T=0$ 
we expect all clones to be activated, in a hierarchical fashion, 
whereby 
the system sends the highest possible signal 
to the clone with maximum number of receptors, 
the clone with second highest number of receptors is signalled 
by the remaining 
$N-q_1 N^{1-\gamma}$ spare T cells and so on. This leads to the 
following heuristic rule for noiseless clone activations
\bea
m_\mu=\prod_{j=1}^{\mu-1}\bigg(1-\frac{q_{j}}{N^\gamma}\bigg),
\quad\quad\mu=1,\dots, P,\label{eq:qhier}
\eea
in agreement with Monte Carlo simulations shown in fig. \ref{fig:qhier}.

\subsubsection{Sequential B-clones activation: critical temperature and interference effects.}
In this section we calculate the critical temperature at which clones 
$\mu>2$ with fewer receptors get activated. We focus on the 
the regime of clonal interference $\delta \geq \gamma$, as 
for $\delta<\gamma$ each clone $\mu$ gets active at its own critical 
temperature $T_\mu=q_\mu$. 
Without loss of generality we can set $P=\alpha N^\gamma$ for 
$\delta\geq \gamma$,  
where $\alpha=1$ for $\delta=\gamma$ and $\alpha\to\infty$ for $\delta>\gamma$.
In the following we will consider $q_1>q_2>\dots>q_P$. Assuming that ${\bf m}$ bifurcates continuously from the pure state, 
we can Taylor expand (\ref{eq:mqmed}) at the steady state for small 
$m_{\nu}$, with $\nu\neq 1$, while $m_1=\order{(1)}$. For $\mu \neq 1$ 
we have 
\bea 
m_{\mu}
&=& \frac{N^{\gamma}}{q_{\mu}}\bigg\langle\xi^{\mu}\tanh\left(\beta(\xi^1 q_1 m_1+ \sum_{\nu=2}^P q_{\nu}\xi^{\nu}m_{\nu})\right)\bigg\rangle_{\bxi} 
\nonumber\\
&=&\bigg\langle\xi^{\mu}\tanh\left(\beta\sum_{\nu=2}^P q_{\nu}\xi^{\nu}m_{\nu}\right)\bigg\rangle_{\bxi,\xi^\mu\neq 0}+\order{(N^{-\gamma})}
\eea
\bea
&\simeq&
\beta \sum_{\nu= 2}^P q_\nu \langle \xi^{\mu}\xi^{\nu}\rangle_{\xi^\mu\neq 0} m_\nu -   \beta^3 \sum_{\rho\neq\nu}^P 
\langle\xi^{\mu}\xi^{\nu}\xi^{\rho}\xi^{\lambda} \rangle_{\xi^\mu\neq 0} m_\nu m_\rho m_\lambda q_{\nu}q_{\rho}q_{\lambda}\ .
\nonumber\\
&\simeq&  \beta q_{\mu}m_\mu - \frac{\beta^3 q_\mu^3 }{3}m_{\mu}^3 -\frac{\beta^3 q_\mu m_{\mu} }{N^{\gamma}}\sum_{\rho\neq\mu}^P q_{\rho}^3 m_{\rho}^2\ .
\eea
The solutions are $m_{\mu}=0$, or
\bea 
1=  \beta q_{\mu}- \frac{\beta^3 q_\mu^3}{3}m_{\mu}^2 -\frac{\beta^3 q_\mu }{N^{\gamma}}\sum_{\rho\neq\mu}^P q_{\rho}^3 m_{\rho}^2\ .
\label{eq:qdep}
\eea 
Hence, for $1-\beta q_\mu>0$, $m_{\mu}=0$ while 
for $1-\beta q_\mu<0$ the $\mu$-th clone may be activated.
Hence, the first state to bifurcate away from the pure state is 
${\bf m}=(m_1,m_2, 0,\dots,0)$ at $T=q_2$. Its amplitude at $\beta q_2=1-\epsilon$ is
\bea 
m_{2}^2 \simeq 3\epsilon -\frac{3q_{1}^3}{q_2^2N^{\gamma}}m_1^2
\eea
{\em i.e.} $m_2^2=3\epsilon +\mathcal{O}(N^{-\gamma})$. 
As $T$ is lowered below $q_2$ we expect that the clones will activate sequentially 
one after the other, each one at its own temperature. 
In particular, assuming $m_{1,2}=\mathcal{O}(1)$, we have for $\mu>2$  
\bea 
m_\mu
&=&\bigg\langle\xi^{\mu}\tanh\left(\beta\sum_{\nu>3}^P q_{\nu}\xi^{\nu}m_{\nu}\right)
\bigg\rangle_{\bxi,\xi^\mu\neq 0}+\order{(N^{-\gamma})}
\eea
and expanding for $m_{\nu>2}$ small at $T<q_2$
shows that $m_3$ becomes non-zero at $T=q_3+\order{(N^{-\gamma)}}$. 
Generalizing for $n\ll N^{\gamma}$ activated clones 
$m_1,\ldots, m_n=\order{(1)}$ we have for $\mu>n$
\bea 
m_\mu
&=&\bigg\langle\xi^{\mu}\tanh\left(\beta\sum_{\nu>n}^P q_{\nu}\xi^{\nu}m_{\nu}\right)
\bigg\rangle_{\bxi,\xi^\mu\neq 0}+\order{(nN^{-\gamma})}
\eea
giving as bifurcation temperature $T_n=q_n+ \mathcal{O}(nN^{-\gamma})$.
As the number of activated clones increases, their cumulative effect on the 
activation temperature of the remaining clones 
increases and can no longer be neglected 
for $n=\mathcal{O}(N^\gamma)$.  
The activation temperature of pattern $n+1$, when 
$n\sim N^\gamma $ clones have been activated, can 
be worked out from
\bea 
m_{n+1}=\bigg\langle\tanh\left(\beta(q_{n+1}m_{n+1}+ \sum_{\nu=1}^nq_{\nu}\xi^{\nu}m_{\nu})\right)\bigg\rangle_{\bxi} \ ,
\eea
that gives, upon insertion of 
$\int {\rm d}z \delta(z-\sum_{\nu=1}^nq_{\nu}\xi^{\nu}m_{\nu})=1$
\bea 
m_{n+1}=\int {\rm d} z P_n(z|\bm,\bq)\tanh(\beta (q_{n+1}m_{n+1}+z))\ .
\eea
with $P_n(z|\bm,\bq)$ defined in (\ref{eq:distriblim}).
Taylor expanding for $m_{n+1}$ small we have, to leading orders
\bea
m_{n+1}= \int {\rm d} zP_n(z|\bm,\bq) \bigg[
(1-\tanh^2(\beta z))\beta q_{n+1}m_{n+1}+\order{(m_{n+1}^2)}
\bigg]
\eea 
where we have used $P_n(z|\bm,\bq)=P_n(-z|\bm,\bq)$. A solution is $m_{n+1}=0$ 
and a non-zero solution is possible for 
\bea 
\beta q_{n+1}=\frac{1}{1-\int{\rm d} z P_n(z|\bm,\bq)\tanh^2(\beta z)}\ .
\label{eq:critican+1}
\eea 
For $n\ll N^{\gamma}$, $P_n(z|\bm,\bq)=\delta(z)$, 
and we retrieve $\beta q_{n+1}=1$ for the temperature at which $m_{n+1}$ 
becomes non-zero. For $P_n(z|\bm,\bq)$ having a small but finite width we 
can use $\tanh(\beta z)\simeq\beta z$
\bea
\beta q_{n+1}\simeq 1+\frac{\langle z^2\rangle}{q_{n+1}^2}\simeq 1+\frac{\sum_{\mu=1}^n q_{\mu}^3m_{\mu}^2}{q_{n+1}^2N^\gamma}\ ,
\eea
showing that $T_{n+1}<q_{n+1}$ and 
deviations from $q_{n+1}$ depend on the promiscuity distribution of 
the activated clones. 
Equation (\ref{eq:critican+1}) shows that as more clones are activated, 
these create an interference, encoded in $P_n(z|\bm,\bq)$, that decreases the 
activation temperature of the inactive ones. 
Furthermore, it suggests that the number $n$ of clones 
that the system can activate
({\em i.e.} the number of $\order{(1)}$ order parameters $m_\mu$) at 
small but finite temperature 
is $\order{(N^\gamma)}$. 

\subsubsection{Numerical examples.}
In this section we test \eqref{eq:critican+1} and look at the 
effect of receptors' heterogeneity 
on the intensity of B clones activation, for 
three simple cases that can be treated analytically, with 
$q_1>q_2$ and $P=\alpha N^\gamma$:
\begin{enumerate}
\item ${\bf q} = (q_1, q_2, \dots,q_2)$: only one clone has a higher number of receptor;
\item ${\bf q} = (q_1,\dots,q_1 ,q_2,\dots,q_2)$: half of the clones have 
promiscuity $q_1$, and half have promiscuity $q_2$;
\item ${\bf q}=(q_1,\dots,q_1,q_2)$: only one clone has a 
smaller number of receptors.
\end{enumerate}
Our goal is to analyse the increasing interference effect due to the 
activated clones with more receptors on the quiescent ones with less 
receptors. 
According to \eqref{eq:critican+1} active clones should play the role of 
interference terms that lower the critical temperature of the quiescent clones.

\begin{enumerate}
\item Clones with the same number of receptors will be activated 
with the same intensity \cite{jphysaas} and the activation vector bifurcating 
away from the pure state will have the form (see Sec. \ref{sec:LSABB} 
for a rigorous derivation)
${\bf m}=(m_1,m_2,\dots,m_2)$.  
The amplitudes $m_1$ and $m_2$ are found from 
\bea
m_1&=&\frac{N^{\gamma}}{q_1}
\bigg\langle\xi^1\tanh\left(\beta(\xi^1 q_1 m_1+ q_2 \sum_{\nu=2}^P 
\xi^{\nu}m_{\nu})\right)\bigg\rangle_{\bxi} 
\nonumber\\
&=&\bigg\langle\tanh\left(\beta(q_1 m_1+ q_2 m_2\sum_{\nu=2}^P 
\xi^{\nu})\right)\bigg\rangle_{\bxi} 
\nonumber\\
&=&\sum_z P_{P-1}(z|\bq)\tanh \beta(q_1 m_1+q_2 m_2 z)
\label{eq:dynq1q1}
\eea
and 
\bea
m_2&=& \sum_z P_{P-2}(z|\bq)
\tanh\beta[q_1 \xi^1 m_1+q_2 m_2 (1+z)]
\nonumber\\
&=&\sum_z P_{P-2}(z|\bq)
\tanh\beta[q_2 m_2 (1+z)] +\order{(N^{-\gamma})}
\label{eq:dynq1q2}
\eea
where 
\bea
P_n(z|\bq)=\langle \delta (z-\sum_1^{n}\xi^{\nu})\rangle_\bxi.
\label{eq:Pnzq}
\eea
Using \eqref{eq:noisedist}, we can write 
$P_P(z|\bq)={\rm e}^{-q_2\alpha} I_z(q_2 \alpha)$ where 
$I_z(x)$ is the modified Bessel function of the first kind 
\cite{Bessel}, and $P_{P-1}(z|\bq)\simeq P_{P-2}(z|\bq)\simeq P_P(z|\bq)$.
The activation temperature of the clones with smaller promiscuity
follows from Taylor expansion of (\ref{eq:dynq1q2}) for small $m_2$ giving 
\bea
m_2 &=&
\sum_z P_{P-2}(z|\bq)[\beta q_2 m_2  (1+z)-\frac{1}{3} (\beta q_2 m_2)^3  (1+z)^3] +\order{(N^{-\gamma})}
\nonumber\\
&=&\beta q_2 m_2-\frac{1}{3} (\beta q_2 m_2)^3  (1+3\bra z^2\ket)] +\order{(N^{-\gamma})}
\nonumber\\
&=&\beta q_2 m_2-\frac{1}{3} (\beta q_2 m_2)^3  (1+3\alpha q_2)] +\order{(N^{-\gamma})}
\label{eq:m2bif1}
\eea
where we used the parity of $P_n(z|\bq)=P_n(-z|\bq)~\forall~n$ and 
$\bra z^2\ket\equiv \sum_z P_P(z|\bq) z^2=\alpha q_2$. 
We retrieve $\beta q_2=1+\order{(N^{-\gamma})}$ for the activation teperature 
of $m_2$, consistently 
with \eqref{eq:critican+1}, for $n=1$. The activation intensity at 
$\beta q_2=1+\epsilon$ 
follows from (\ref{eq:m2bif1}) as
\bea
m_2=\frac{3\epsilon}{1+3q_2\alpha},
\eea 
as one expects for homogeneous clones
in the absence of clone $\mu=1$ \cite{jphysaas}.
In fig. \ref{fig:mediusteady} we plot the amplitudes $m_1, m_2$ 
resulting from
\eqref{eq:dynq1q1}, \eqref{eq:dynq1q2}, as a function of the temperature (left) 
and those resulting from Monte Carlo simulations, as a function of time 
(right). The latters are seen to relax to the theoretically predicted steady 
state.
\begin{figure}[htb!]
\centering
\includegraphics[width=0.45\textwidth]{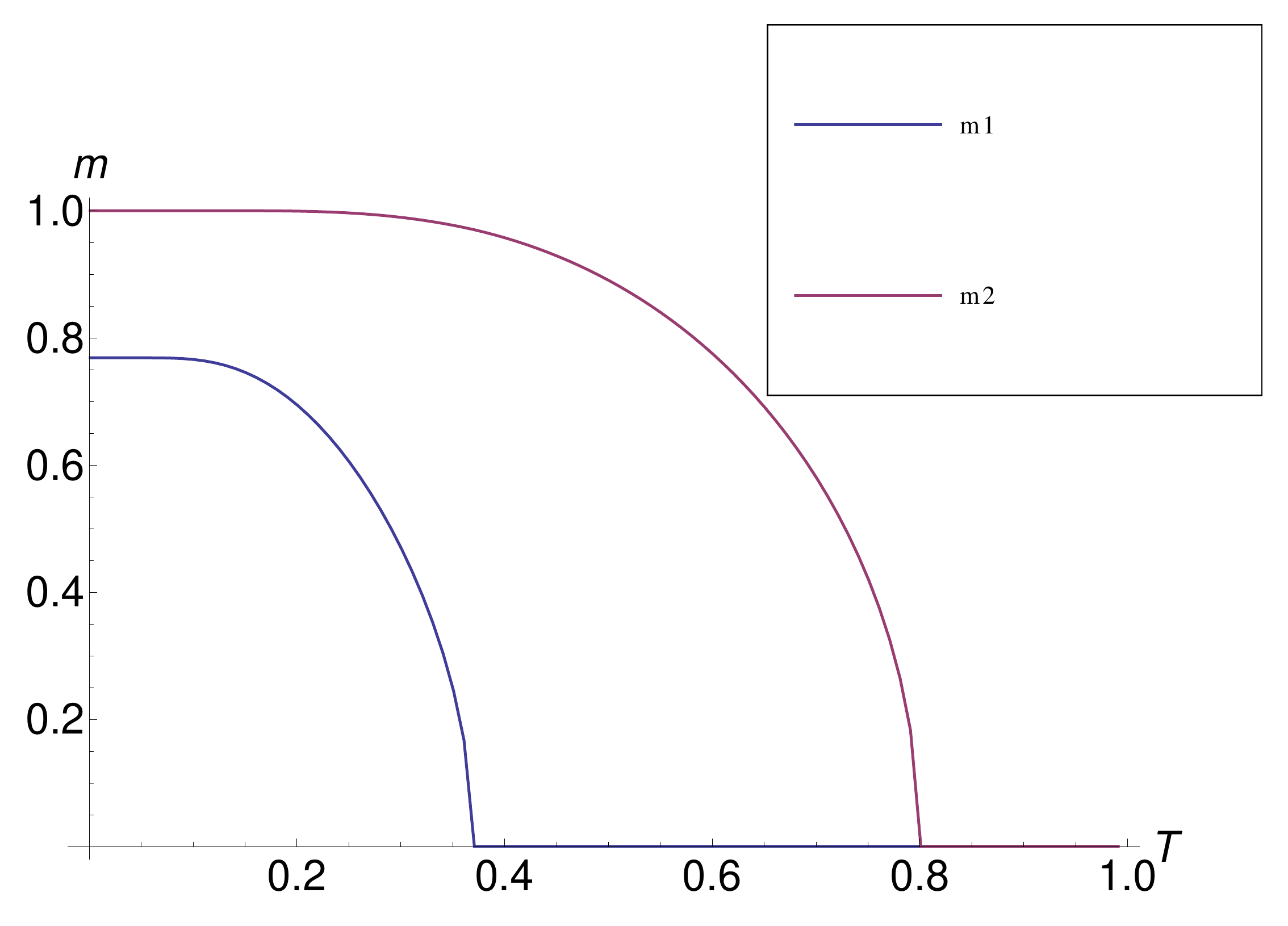}
\includegraphics[width=0.46\textwidth]{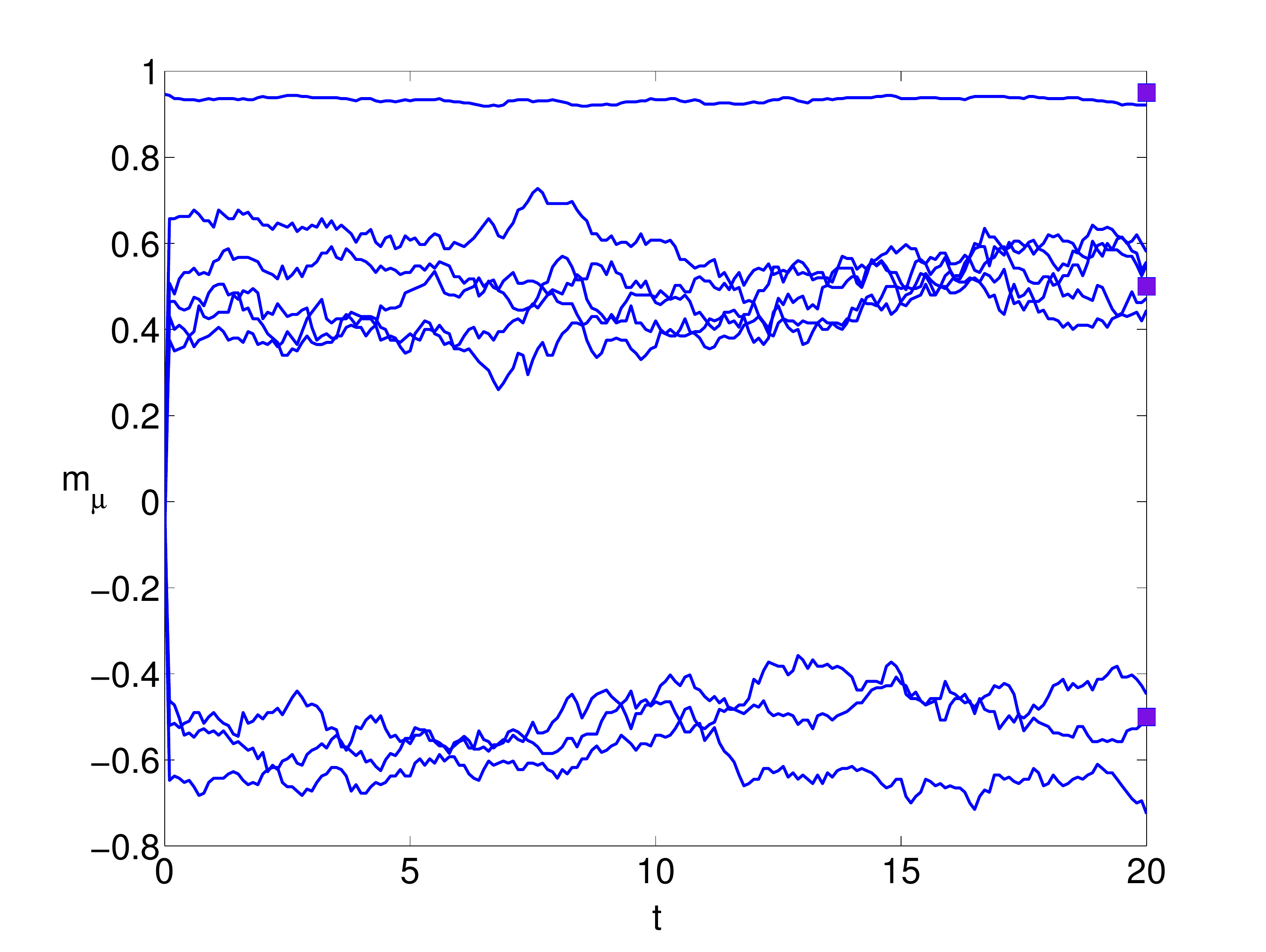}
\caption{Left: steady state solutions $m_1, m_2$ of the dynamical system 
\eqref{eq:dynq1q1}, \eqref{eq:dynq1q2}, as function of the temperature $T$, 
for $q_1=0.8, q_2=0.4$. Right: Monte Carlo simulations with $N=10^4$ spins, 
$\delta=\gamma=0.25$, $T=0.3$, $q_1=0.8$, $q_2=0.4$. The markers are the 
theoretically predicted steady state activations \eqref{eq:dynq1q1}, 
\eqref{eq:dynq1q2}.}
\label{fig:mediusteady} 
\end{figure}
A contour plot of $m_2=0$ in the $T-q_2$ plane, 
computed numerically from (\ref{eq:dynq1q2}),   
is shown in fig. \ref{fig:bifm2}.  
Deviations from the line 
$T=q_2$ are consistent with finite size effects $N^{-\gamma}$.
\begin{figure}
\centering
\includegraphics[width=0.4\textwidth]{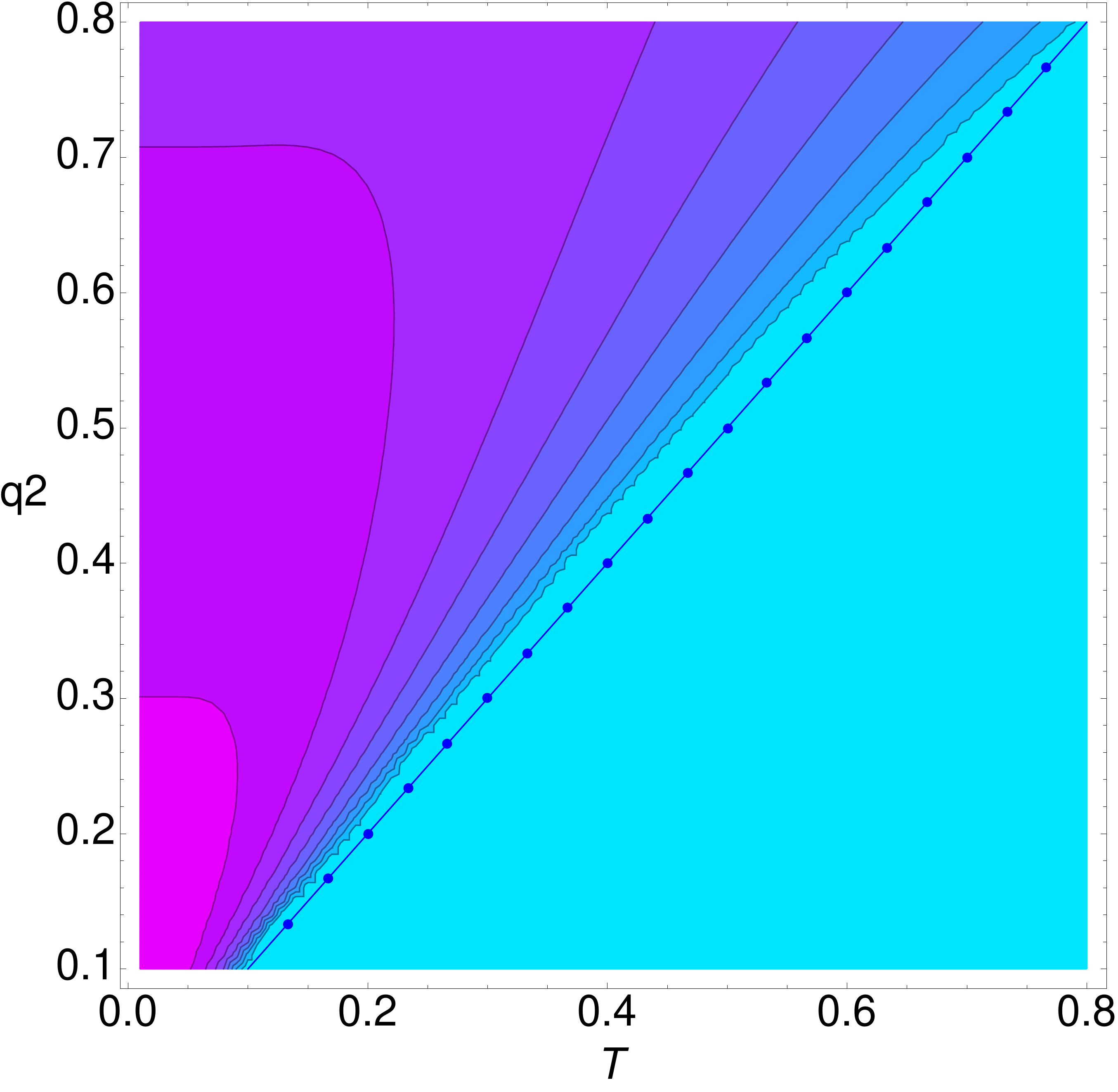}
\caption{Contours of constant $m_2$ with 
${\bf q}=(q_1, q_2, \dots,q_2)$. The contour indicating the 
on-set of a non-zero value of $m_2$ is very close to the blue dotted 
line $T=q_2$ theoretically predicted. Deviations are within finite size 
effects $\order{(N^{-\gamma})}$.
} \label{fig:bifm2}
\end{figure}

In conclusion, the presence of one clone with a higher number of receptors, 
does not affect, in the thermodynamic limit, 
the activation temperature nor the activation intensity of clones 
with fewer receptors.

\item
In this case we expect a transition from the state 
${\bf m}=(m_1,\dots,m_1,0,\dots,0)$ to 
${\bf m}=(m_1,\dots,m_1,m_2,\dots, m_2)$,
with 
\bea 
m_1=\sum_{z_1,z_2}P_{P/2}(z_1|\bq)P_{P/2}(z_2|\bq)\tanh(\beta(q_1 m_1(1+z_1)+q_2m_2z_2))\ ,
\nonumber\\
m_2=\sum_{z_1,z_2}P_{P/2}(z_1|\bq)P_{P/2}(z_2|\bq)\tanh(\beta(q_2m_2(1+z_2)+q_1m_1z_1))\ ,
\label{eq:m2hh}
\eea
The temperature at which the on-set of non-zero $m_2$ occurs 
is found by Taylor expanding (\ref{eq:m2hh}) for small $m_2$ 
with $m_1=\order{(1)}$
\bea
\hspace*{-2cm}
m_2&=&\sum_{z_1,z_2}P_{P/2}(z_1|\bq)P_{P/2}(z_2|\bq)
\left[
\tanh(\beta q_1m_1z_1)+(1-\tanh^2(\beta q_1m_1z_1)) \beta q_2m_2(1+z_2)
+\order{(m_2^2)}\right]
\nonumber\\
\hspace*{-2cm}
&=& \beta q_2m_2\sum_{z_1}P_{P/2}(z_1|\bq)(1-\tanh^2(\beta q_1m_1z_1)
\eea
giving $m_2\neq 0$ for 
\bea
\beta q_2=\frac{1}{1-\sum_z P_{P/2}(z|\bq)\tanh^2(\beta q_1m_1 z)}\ .
\label{eq:theobifhh}
\eea
\begin{figure}[htb!]
\centering
\includegraphics[width=0.4\textwidth]{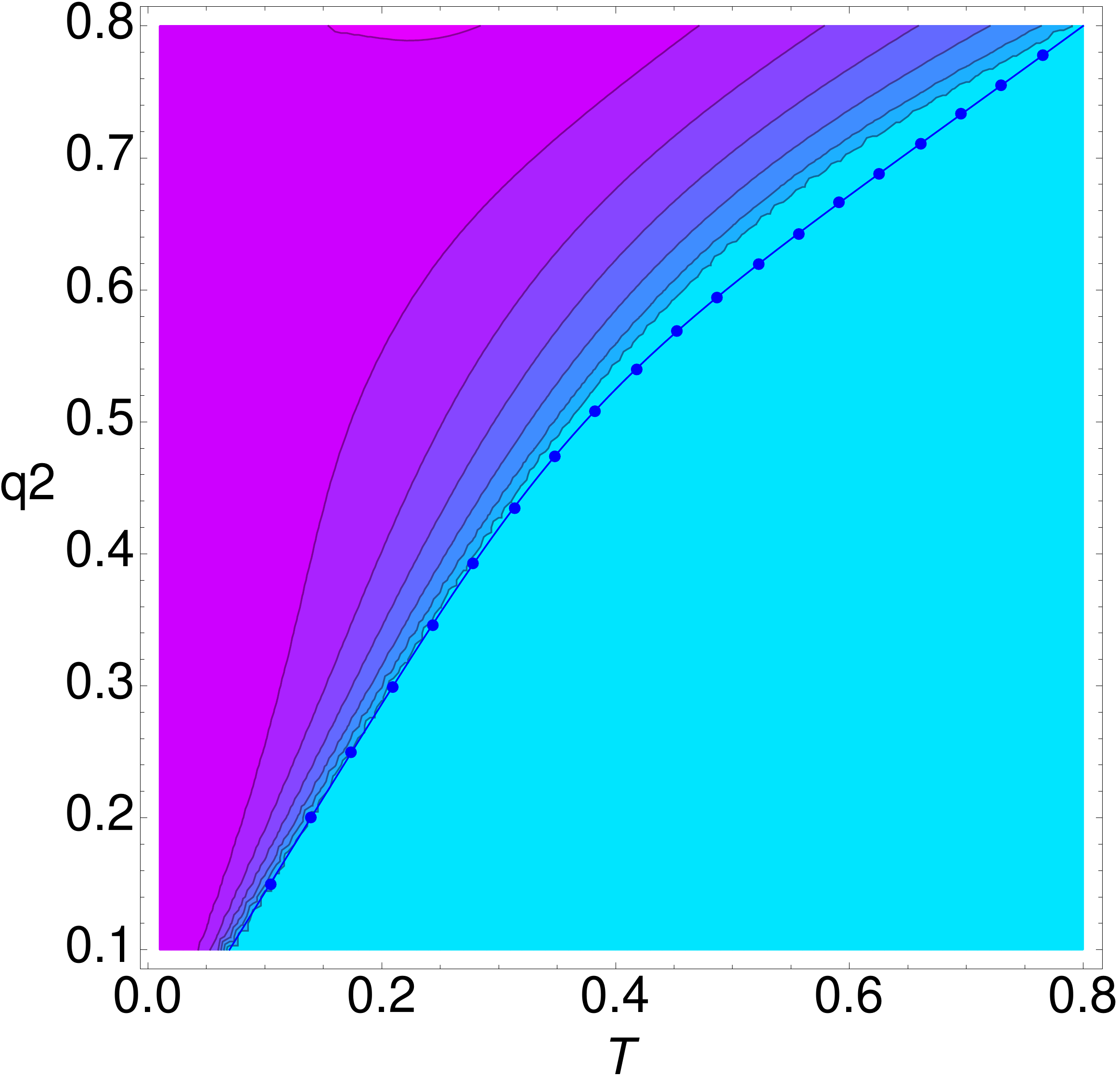}
\includegraphics[width=0.4\textwidth]{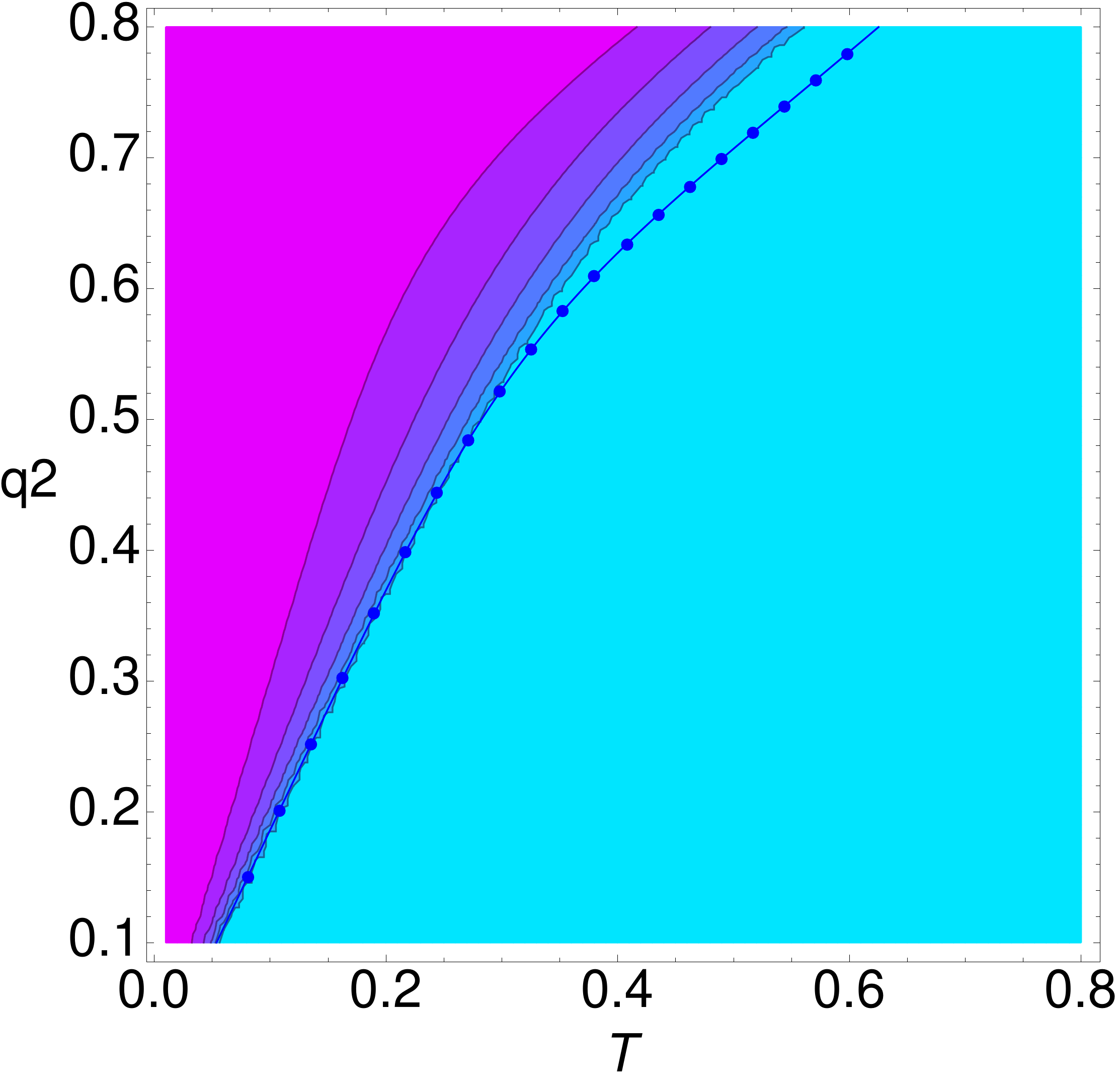}
\caption{Contour plot of $m_2$ in the $(T,q_2)$ plane for ${\bf q}=(q_1,\dots,q_1, q_2,\dots,q_2)$ (left) and ${\bf q}=(q_1,\dots,q_1, q_2)$ (right). The blue dotted line represents the theoretical critical temperature line computed using respectively the self-consistent equations \eqref{eq:m2hh} (left) and \eqref{eq:sysq1q2}(right) together with the theoretical predictions for the critical temperature \eqref{eq:theobifhh}(left) and \eqref{eq:theobif}(right). Deviations from the numerical results when $q_2\simeq q_1$ are due to the fact that \eqref{eq:critican+1} is obtained assuming $m_2\ll m_1$, condition which is not satisfied when $q_2\simeq q_1$. }
\label{fig:theobifu}
\end{figure}
The theoretically predicted critical line \eqref{eq:theobifhh} is in excellent agreement with the contour plot of 
$m_2\neq 0$ in the $T-q_2$ plane, shown in fig. \ref{fig:theobifu} (left), computed numerically from (\ref{eq:m2hh}). 
The plot shows that in the presence of $\order{(N^\gamma)}$ 
clones with higher numbers of receptors, 
the activation temperature of those with smaller promiscuity 
$q_2$ will deviate from the line $T=q_2$.  
\item This is the case where deviations from the line $T=q_2$ are 
expected to be largest.
From \eqref{eq:mqmed} we have
\bea
m_1&=&\frac{N^{\gamma}}{q_1}
\bigg\langle\xi^1\tanh\left(\beta(\xi^1 q_1 m_1+ \sum_{\nu=2}^P q_\nu 
\xi^{\nu}m_{\nu})\right)\bigg\rangle_{\bxi} 
\nonumber\\
&=&
\bigg\langle\tanh\left(\beta(q_1 m_1+ \sum_{\nu=2}^P q_\nu 
\xi^{\nu}m_{\nu})\right)\bigg\rangle_{\bxi} 
\nonumber\\
&=&\sum_z P_{P-2}(z|\bq)\tanh[ \beta(q_1 m_1(1+z) +q_2 m_2 \xi^n)]
\nonumber\\
&=&\sum_z P_{P-2}(z|\bq)\tanh [\beta q_1 m_1(1+z)] +\order{(N^{-\gamma})}
\label{eq:sysq1q2}
\eea
and 
\bea
m_2&=& \bigg\langle\tanh\left(\beta(q_2 m_2+ \sum_{\nu\neq 2}^P q_\nu 
\xi^{\nu}m_{\nu})\right)\bigg\rangle_{\bxi} 
\nonumber\\
&=&\sum_z P_{P-1}(z|\bq)
\tanh[\beta(q_2 m_2+q_1 m_1 z)]
\eea
When clones with fewer receptors get active, we expect 
$m_1=\order{(1)}$ and $m_2$ small. 
Taylor expanding for $m_2$ small and using the parity of 
$P_n(z|\bq)=P_n(-z|\bq)~\forall ~n$
\bea 
m_2&=&\sum_z P_{P-1}(z|\bq)
[\tanh(\beta q_1 m_1 z)+(1-\tanh^2 (\beta q_1 m_1 z)\beta q_2 m_2)+\order{(m_2^2)}
\nonumber\\
&=&\beta q_2 m_2[1-\sum_z P_{P-1}(z|\bq)\tanh^2 (\beta q_1 m_1 z)]+\order{(m_2^2)}
\eea
we obtain for the activation temperature of the only clone with lowest number of receptors
\bea
\beta q_2=\frac{1}{1-\sum_z P_{P-1}(z|\bq)\tanh^2(\beta q_1m_1 z)}
\label{eq:theobif}
\eea
in agreement with the contour plot of $m_2\neq 0$ in the $T-q_2$ found numerically, shown 
in fig. \ref{fig:theobifu} (right). Deviations are compatible with finite size effects and are largest 
for $q_2\simeq q_1$ where the assumption $m_2\ll m_1$ is no longer valid.
Monte Carlo simulations with $10$ B clones, one of which has a lower promiscuity $q_2=0.6$, are shown in fig. \ref{fig:simbifu}:
at $T=q_2$ the pattern with lowest receptors is inactive and it is activated only at a much lower temperature $T\simeq 0.2$.
\begin{figure}[htb!]
\centering
\includegraphics[width=0.48\textwidth]{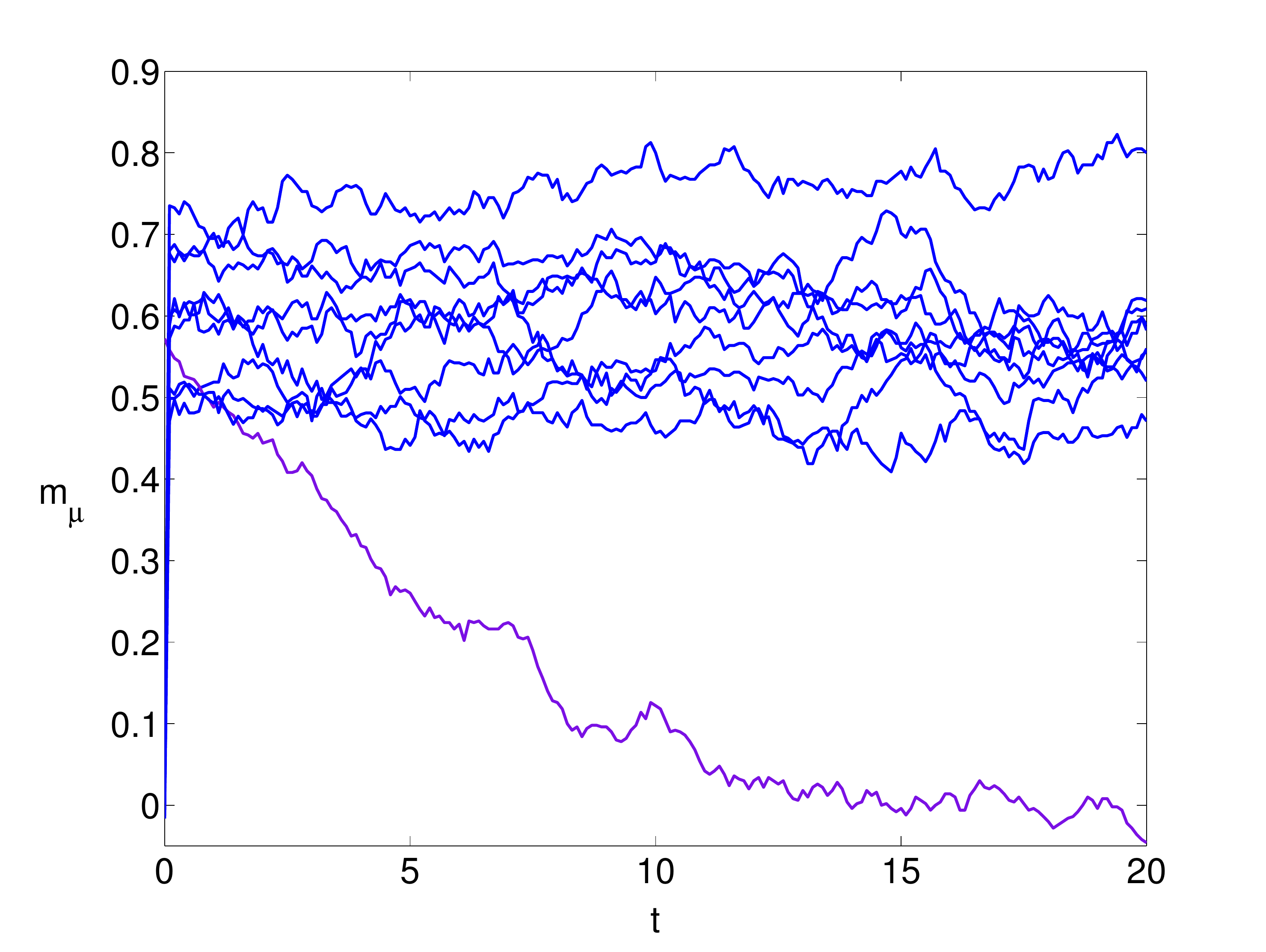}
\includegraphics[width=0.5\textwidth]{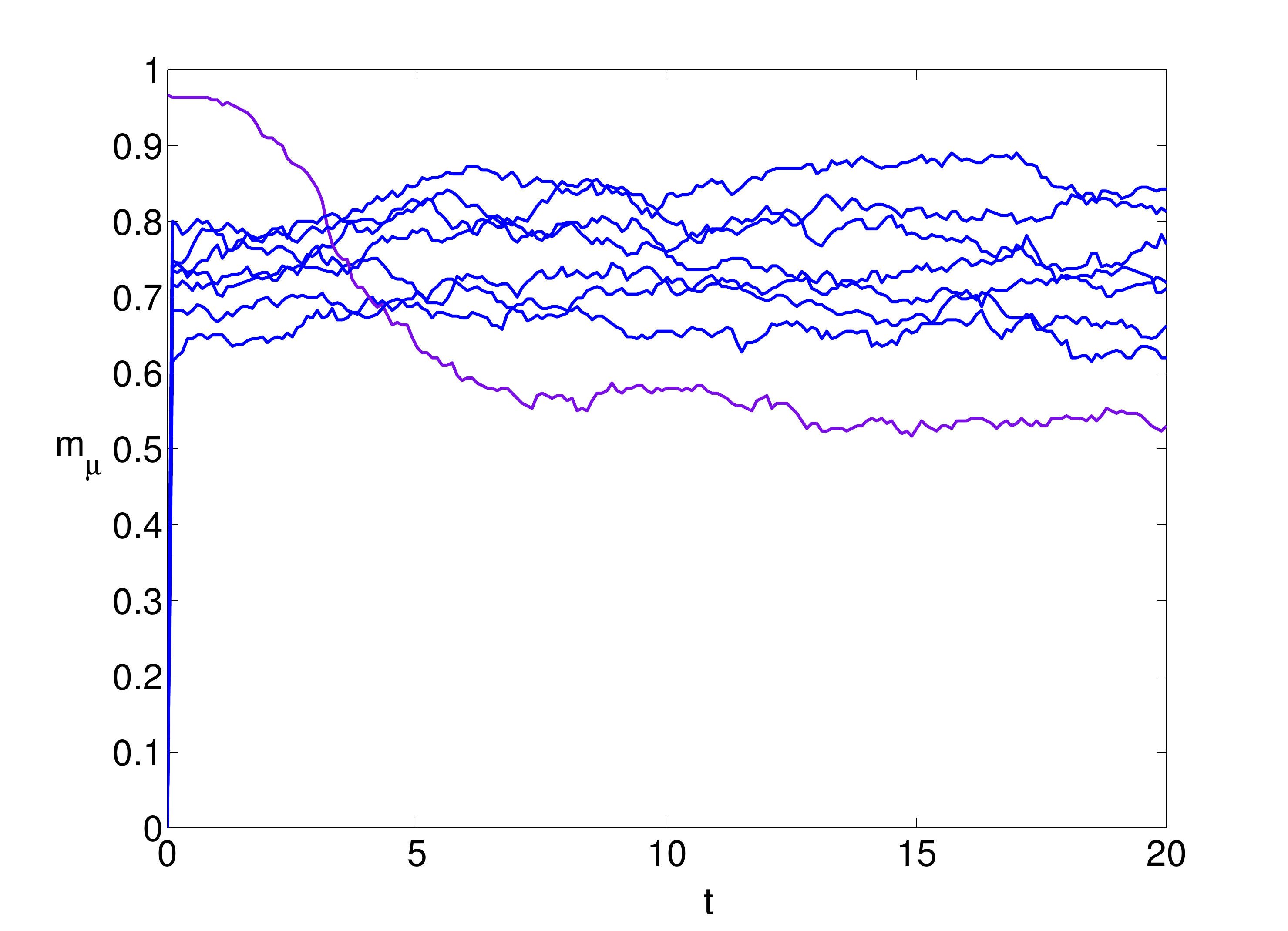}
\caption{Time evolution of B clones activation in Monte Carlo simulations with $10^4$ spins and ${\bf q}= (q_1,\dots,q_1, q_2)$ (case (iii)) with $q_1=0.8$ (blue) and $q_2=0.6 $ (violet). The activation for the clone with less receptors (violet) decays to $0$ for $T=0.5$ (left) and stays non-zero for $T=0.2 $ (right), {\em i.e.} at temperatures considerably lower than that in the absence of 
clonal interference $T=q_2$.}
\label{fig:simbifu}
\end{figure}

Finally in fig. \ref{fig:comparem_2} we plot $m_2$ as a 
function of the temperature in the three different cases with promiscuity 
(i) ${\bf q}=(q_1,q_2,\dots, q_2)$, 
(ii) ${\bf q}=(q_1,\dots,q_1, q_2,\dots,q_2)$, 
(iii) ${\bf q}=(q_1,\dots,q_1, q_2)$. 
The presence of clones with higher promiscuity $q_1$ and activation intensity 
$m_1$, does not only affect the activation temperature of 
clones with lower promiscuity $q=q_2$, but also the intensity 
$m_2$ of their immune responses. 
In conclusion, clones with fewer triggered receptors are 
activated at a lower temperature and will produce a weaker response 
than clones with a higher number of active receptors. 
\begin{figure}
\centering
\includegraphics[width=0.5\textwidth]{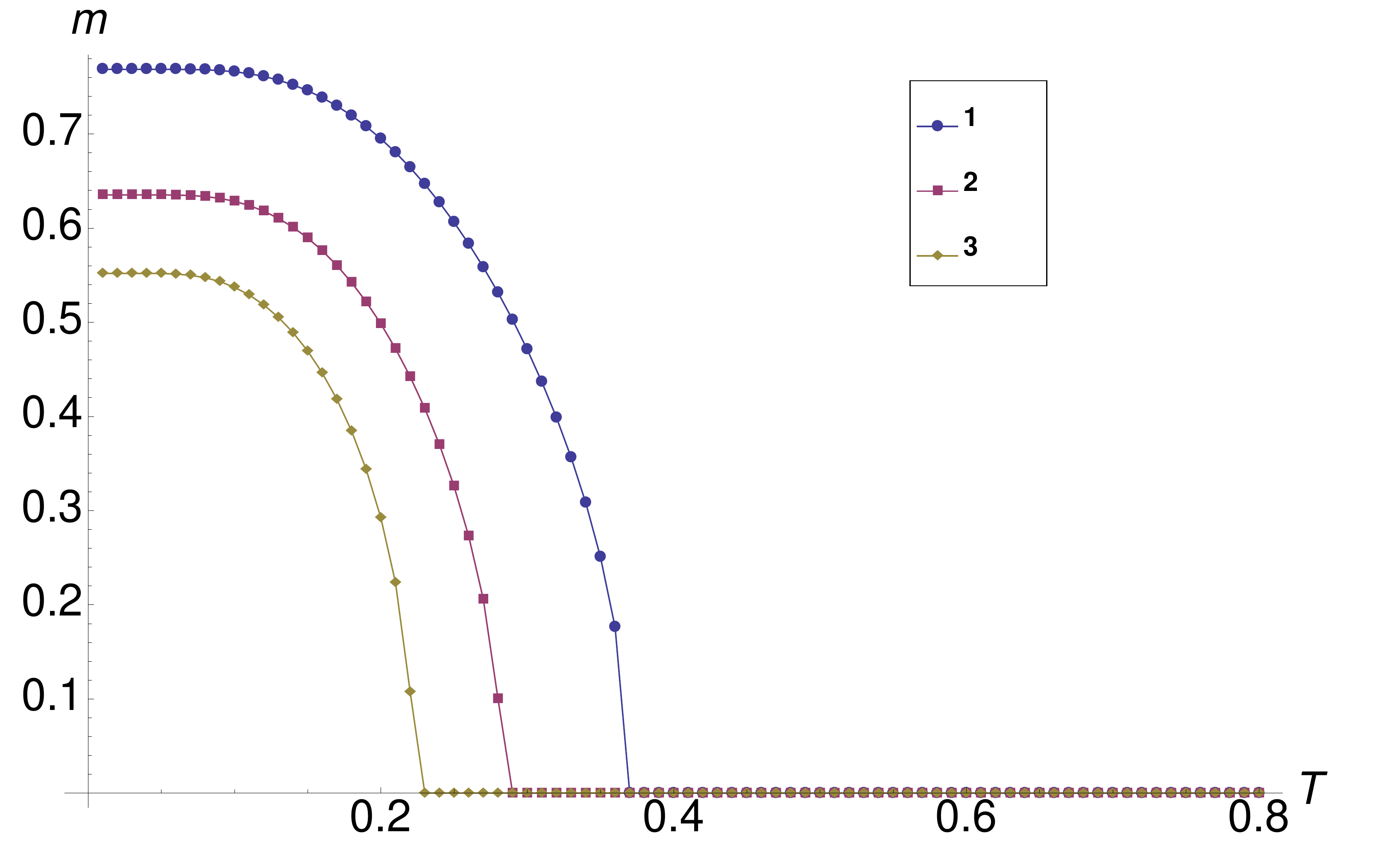}
\caption{Plot of $m_2$ as a function of $T$ in the different cases: (1) ${\bf q}=(q_1,q_2,\dots, q_2)$, (2) ${\bf q}=(q_1,\dots,q_1, q_2,\dots,q_2)$, (3) ${\bf q}=(q_1,\dots,q_1, q_2)$. Increasing the interference due 
to clones activated at a higher temperature, the $m_2$ intensity decreases.}
\label{fig:comparem_2}
\end{figure}
\end{enumerate}
%

\section{Idiotypic interactions}\label{sec:BB}
In this section we study the effect of the idiotypic interactions on the 
ability of the system to activate multiple clones in parallel. 
For simplicity, here we 
will assume homogeneous receptor promiscuities 
{\em i.e.} $q_\mu=c, \ \forall\mu$, so that the dynamical 
equations \eqref{eq:mm} read
\bea 
\frac{{\rm d} m_{\mu}}{{\rm d} t}= \frac{N^\gamma}{c}\langle\xi^{\mu} \tanh(\beta c\sum_{\rho\nu}\xi^{\nu}(\bA^{-1})_{\rho\nu}m_{\rho})\rangle_{\bxi} - m_{\mu}
\label{eq:sysA}
\eea
Matrix $A$ is positive definite for $k\in (0,1)$ and symmetric, hence the 
free-energy of the system is a Lyapunov function for the dynamics
\cite{Lyapunov} and the system will converge to a steady state.
Next, we compute the critical temperature at which clonal activation 
emerges in the steady state 
\bea 
m_{\mu}= \frac{N^\gamma}{c}\langle\xi^{\mu} \tanh(\beta c\sum_{\rho\nu}\xi^{\nu}(\bA^{-1})_{\rho\nu}m_{\rho})\rangle_{\bxi}. 
\label{eq:BBsteadyM}
\eea
By summing over $\sum_{\mu\lambda}m_{\lambda}(\bA^{-1})_{\lambda\mu}$
\bea
\sum_{\mu\lambda}m_{\lambda}(\bA^{-1})_{\lambda\mu}m_{\mu}= \frac{N^\gamma}{c} \langle\sum_{\mu\lambda}m_{\lambda}(\bA^{-1})_{\lambda\mu} \xi^{\mu}\tanh(\beta c\sum_{\rho\nu}\xi^{\nu}(\bA^{-1})_{\rho\nu}m_{\rho})\rangle_{\bxi}
\eea
using the inequality $|\tanh(x)|\leq|x|$ and averaging over the disorder we
obtain
\bea \sum_{\mu\lambda}m_{\lambda}(\bA^{-1})_{\lambda\mu}m_{\mu}
\leq \beta c \sum_{\rho\lambda}m_{\lambda}(\bA^{-2})_{\lambda\rho}m_{\rho}
\eea
Next we diagonalise the matrix ${\bf A}$ by means of the similarity 
transformation $\bD=\bP^{-1}\bA \bP$, 
where $\bD$ is the diagonal matrix constructed 
from the eigenvalues $\{\mu_\rho\}_{\rho=1}^P$ of $\bA$ and $\bP$ is the 
matrix of eigenvectors, which is unitary, {\em i.e.} $\bP^{-1}=\bP^{T}$, since 
$\bA$ is symmetric. Hence, we can rewrite the equation above 
for the transformed vector $\bv=\bP^{-1}\bm $ as 
\bea
\bv^T\bD^{-1}\bv-\beta c \bv^T \bD^{-2}\bv\leq 0
\eea
which gives 
\bea
\sum_{\nu} \left(\frac{v_{\nu}}{\mu_\nu}\right)^2
(\beta c -\mu_\nu)\geq 0
\eea
yielding ${\bf v}={\bf 0}$ for $\beta c < \mu_{\rm min}$, where $\mu_{\rm min}=\min_{\nu} 
\mu_{\nu}$. The eigenvalues of ${\bf A}$ are
\bea
\mu_1=1-k,\quad\quad deg(\mu_1)=P/2\\
\mu_2=1+k,\quad\quad deg(\mu_2)=P/2
\label{eq:eigenA}
\eea 
hence, above the critical temperature
$T_c=c(\mu_{\rm min})^{-1}=c/(1-k)$ we have $\bm={\bf 0}$, 
whereas below criticality $\bm\neq {\bf 0}$ is possible (and expected). 
Remarkably, the critical temperature 
increases with $k$, meaning that idiotypic 
interactions enhance immune system's activation. 
Next, we investigate the structure of the states bifurcating 
away from $\bm={\bf 0}$ below $T_c$ 
and their stability. 

\subsection{Dynamical equations for two B clones}
\begin{figure}
\centering
\includegraphics[width=0.4\textwidth]{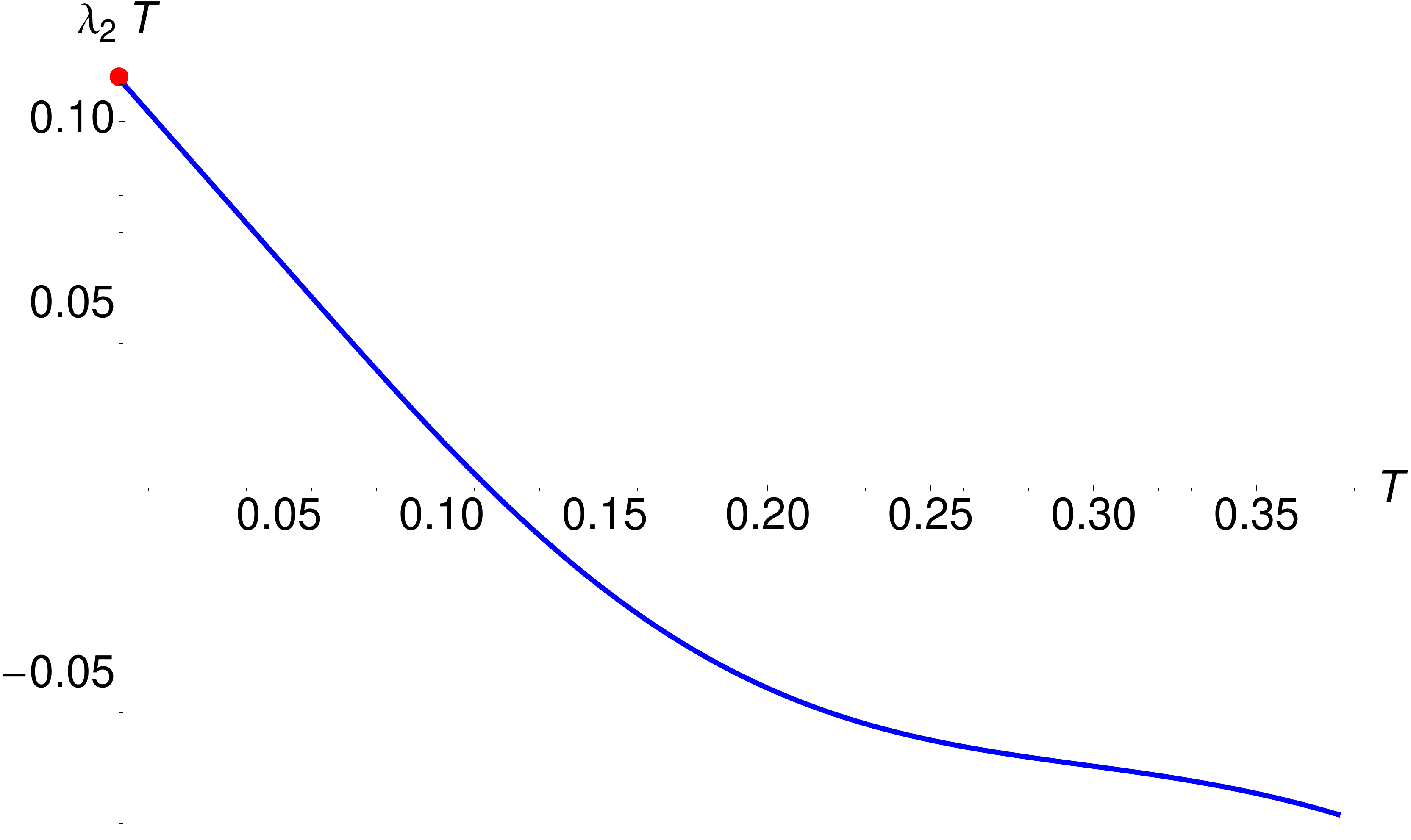}
\includegraphics[width=0.57\textwidth]{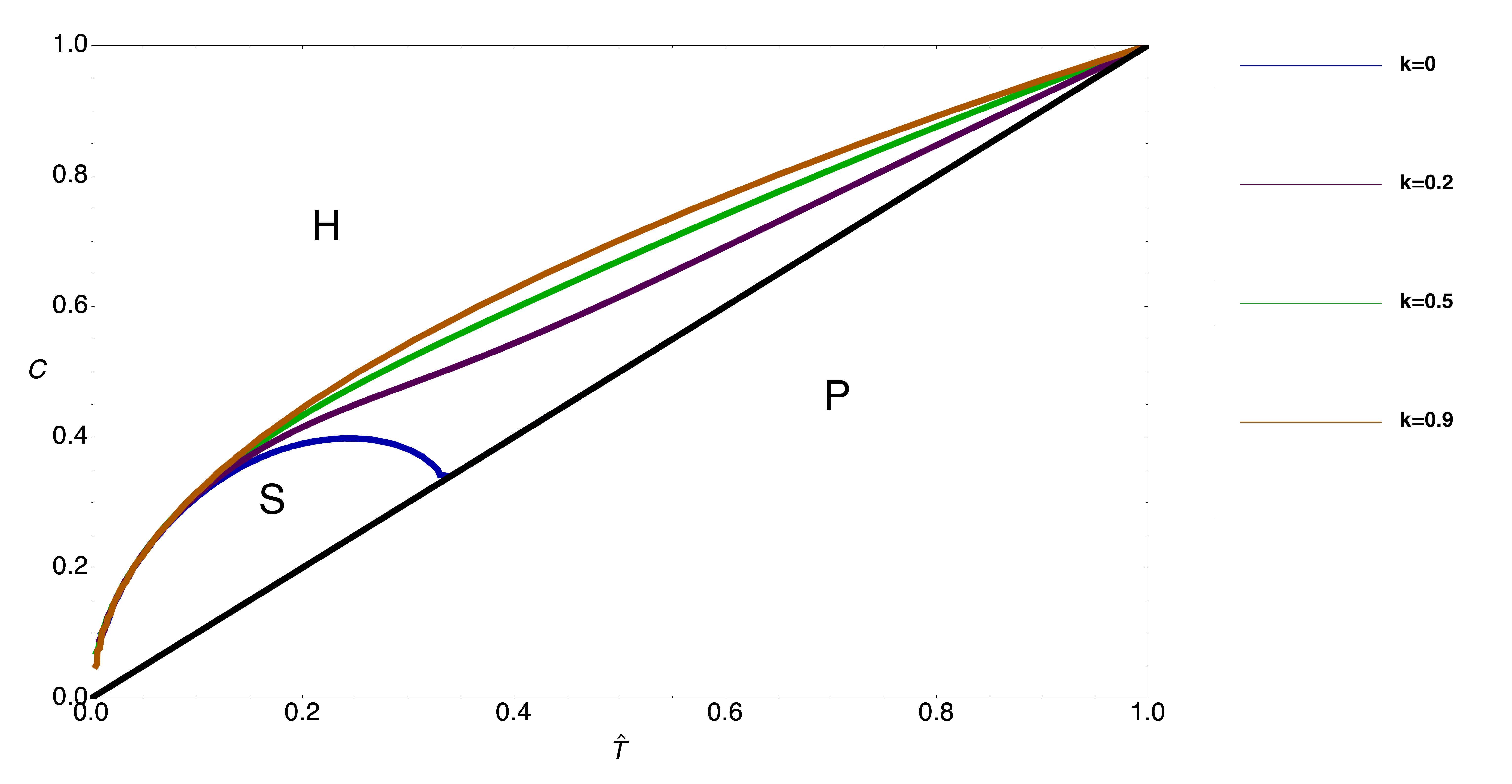}
\caption{Left: $\lambda_2 T$ as a function of $T$ for 
$c=0.3$ and k=0.2. The value at $T=0$ matches the analytical prediction 
$\lambda_2 T=c^2/1-k$ (red marker).
Right: Contour plot of $\lambda_2=0 $ with different interaction strengths $k=0.8,k=0.5, k=0.2$ in the parameter space $(T,c)$. The region where clones are activated in parallel with the same intensity ({\bf S}) becomes wider as $k$ increases.}
\label{fig:phaseBB}
\end{figure}

As before, it is useful to consider first the toy model with 
$P=2$ clones. 
For $k\neq 0$, the dynamics of different clones is always coupled, for all $\gamma \geq 0$. 
We first look at the case $\gamma=0$, where the system evolves according to the equations
\bea 
\frac{{\rm d} m_1}{{\rm d} t}&=&(1-c)\tanh\left(\frac{\beta c}{1-k^2}(m_1 +k m_2)\right) +\frac{c}{2} \bigg[\tanh\left(\frac{\beta c}{1-k}(m_1+m_2)\right)
\nonumber\\
&&+\tanh\left(\frac{\beta c}{1-k}(m_1-m_2)\right)\bigg]-m_1
\nonumber\\
\frac{{\rm d} m_2}{{\rm d} t}&=&(1-c)\tanh\left(\frac{\beta c}{1-k^2}(k m_1 + m_2)\right) +\frac{c}{2} \bigg[\tanh\left(\frac{\beta c}{1-k}(m_1+m_2)\right)
\nonumber\\
&&-\tanh\left(\frac{\beta c}{1-k}(m_1-m_2)\right)\bigg]-m_2
\nonumber\\
\label{eq:BBsys}
\eea
The state ${\bf m}= (0,0)$ is a fixed point of the dynamics at all 
temperatures, but we expect it to become unstable below $T_c=c/(1-k)$. 
In order to inspect the structure of the bifurcating state 
it is convenient to analyse the steady state equations in terms 
of the variables $x=m_1+m_2$ and $y=m_1-m_2$
\begin{figure}
\centering
 \includegraphics[width=0.35\textwidth]{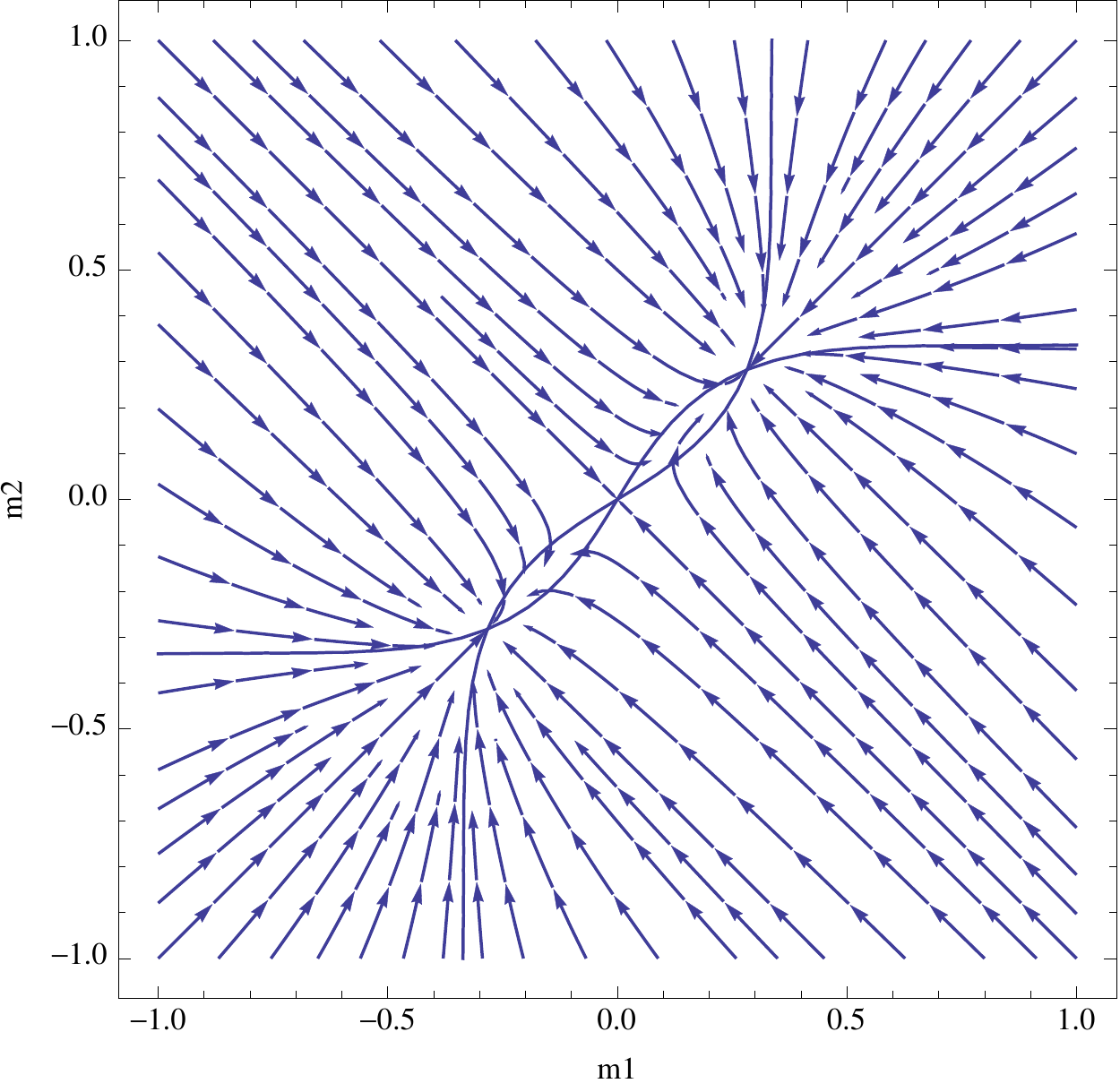}
  \includegraphics[width=0.53\textwidth]{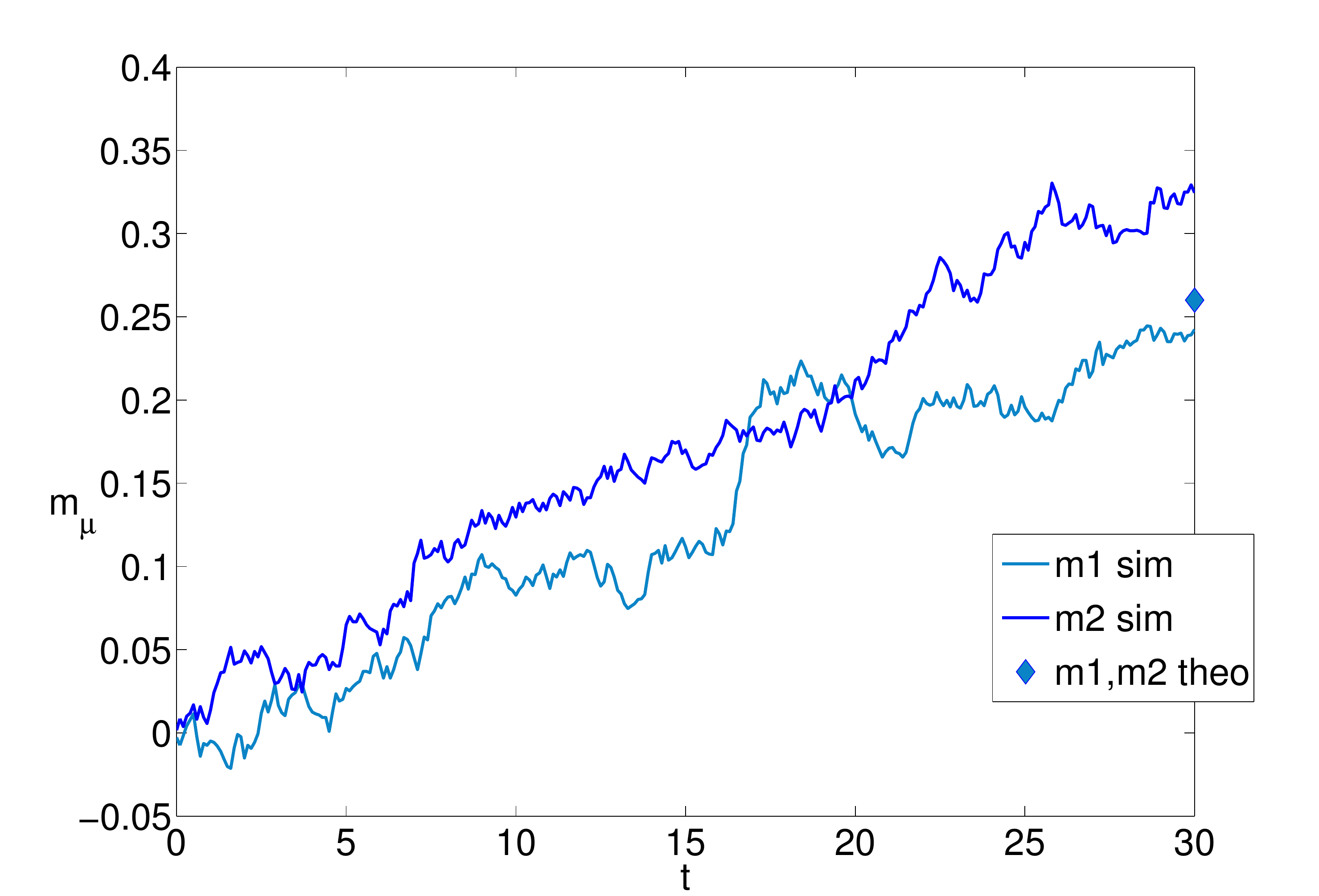}
 \caption{Flow diagram in the plane $(m_1,m_2)$ (left) and Monte Carlo simulations with $10^4$ spins (right) for the dynamical system \eqref{eq:BBsys} with 
$T=1.7, c=0.55, k=0.7$ ($T_c=1.83$). }
 \label{fig:BBsim}
 \end{figure}
\begin{figure}
\centering
\includegraphics[width=0.35\textwidth]{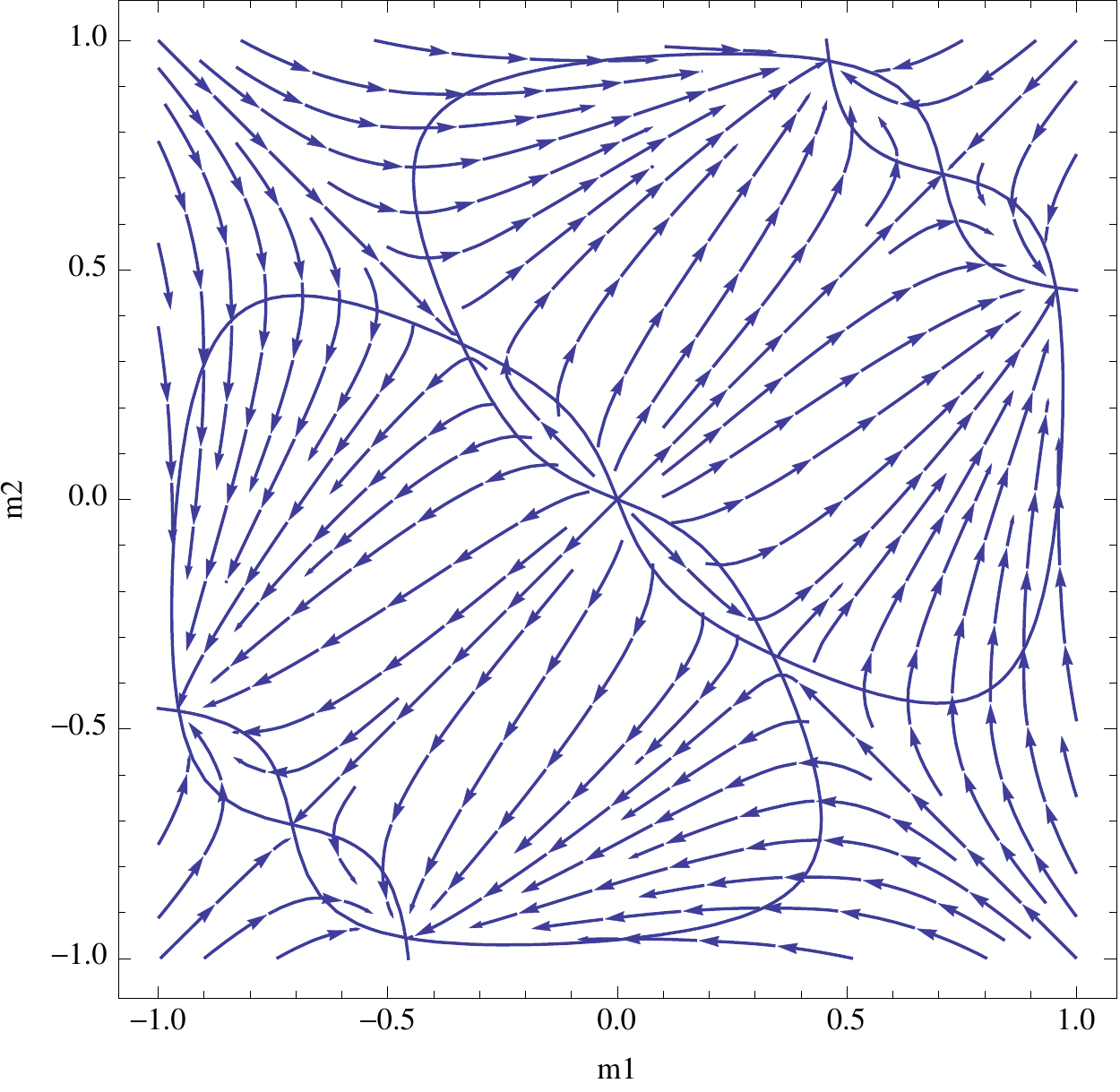}  \includegraphics[width=0.51\textwidth]{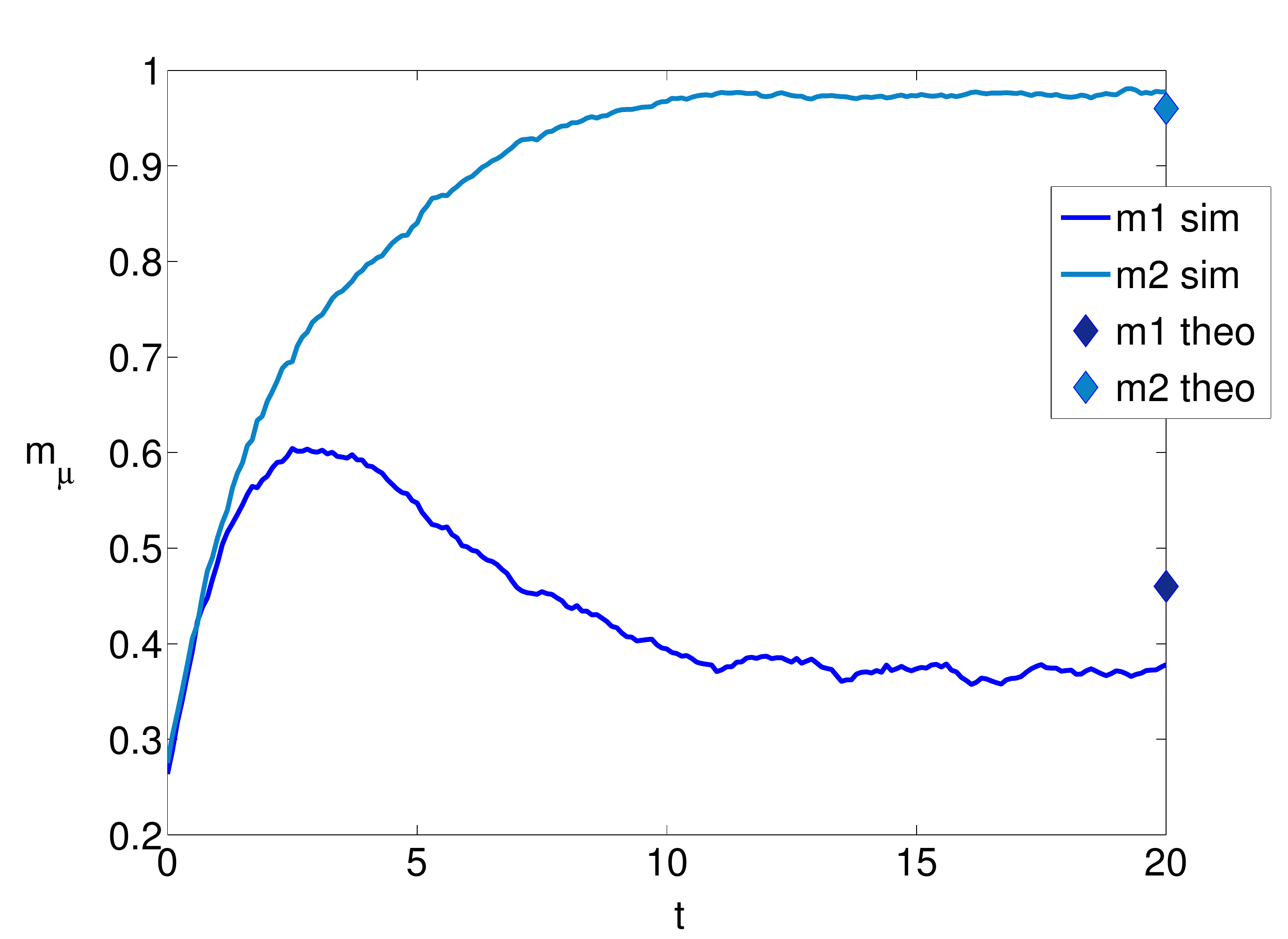}
\caption{Flow diagram in the plane $(m_1,m_2)$ (left) and Monte Carlo simulations with $10^4$ spins (right) for the dynamical system \eqref{eq:BBsys}, for $T=0.65, c=0.55, k=0.7$  ($T_c=1.83$). }
\end{figure}
\bea 
x&=&c \tanh\bigg(\frac{\beta c x}{1-k}\bigg) +(1-c)\bigg[\tanh\bigg(\frac{\beta c }{1-k^2}(x+(k-1)m_2)\bigg) 
\nonumber\\
&&+\tanh\bigg(\frac{\beta c }{1-k^2}(x+(k-1)m_1)\bigg)\bigg]
\nonumber
\\ 
y&=&c \tanh\bigg(\frac{\beta c y}{1-k}\bigg) +(1-c)\bigg[\tanh\bigg(\frac{\beta c }{1-k^2}(x+(k-1)m_2)\bigg)
\nonumber\\
&&-\tanh\bigg(\frac{\beta c }{1-k^2}(x+(k-1)m_1)\bigg)\bigg].
\nonumber
\eea 
Assuming continuos bifurcations, we Taylor expand for small ${\bf m}$ 
near $\beta c=1-k$ {\em i.e.} at 
$\hat{\beta}\equiv \frac{\beta c}{1-k}=1+\epsilon$, obtaining, to 
leading orders
\bea 
x&=& c\hat{\beta} x -\frac{1}{3}c \hat{\beta}^3 x^3 +(1-c)\hat{\beta} x -\frac{1-c}{3(1+k)^2} \hat{\beta}^3 x\bigg[(x+(k-1)m_2)^2 
\nonumber\\ 
&&+( x+(k-1)m_1)^2 -(x+(k-1)m_1)(x+(k-1)m_2)\bigg]
\label{eq:xcrit}
\\
y&=& c\hat{\beta} y -\frac{1}{3}c \hat{\beta}^3 y^3 +\frac{(1-c)(1-k)}{(1+k)}\hat{\beta} y -\frac{(1-c)(1-k)}{3(1+k)^3} \hat{\beta}^3 y\bigg[(x+(k-1)m_2)^2 
\nonumber\\
&&+( x+(k-1)m_1)^2 -(x+(k-1)m_1)(x+(k-1)m_2)\bigg].
\label{eq:ycrit}
\eea
We note that $x=y=0$ is always a solution 
(corresponding to ${\bf m}=(0,0)$). A solution $y\neq 0$ is not possible as 
in \eqref{eq:ycrit} terms $\order{(\epsilon^0)}$
do not simplify. Hence, $y=0$ is the only solution, implying $m_1=m_2$. In contrast, in \eqref{eq:xcrit} first order terms simplify, hence we can have $x\neq0$.
Therefore, mixtures bifurcate from ${\bf m}=(0,0)$ in a symmetric fashion ${\bf m}=m(1,1)$. We can compute the amplitude $m$, Taylor expanding  \eqref{eq:BBsys} at the steady state near $T_c$, for small $m$,
\bea
m=\hat{\beta}m-\frac{1}{3} \hat{\beta}^3m^3 -c \hat{\beta}^3 m^3
\eea
yielding, at $\hat{\beta}= 1+\epsilon$,  
\bea
m^2=\frac{3\epsilon}{1+3c}
\eea
or the trivial solution $m=0$.
Flow diagrams and Monte Carlo simulations are in agreement with theoretical 
predictions, showing that close to criticality both B clones are activated 
with the same intensity (fig. \ref{fig:BBsim}).
Next we investigate the stability 
of symmetric solutions by computing the eigenvalues of the Jacobian
\bea
\hspace{-0.7cm}J_{\mu \nu}=\frac{\partial F_{\mu}^{(1)}({\bf m})}{\partial m_{\nu}}\bigg|_{{\bf m}={\bf m}^\star},
\quad\quad F_{\mu}^{(1)}({\bf m})=\frac{N^\gamma}{c}\langle\xi^{\mu} \tanh(\beta c\sum_{\rho\lambda}\xi^{\lambda}(\bA^{-1})_{\rho\lambda}m_{\rho})\rangle_{\bxi} - m_{\mu} \ ,
\label{eq:JacobianB}
\eea
One has 
\bea 
J_{\mu \nu}=\beta c (\bA^{-1})_{\mu\nu}( 1-\langle\tanh^2(\beta c \sum_{\rho\lambda}\xi^{\lambda}(\bA^{-1})_{\rho\lambda}m_\rho^\star )\rangle_{\bxi})-
\delta_{\mu\nu} +\order{(N^{-\gamma})}
\label{eq:JacobBB}
\eea
and substituting ${\bf m^{*}}=m(1,1)$ we find for the eigenvalues
\bea 
\lambda_1= \frac{\beta c}{1 -k } - 
  1 - \frac{\beta c (1 - c)}{1 - k} \tanh\bigg(\frac{m \beta c}{1 - 
        k}\bigg)^2 - \beta \frac{c^2}{1 - k} \tanh\bigg(\frac{2 m \beta c}{1 - k}\bigg)^2 \ ,\\
\lambda_2=  \frac{\beta c (1 - c)}{1 + k} \bigg(1 - 
     \tanh\bigg(\frac{m \beta  c}{1 - k}\bigg)^2\bigg) + \frac{ \beta c^2}{1 - k} - 1\ .
\eea
These can be calculated analytically near criticality {\em i.e.} 
at $\beta c/1-k=1+\epsilon$ where
\bea 
\lambda_1=\frac{-2\epsilon -6\epsilon c}{1+3c} \ ,\\
\lambda_2= \frac{-2\epsilon +6\epsilon c -2k+2k\epsilon -4 kc -4 kc\epsilon}{(1+3c)(1+k)}\ ,
\eea
and for $T\to 0$, where
\bea
\lambda_1= -1 \ ,\\
\lambda_2=\frac{ \beta c^2}{1-k}-1\ .
\eea
We deduce that $\lambda_1$ is always negative, hence the stability of symmetric 
solutions is determined by $\lambda_2$. 
A plot of $\lambda_2$ as a function of $T$ is shown in 
fig. \ref{fig:phaseBB} (left) for fixed dilution and B-B interaction 
strength $k$. The critical line where $\lambda_2=0$ in the $T-c$ 
plane for different values of $k$ is shown in fig. \ref{fig:phaseBB} (right). 
Remarkably, the region ({\bf S}) where activation of parallel immune responses 
is accomplished with the same intensity gets wider as $k$ increases.  
For $\gamma>0$, the system evolves according to the equations
\bea 
\frac{{\rm d} m_1}{{\rm d} t}&=&\tanh\left(\frac{\beta c}{1-k^2}(m_1 +k m_2)\right) -m_1
\nonumber\\
\frac{{\rm d} m_2}{{\rm d} t}&=&\tanh\left(\frac{\beta c}{1-k^2}(k m_1 + m_2)\right) -m_2
\nonumber\\
\label{eq:BBdil}
\eea
One can show again that $\bm=(0,0)$ is the only fixed point above 
$T_c=c/(1-k)$, 
and symmetric mixtures $\bm=m(1,1)$ bifurcate away from $\bm=(0,0)$ at $T_c$. 
Now, however, symmetric mixtures remain stable for any $T<T_c$ 
(see Sec. \ref{sec:LSABB}), with the intensity 
$m$ of the symmetric activation found from 
\bea
m=\tanh\left(\frac{\beta cm}{1-k}\right)
\eea
Hence for $\gamma>0$ the system reduces to independent 
Curie-Weiss ferromagnets, even in the presence of idiotypic interactions, 
with the critical temperature $T_c=c/(1-k)$
increasing with the strength of interactions.

\subsection{Generalisation to P clones}
\label{sec:BBP}
In this section we turn to the case of 
$P= N^\delta$, $0<\delta<1$ B clones.
Again, we expect clones to be activated at 
$T_c=c/(1-k)$ in a symmetric fashion. For $\delta< \gamma$ we expect 
symmetric solutions to be stable for all $T<T_c$, whereas for $\delta\geq 
\gamma$ 
we expect them to destabilise at low temperature. 
Without loss of generality, we can set 
$P=\alpha N^\gamma$ with $\alpha=1$ for $\delta=\gamma$ and the cases 
$\delta<\gamma$ and $\delta>\gamma$
retrieved in the limits $\alpha\to 0$ and $\alpha\to\infty$ respectively.

To inspect the behavior of the system near criticality, it is convenient 
to write the steady state equations in terms of the rescaled matrix 
$\hat A=A/(1-k)$, 
with eigenvalues 
\bea 
\mu_1=1\ ,\quad\quad && deg(\mu_1)=P/2\label{eq:eigenris1} \\
\mu_2=\frac{1+k}{1-k}\ , \quad\quad && deg(\mu_2)=P/2 \ . 
\label{eq:eigenris} \eea 
Taylor expanding the steady state equations \eqref{eq:BBsteadyM} for small $m$ 
near $T_c$, we have, with $\hat{\beta}=\frac{\beta c}{1-k}$, 
\begin{align}
m_{\mu}&\simeq \frac{N^{\gamma}}{c}\hat{\beta}\sum_{\rho\nu}(\hat{A})^{-1}_{\rho\nu}m_\rho\langle\xi^{\mu}\xi^{\nu}\rangle -\frac{\hat{\beta}^3}{3}\frac{N^\gamma}{c}\bigg[\sum_{\rho_1,\nu_1,\rho_2,\nu_2,\rho_3,\nu_3}\langle\xi^{\mu}\xi^{\nu_1}\xi^{\nu_2}\xi^{\nu_3}\rangle
m_{\rho_1}\hat{A}_{\rho_1\nu_1}m_{\rho_2}\hat{A}_{\rho_2\nu_2}m_{\rho_3}\hat{A}_{\rho_3\nu_3}\bigg]\ 
\end{align}
and averaging over the disorder gives 
\begin{align}
m_{\mu}&=\hat{\beta}\sum_{\rho}(\hat{A})^{-1}_{\rho\mu}m_\rho
 -\frac{\hat{\beta}^3}{3}\sum_\rho (\hat{A})_{\rho\mu}^{-3} m_\rho^3 -\hat{\beta}^3 \frac{c}{N^\gamma}\sum_{\rho}(\hat{A})^{-1}_{\rho\mu}m_\rho\sum_{\tau\pi}(\hat{A})^{-1}_{\tau\pi}m_\pi^2 \ .
\end{align}
To make progress, we project $m_\mu$ on the eigenvectors of 
${\bf \hat{A}}^{-1}$: ${\bf \hat{A}}^{-1} {\bf v}^{(j)}=\mu_j{\bf v}^{(j)}$
\bea 
\sum_{j=1}^Pa_j v^{(j)}_\mu=\hat{\beta}\sum_{j=1}^Pa_j\mu_j v^{(j)}_\mu -\frac{\hat{\beta}^3}{3}\sum_{i,j,k}a_ia_ja_k \mu_i \mu_j \mu_k v^{(i)}_\mu v^{(j)}_\mu v^{(k)}_\mu\nonumber\\ -\hat{\beta}^3 \frac{c}{N^\gamma}
\sum_ia_i\mu_i v^{(i)}_\mu\sum_{\mu\neq\rho}\sum_{j,k}  a_j a_k\mu_j \mu_k  v^{(j)}_\rho v^{(k)}_\rho 
\label{eq:bbexp}\eea
Next we split the sum into $j=1,\dots,P/2$ and $j=P/2+1,\dots,P$, 
substitute the eigenvalues \eqref{eq:eigenris1}, \eqref{eq:eigenris}, and 
equating linear terms we get
\bea
\sum_{j=1}^{P/2}a_j v^{(j)}_\mu (1-\hat{\beta})+\sum_{j=P/2+1}^{P}a_j v^{(j)}_\mu \bigg(1-\hat{\beta}\frac{1-k}{1+k}\bigg)=0
\eea
At $\hat{\beta}=1$ linear terms cancel only if 
$\sum_{j=P/2+1}^{P}a_j v^{(j)}_\mu =0~\forall ~\mu$, 
which implies $v_\mu^{(j)}=0~\forall~j=\frac{P}{2}+1,\ldots,P, ~\forall~\mu$.
We note that the eigenvectors of ${\bf A}^{-1}$ are in the form
\bea
v_\mu^{(j)} =\frac{1}{2}(m_\mu +m_{\mu +P/2})\ ,\quad\quad j=1,..,P/2\ ,\\
v_\mu^{(j)}=\frac{1}{2}(m_\mu-m_{\mu+ P/2})\ ,\quad\quad j=P/2+1,..,P\ ,
\eea
Therefore we have, for all $\mu$, $m_\mu=m_{\mu+p/2}$ and $ v^{(j)}_\mu=m_\mu ~
\forall~j=1,\ldots,P/2$. This implies 
$m_\mu=\sum_{j=1}^Pa_j v^{(j)}_\mu=\sum_{j=1}^{P/2}a_j m_\mu$, hence 
$\sum_{j=1}^{P/2}a_j=1$.
If we plug these conditions into \eqref{eq:bbexp}
\bea
m_\mu(-1+\hat{\beta})-\frac{\hat{\beta}^3}{3}
m_\mu^3 -\hat{\beta}^3 \frac{c}{N^\gamma}
m_\mu\sum_{\rho\neq\mu}m_\rho^2=0\ ,
\eea
we have at $\hat{\beta}=1+\epsilon$,
\bea
m_\mu\epsilon-\frac{1}{3} m_\mu ^3 -\frac{c}{N^\gamma}
m_\mu|{\bf m}|^2 =0\ ,
\eea
where $|{\bf m}|^2= \sum_\rho m_\rho^2$, yielding 
$m_\mu=0$ or 
\bea
m_\mu ^2 = 3\epsilon -\frac{3c}{N^\gamma}
|{\bf m}|^2.
\eea
Each clone activation $m_\mu$ depends on the whole vector ${\bf m}$, 
hence all clones will have the same activation strength $m_\mu\in(-m,+m, 0)$. 
Assuming the non-zero components are a fraction $\phi\leq \alpha$ 
of the total number 
of components $P$, we have 
$|{\bf m}|^2= \alpha \phi N^\gamma m^2$ 
yielding 
\bea
m^2= \frac{3\epsilon}{1+3c\alpha\phi}
\eea 
We will see in the next section that while for $k=0$ stability of 
symmetric mixtures is only ensured for $\phi=1$, in the presence 
of idiotypic interactions, {\em i.e.} for $k\neq 0$, values $\phi\neq 1$ are possible.

\subsubsection{Linear stability analysis and phase diagram}
\label{sec:LSABB}
In this section we study the stability of symmetric 
clonal activation, ${\bf m}=m(1,\dots,1,0,\dots0)$ 
with $n=\alpha \phi N^\gamma$ activated clones, below criticality, in the 
presence of idiotypic interactions.
To this purpose, we study the eigenvalues of the 
Jacobian of the dynamical system \eqref{eq:sysA}, which has a block structure 
with diagonal elements given for $\mu<n$ by
\bea 
J_{\mu \mu}=\beta (\bA^{-1})_{\mu\mu}( 1-\langle\tanh^2 [\beta
(\sum_\rho (\bA^{-1})_{\rho\mu}m_{\rho} +
\sum_{\rho,\lambda\neq\mu}\xi^{\lambda}
(\bA^{-1})_{\rho\lambda}m_\rho )]\rangle_{\bxi})-1,
\eea
and for $\mu> n$ by
\bea 
J_{\mu \mu}=\beta (\bA^{-1})_{\mu\mu}[ 1-
\langle\tanh^2 \beta
(\sum_{\rho,\lambda\neq\mu}\xi^{\lambda}
(\bA^{-1})_{\rho\lambda}m_\rho)]\rangle_{\bxi}-1,
\eea
while off-diagonal elements are, for $\mu,\nu\leq n$ 
\bea 
J_{\mu \nu}=-\frac{\beta
(\bA^{-1})_{\mu\nu}}{N^{\gamma}}
\langle\tanh^2
[\beta(\sum_{\rho,\lambda\neq\mu}\xi^{\lambda}(\bA^{-1})_{\rho\lambda}
m_\rho)\rangle_{\bxi}-1]\ 
\eea
and null otherwise.
For $N\to\infty$ the matrix becomes diagonal 
and eigenvalues are given by the diagonal elements. 
At the symmetric fixed point, using 
$\sum_{\rho}(\bA^{-1})_{\rho\lambda}=1/(1-k)~\forall~\lambda$, we get
\bea
\lambda_1= \frac{\hat{\beta}}{1+k}\left( 1-\sum_zP_n(z) \tanh^2(\hat{\beta}(m(1+z))\right)-1 \quad\quad && deg(\lambda_1)=n\ ,
\label{eq:l1bb}\\
\lambda_2=\frac{\hat{\beta}}{1+k} \left( 1-\sum_z P_n(z)\tanh^2(\hat{\beta}m z)\right)-1 \quad\quad && deg(\lambda_1)=P-n
\label{eq:l2bb}
\eea
with 
$P_n(z)$ defined by (\ref{eq:Pnzq}) with $\bq=c(1,\dots,1)$
and given by
$P_n(z)={\rm e}^{-\alpha\phi c} I_z(\alpha \phi c)$, 
where $I_z(x)$ is the modified Bessel function of the first kind.
Close to criticality, by Taylor expanding in power of 
$\epsilon=\hat{\beta}-1$ we obtain  
\bea
\lambda_1= \frac{-2\epsilon -k(1+3c\alpha\phi)}{(1+k)(1+3c\alpha\phi)}\\
\lambda_2= \frac{\epsilon -k(1+3c\alpha\phi)}{(1+k)(1+3c\alpha\phi)} .
\eea
In contrast to the case $k=0$ where $\lambda_2>0$ and symmetric activation
is stable only if it involves {\it all} clones, 
in the presence of idiotypic interactions, {\em i.e.} for $k\neq0$, 
both eigenvalues are negative at criticality, 
showing the emergence of local minima where not all the clones are activated. 
In the opposite limit, {\em i.e.} $T\to 0$ we get 
\bea
\lambda_1=\frac{\hat{\beta}}{1+k}e^{-\alpha\phi c}I_1(\alpha\phi c)-1 \\
\lambda_2=\frac{\hat{\beta}}{1+k}e^{-\alpha \phi c}
I_0(\alpha\phi c)-1 
\eea
For $\delta<\gamma$, $\alpha=0$, hence using the properties of Bessel functions 
$I_1(0)=0$ and $I_0(0)=1$, we get $\lambda_1<0$ and $\lambda_2>0$.
Since $\lambda_2$ has degeneracy $P-n$, symmetric mixtures 
$\bm=m(1,\ldots,1)$
{\em i.e.} with $n=P$ will be stable for all $T<T_c$.
In contrast, for $\delta=\gamma$ {\em i.e.} $\alpha\neq 0$, $\lambda_1>0$ as 
$I_1(x)>0$ for any $x>0$, meaning that symmetric mixtures 
are unstable at low temperature for any $n$. 
Finally, for $\delta>\gamma$ one has 
$\alpha\to \infty$, in the thermodynamic limit so $I_n(\alpha \phi c)
\simeq (\alpha \phi c)^{-1/2}~\forall~n$, and 
symmetric mixtures will gain stability at low temperature
as $N\to \infty$.

\begin{figure}[htb!]
\centering
\includegraphics[width=0.63\textwidth]{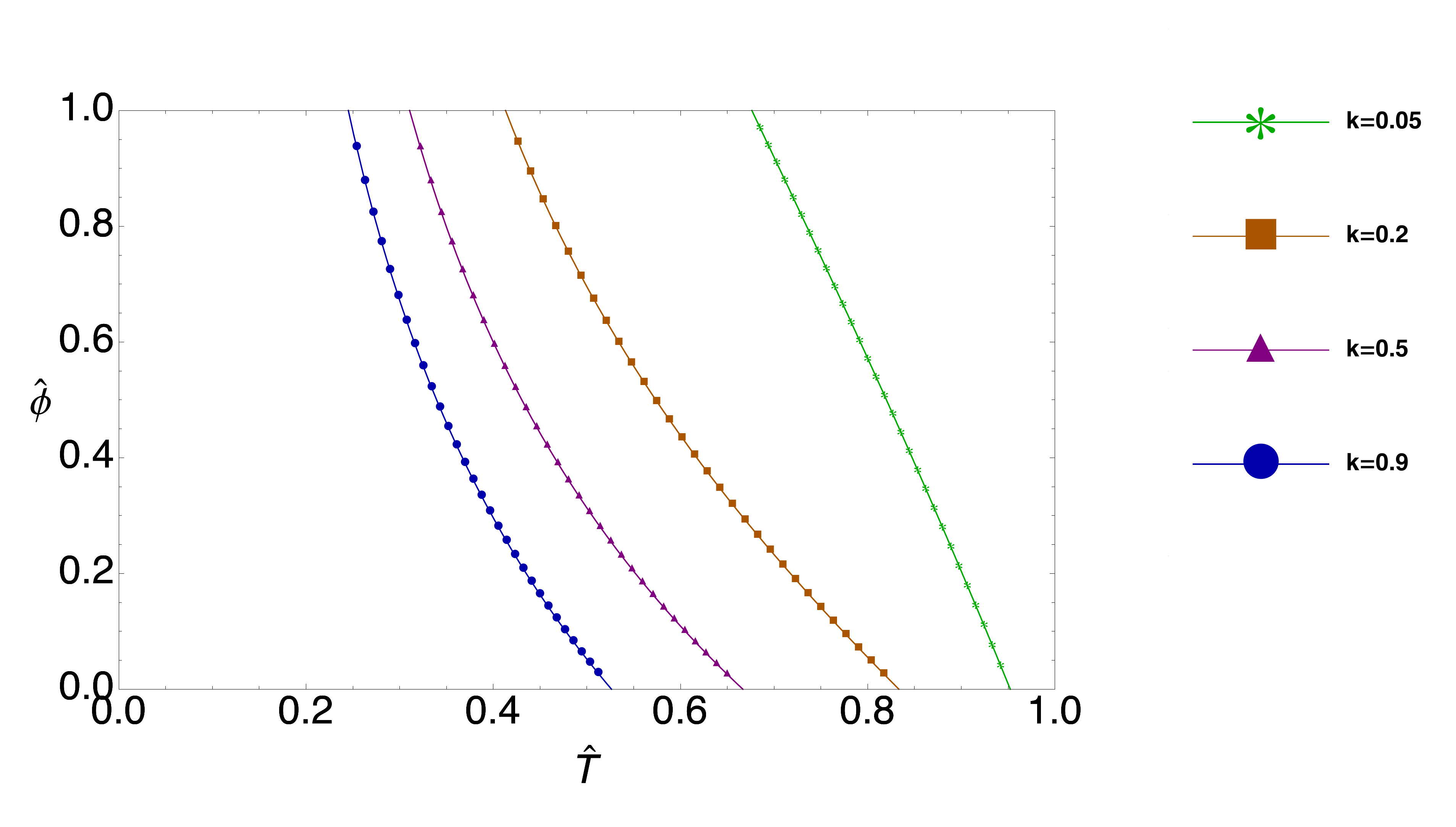}
\includegraphics[width=0.31\textwidth]{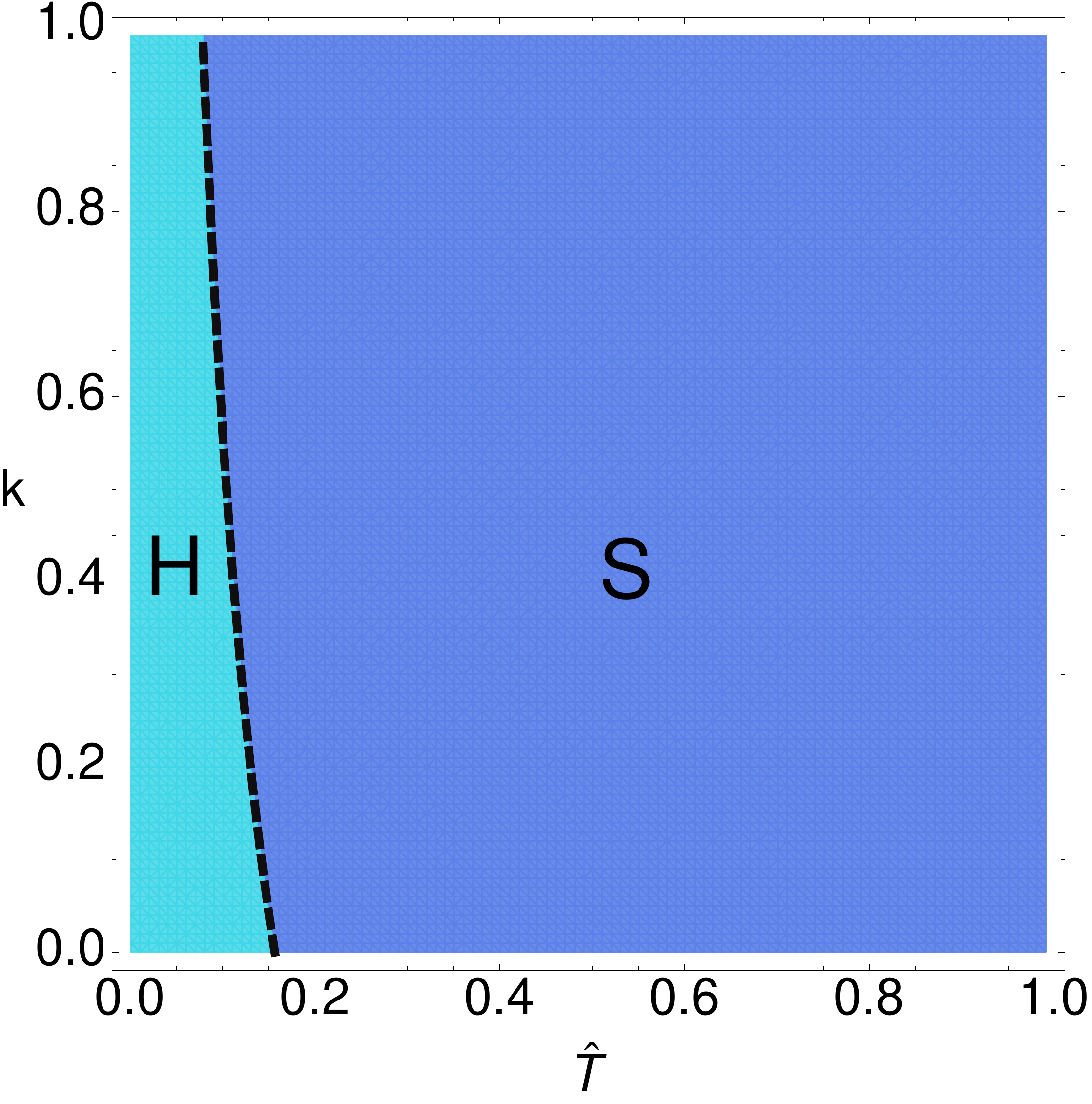}
\caption{Left: Contour plot of $\lambda_2=0$ (obtained from \eqref{eq:l2bb}) 
for $\alpha=1$, 
as a function of the scaled parameters
$(\hat{T}=T/c, \hat{\phi}=\phi c$, for different value of B-B interaction 
strength $k$.
Increasing $k$ the region where symmetric mixtures are stable (to the right 
of the critical line) becomes wider. Right: 
Contour plot of $\lambda_1=0$ (obtained from \eqref{eq:l1bb}) 
for $\alpha=\phi=1$, 
as a function of the scaled temperature
$\hat{T}=T/c$ and strength $k$ of the idiotypic interactions.}
\label{fig:BBmedphase}
\end{figure}

Let us now compute the critical line in the 
phase diagram where symmetric mixtures become unstable.
We note that $\lambda_2>\lambda_1$ for $T\to 0$ 
and for $T\to T_c$, and deduce that $\lambda_2 >\lambda_1$ for all $T<T_c$. 
Hence, stability of $n$-mixtures $\bm=m(1,\ldots,1,0,\ldots,0)$ 
is given by the region where $\lambda_2<0$.
In fig. \ref{fig:BBmedphase} (left) we show the critical lines where 
$\lambda_2$ gets zero in the space of scaled parameters 
$\hat{T}=T/c,\hat{\phi}=\phi c$  for different values of $k$ and  
$\alpha=1$. 
As $k$ increases 
the region where clones are activated with the same intensity 
gets wider.
When $\lambda_2$ destabilises, symmetric mixtures can only be stable 
for $n=P$, {\em i.e.} for $\phi=1$, in the region where $\lambda_1<0$.  
A contour plot of $\lambda_1=0$ for $\phi=1$ in the $T-k$ plane is shown in
fig. \ref{fig:BBmedphase} (right).
%

\section{ Antigen effect}\label{sec:ant}
In this section we investigate the effect of antigens on the basal activity 
of the immune system analysed in the previous sections. We will 
carry out the analysis for homogeneous promiscuities and 
in the absence of idiotypic interactions. We suppose to have $A=N^a$ 
antigens, in the presence of which 
dynamical equations \eqref{eq:m} become
\bea
\frac{{\rm d} m_\mu}{{\rm d} t}=\frac{N^\gamma}{c} \langle \xi^{\mu} \tanh[\hat{\beta}\sum_{\nu=1}^P\xi^\nu(m_\nu+\psi_\nu)]\rangle_{\bxi} -m_\mu \ ,
\label{eq:fieldeq}
\eea 
where $\hat{\beta}=\beta c$ and $\psi_\nu$ is the antigenic field associated 
to B clone $\nu$. 
At the steady state we have two sets of equations, 
those for 
clones $\mu=1,\dots,A$ complementary 
to the incoming viruses
and those for non-activated clones  
$\nu=A+1, \dots,P$, performing basal activities:
\bea
m_\mu=\langle\tanh(\hat{\beta}(m_\mu +\psi_\mu
+\sum_{\rho\neq\mu}^A\psi_\nu\xi^\nu +\sum_{\rho\neq\mu}^P m_\nu\xi^\nu )
\rangle_\bxi\ ,\quad\quad &&\mu=1,\dots,A\ ,\\
m_\nu=\langle\tanh(\hat{\beta}(m_\nu +
+\sum_{\rho\neq\mu}^A\psi_\nu\xi^\nu +\sum_{\rho\neq\mu}^P m_\nu\xi^\nu
\rangle_\bxi\ ,\quad\quad &&\nu=A+1,\dots, P. 
\eea
We first look at the case $A\ll N^\gamma$, where the equations can be written as
\bea
m_\mu=\int {\rm d}z  P(z|\bm)
\tanh(\hat{\beta}(m_\mu +\psi_\mu +z ))\ ,\quad\quad 
&&\mu=1,\dots,A \ ,\label{eq:fewant1}\\
m_\nu=\int {\rm d} z  P(z|\bm)\tanh(\hat{\beta}(m_\nu +z) )\ ,
\quad\quad 
&&\nu=A+1,\dots, P \ ,
\label{eq:fewant2}
\eea
where $P(z|\bm)$ is the large $N$ limit of 
$P_\mu(z|\{m_\rho,1\})$ which was defined in 
(\ref{eq:noisedist}), and
only depends on the vector $\bm$ 
of basal activation.

For
$P\ll N^\gamma$, we have $P(z|{\bf m})\equiv 
\bra \delta(z-\sum_\nu^P \xi^\nu m_\nu)\ket_\bxi\simeq \delta(z)$ and the 
equations decouple, hence there is no interference between 
infected and non-infected clones
\bea
m_\mu=\tanh(\hat{\beta}(m_\mu +\psi_\mu))\ ,\quad\quad &&\mu=1,\dots,A\ ,\\
m_\nu=\tanh(\hat{\beta}m_\nu)\ , \quad\quad &&\nu=A+1,\dots, P\ .
\eea
For the infected clones, the presence of the field induces hysteresis 
effect in the clonal activation \cite{CWfield}. 
This may explain immunological memory effects \cite{Martin,memory}, without 
the requirement of dedicated memory cells: after an infection, the 
responsive B cells, may retain a non-zero activation 
as the antigen is fought and its concentration is sent to zero, 
and on a successive encounter with the same 
antigen they will provide a higher and faster response.

For $P=\order{(N^\gamma)}$, $P(z|\bm)$ in \eqref{eq:fewant1}, 
\eqref{eq:fewant2}, has a finite width, and
both antigen-induced and basal activities reduce, due to clonal interference, 
and hysteresis cycles become smaller.
However, the presence of antigens does not 
affect the basal activity of uninfected 
clones $\nu=A+1,\dots, P$, as long as $A\ll N^\gamma$.

Next we investigate the scenario where the number of antigens is
$A=\order{(N^\gamma)}$ and ask whether the 
system is able to fight against all of them in parallel.  
For simplicity, we set $P=N^\gamma$ and $A=\phi_1 N^\gamma$, with $\phi_1$ 
denoting the fraction of infected clones and we assume that  all 
viruses have the same 
concentrations {\em i.e.} $\bpsi=\psi(1,\dots,1,0,\dots,0)$.
This leads to the steady state equations
\bea
m_\mu= \langle\tanh(\hat{\beta}(m_\mu+\psi+\sum_{\nu=1}^{A}(m_\nu+\psi)\xi^\nu +\sum_{\nu=A+1}^{P}m_\nu\xi^\nu))\rangle_\bxi\quad\quad 
&&\mu=1,\dots,A\nonumber
\\
m_\nu=\langle\tanh(\hat{\beta}(m_\nu +\sum_{\rho=1}^{A}(m_\rho+\psi)\xi^\rho +\sum_{\rho=A+1}^{P}m_\rho\xi^\rho))\rangle_\bxi 
\quad\quad 
&&\nu=A+1, \dots, P
\nonumber\\
\label{eq:partialfield}
\eea
Now antigen interference on the basal activity is relevant 
and will affect the activation of the non-infected clones.
In the small field limit, we can Taylor expand \eqref{eq:partialfield} 
near $\hat{\beta}=1+\epsilon$ and small $m_\nu$, obtaining 
\bea
m_\mu \simeq (1+\epsilon) m_\mu +\psi_\mu (1+\epsilon)-\frac{1}{3}\bigg( (m_\mu+\psi_\mu)^3+\frac{3c}{N^\gamma}(m_\mu+\psi_\mu)\sum_{\rho\neq\mu}(m_\rho+\psi_\rho)\bigg)
\eea
For $\mu<A$ (infected clones) we have for $\psi\ll\epsilon$
\bea
m_\mu=\frac{-\psi(1+\epsilon)}{\epsilon}\equiv m_1\ , \quad\quad \mu=1,\dots, A \ .
\label{eq:inf}
\eea
where the expansion holds for field $\psi\ll \epsilon$, otherwise 
$m_\mu=\order{(1)}$. 
For $\mu>A$ ($\psi_\mu=0$) we have
\bea
m_\mu \simeq (1+\epsilon) m_\mu-\frac{1}{3}\bigg(m_\mu^3+\frac{3c}{N^\gamma}m_\mu\sum_{\rho\neq\mu}(m_\rho+\psi_\rho)^2\bigg)
\label{eq:noninf}
\eea
hence $m_\mu=0$ is always a solution 
(uninfected clones may not be activated) together with
\bea
m_\mu^2&=&3\epsilon-\frac{3c}{N^\gamma}\sum_\rho(m_\rho+\psi_\rho)^2=
\nonumber
\\
&=&3\epsilon-\frac{3c}{N^\gamma}(\sum_{\rho=A+1}^{P}m_\rho^2 + \sum_{\rho=1}^{A}(2m_1\psi +\psi^2))\ , \quad\quad\mu=A+1,\dots,P \ .
\eea
This shows that 
uninfected clones are symmetrically activated, each one with intensity
\bea
m_2^2=\frac{3\epsilon-3c\phi_1\psi^2(2(1+\epsilon)/\epsilon +1)}{1+3c\phi_2}=\frac{3\epsilon(1-2c\phi_1(\frac{\psi}{\epsilon})^2)}{1+3c\phi_2} \ ,
\label{eq:m2psi}
\eea
where $\phi_2\leq 1-\phi_1$ is the fraction of active uninfected 
clones.
Hence, for $\psi\ll \epsilon$, 
clonal activation close to criticality will have the form 
${\bf m^{*}}=(m_1,\dots,m_1,m_2,\dots,m_2,0,\dots,0)$. 
However, upon increasing the fraction $\phi_1$ 
of infected clones or the antigenic field $\psi$, 
(\ref{eq:m2psi}) shows that
non-zero values of $m_2$ may become impossible
and uninfected clones may get activated at a lower temperature 
(similarly to clones with smaller numbers 
of triggered receptors we dealt with in Sec. \ref{sec:q}). 
\begin{figure}[htb!]
\centering
\includegraphics[width=0.43\textwidth]{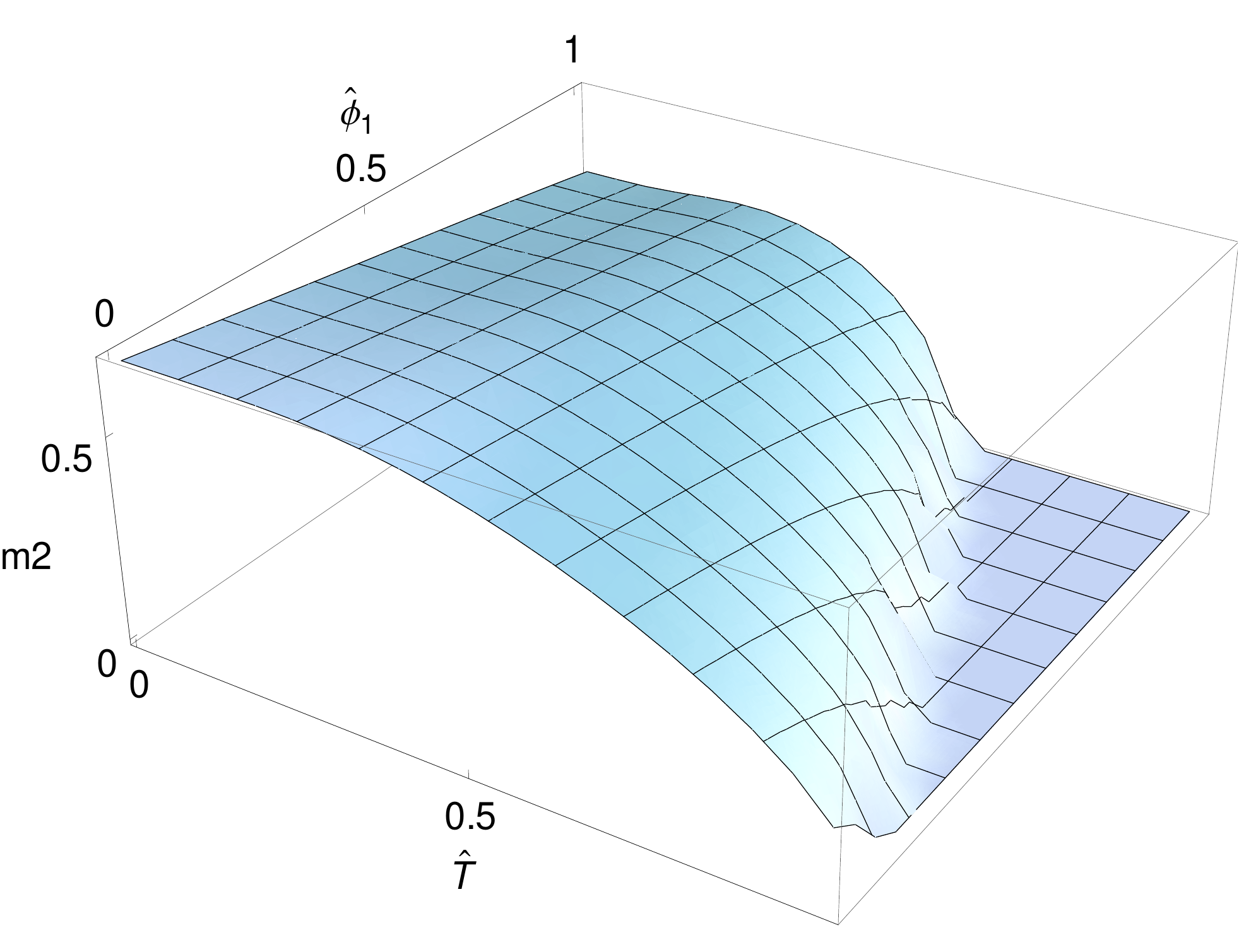}
\caption{3D plot of the activation $m_2$ of non-infected clones 
versus the fraction of infected clones $\hat{\phi}_1=\phi_1 c$ and the scaled
temperature $\hat{T}= T/c$, for fixed $\psi=0.1$. 
Increasing $\hat{\phi}_1$ both the intensity and the critical activation 
temperature decrease, due to the antigenic interference.}
\label{fig:m2noninf}
\end{figure}
This is confirmed by numerical results in 
Fig. \ref{fig:m2noninf}, showing that the response of uninfected B 
clones and their activation temperature decrease for increasing 
fractions of infected clones. This results in a reduced basal activity 
of the immune system, which is vital to keep cells signaled and 
accomplish homeostasis \cite{saturation}. 
\begin{figure}[htb!]
\centering
\includegraphics[width=0.8\textwidth]{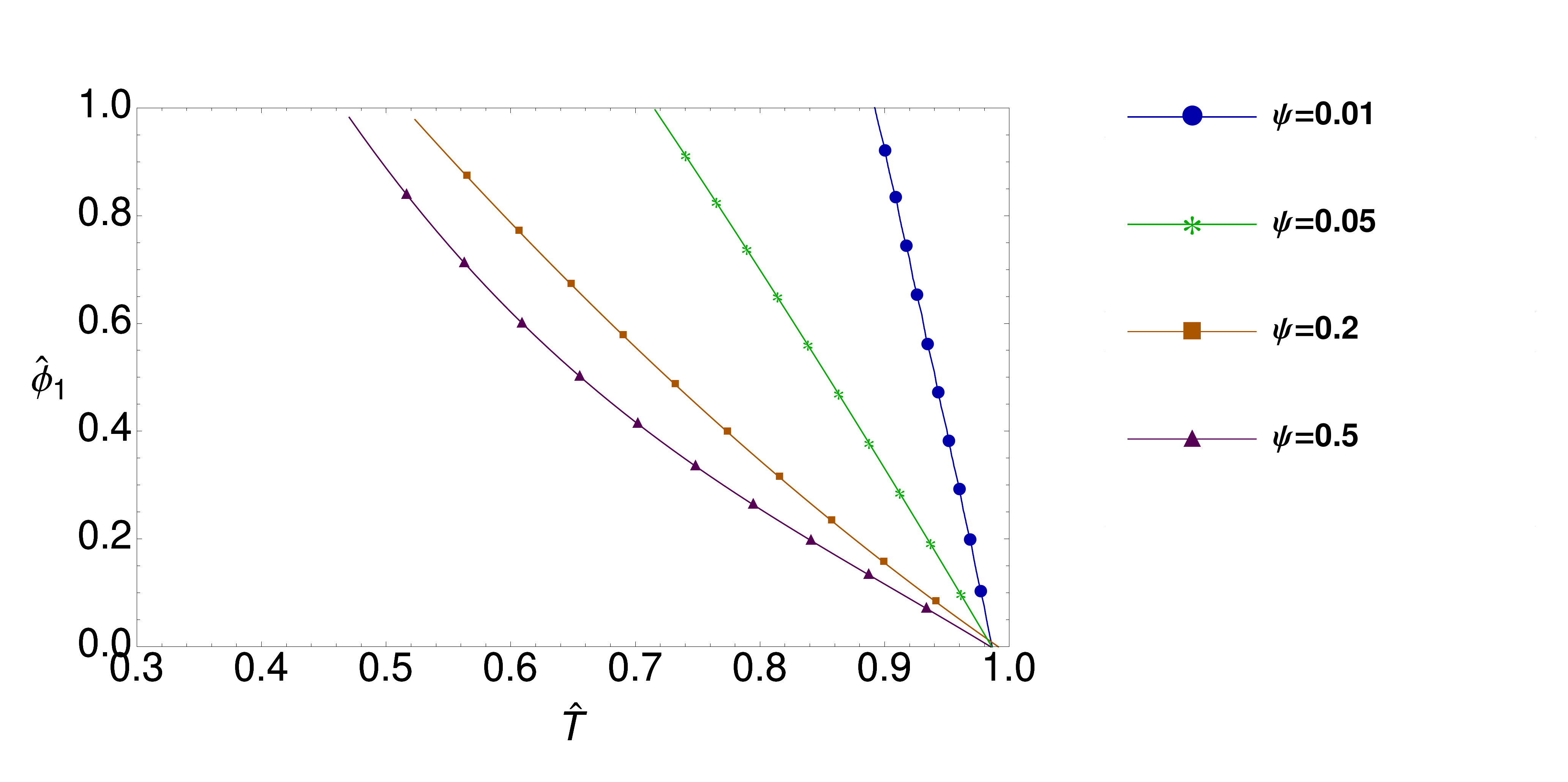}
\label{eq:mbifpsi}
\caption{Critical line for the activation of uninfected clones 
in the space of scaled parameters 
$\hat T=T/c$, $\hat\phi=\phi c$, otained from the condition
$m_2\neq 0$ \eqref{eq:noninf},
for different values of the antigenic field 
$\psi$. Increasing $\psi$, the region where uninfected 
clones are signaled shrinks.}
\label{fig:mfield}
\end{figure}
In fig. \ref{fig:mfield} we study the impact of antigen concentration 
on the critical temperature at which uninfected B clones 
become responsive, by plotting the critical temperature versus 
the fraction of infected clones, for different values of antigen 
concentration. This shows that as the fraction of infected clones 
and the field increase, the basal 
activity is more and more compromised. 

\begin{figure}
\centering
\includegraphics[width=0.45\textwidth]{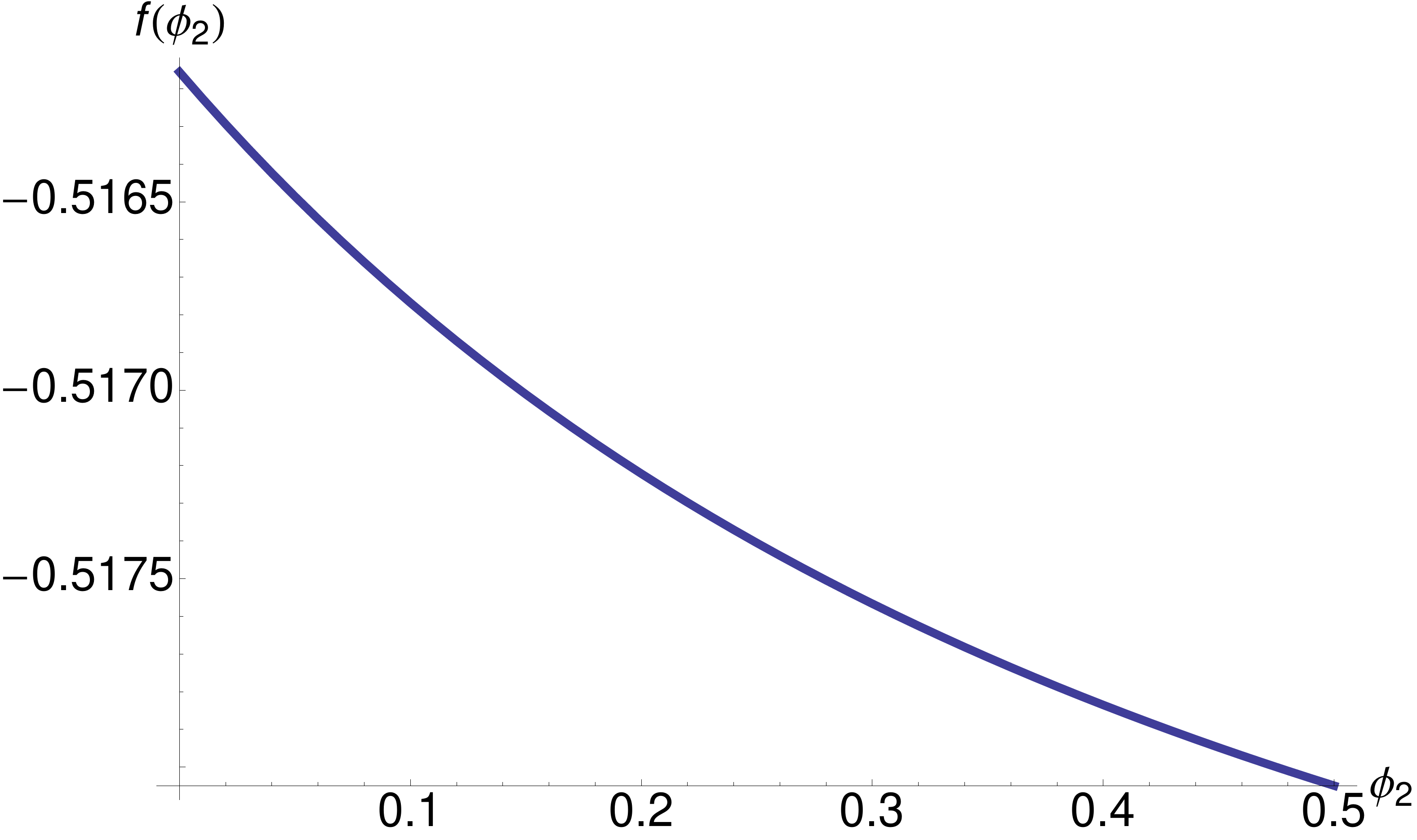}
\caption{Free energy as a function of the fraction of active uninfected clones ($\phi_2$) for $\phi_1=0.5$, $T=0.7, c=0.2, \psi=0.05$.}
\label{fig:free}
\end{figure}

Next, we inspect the stability region of 
${\bf m^{*}}= (m_1,\dots,m_1, m_2,\dots,m_2,0,\dots,0)$ by looking at the 
eigenvalues of the Jacobian of the dynamical system \eqref{eq:fieldeq}. 
This has a diagonal structure, in the thermodynamic limit, where 
off-diagonal elements become negligible and diagonal terms are  
for $\mu< n_1=\phi_1 N^\gamma$
\bea
J_{\mu\mu}=\hat{\beta}(1-\langle\tanh^2(\hat{\beta}(m_\mu+\psi+\sum_{\nu=n_1+1}^{n_2} m_\nu \xi^\nu+ \sum_{\nu=1}^{n_1}(m_\nu+\psi)\xi^\nu))\rangle_\bxi)
\eea
for $n_1<\mu< n_2=\phi_2 N^\gamma$
\bea
J_{\mu\mu}=\hat{\beta}(1-\langle\tanh^2(\hat{\beta}(m_\mu+\sum_{\nu=n_1+1}^{n_2} m_\nu \xi^\nu+ \sum_{\nu=1}^{n_1}(m_\nu+\psi)\xi^\nu))\rangle_\bxi)
\eea
and for $\mu>n_2$
\bea
J_{\mu\mu}=\hat{\beta}(1-\langle\tanh^2(\hat{\beta}(\sum_{\nu=n_1+1}^{n_2} m_\nu \xi^\nu+ \sum_{\nu=1}^{n_1}(m_\nu+\psi)\xi^\nu))\rangle_\bxi)
\eea
Evaluating the Jacobian at the symmetric fixed point 
${\bf m}^\star$ and introducing the 
distribution 
$P_{n}(z)=\langle\delta(z-\sum_{\nu=l}^{\ell+n}\xi^\nu)\rangle\ ,
\forall ~\ell $, 
gives
\bea
\hspace{-2cm}
\lambda_1\!=\!\hat{\beta}\{1\!\!-\!\!\sum_{z_1,z_2}\!\!
P_{n_1}(z_1)P_{n_2}(z_2)\tanh^2[\hat{\beta}((\psi+m_1)(1+z_1) +m_2 z_2)]\}
\!-\!1 \quad &&deg(\lambda_1)\!=\!n_1\label{eq:l1A}\\
\hspace{-2cm}
\lambda_2\!=\!\hat{\beta}\{1\!\!-\!\!
\sum_{z_1,z_2}\!\!P_{n_1}(z_1)P_{n_2}(z_2)\tanh^2[\hat{\beta}((\psi+m_1)z_1 +m_2(1+z_2)]\}-\!1\!\quad &&deg(\lambda_2)\!=\!n_2\label{eq:l2A}\\ 
\hspace{-2cm}
\lambda_3\!=\!\hat{\beta}\{1\!\!-\!\!
\sum_{z_1,z_2}\!\!P_{n_1}(z_1)P_{n_2}(z_2)\tanh^2[\hat{\beta}((\psi+m_1)z_1 +m_2z_2)]\}-\!1\!\quad &&
deg(\lambda_3)\!=\!P\!\!-\!\!n_1\!\!-\!\!n_2\label{eq:l3A}
\eea 
with $m_1,m_2$ following from \eqref{eq:inf}, \eqref{eq:noninf} as
\bea
m_1=\sum_{z_1,z_2}P_{n_1}(z_1)P_{n_2}(z_2)\tanh[
\hat{\beta}((\psi+m_1)(1+z_1) +m_2z_2)]
\nonumber\\
m_2=\sum_{z_1,z_2}P_{n_1}(z_1)P_{n_2}(z_2)\tanh[
\hat{\beta}((\psi+m_1)z_1 +m_2(1+z_2))]
\eea
and $P_n(z)={\rm e}^{\phi c}I_z(\phi c)$ for $n=\phi N^\gamma$, where $I_z(x)$ is the modified Bessel function of the first kind.
\begin{figure}[htb!]
\centering 
\includegraphics[width=0.45\textwidth]{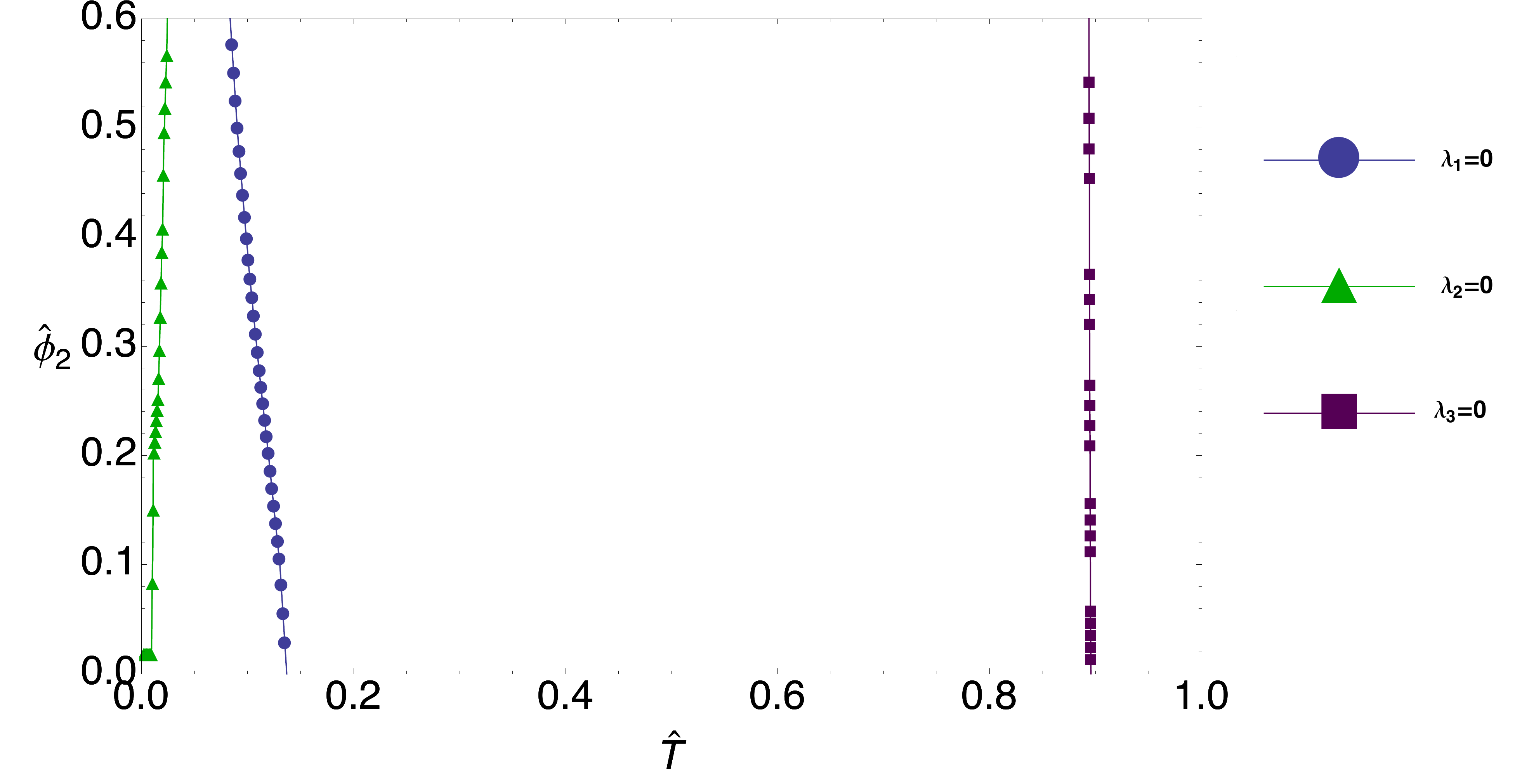}
\includegraphics[width=0.53\textwidth]{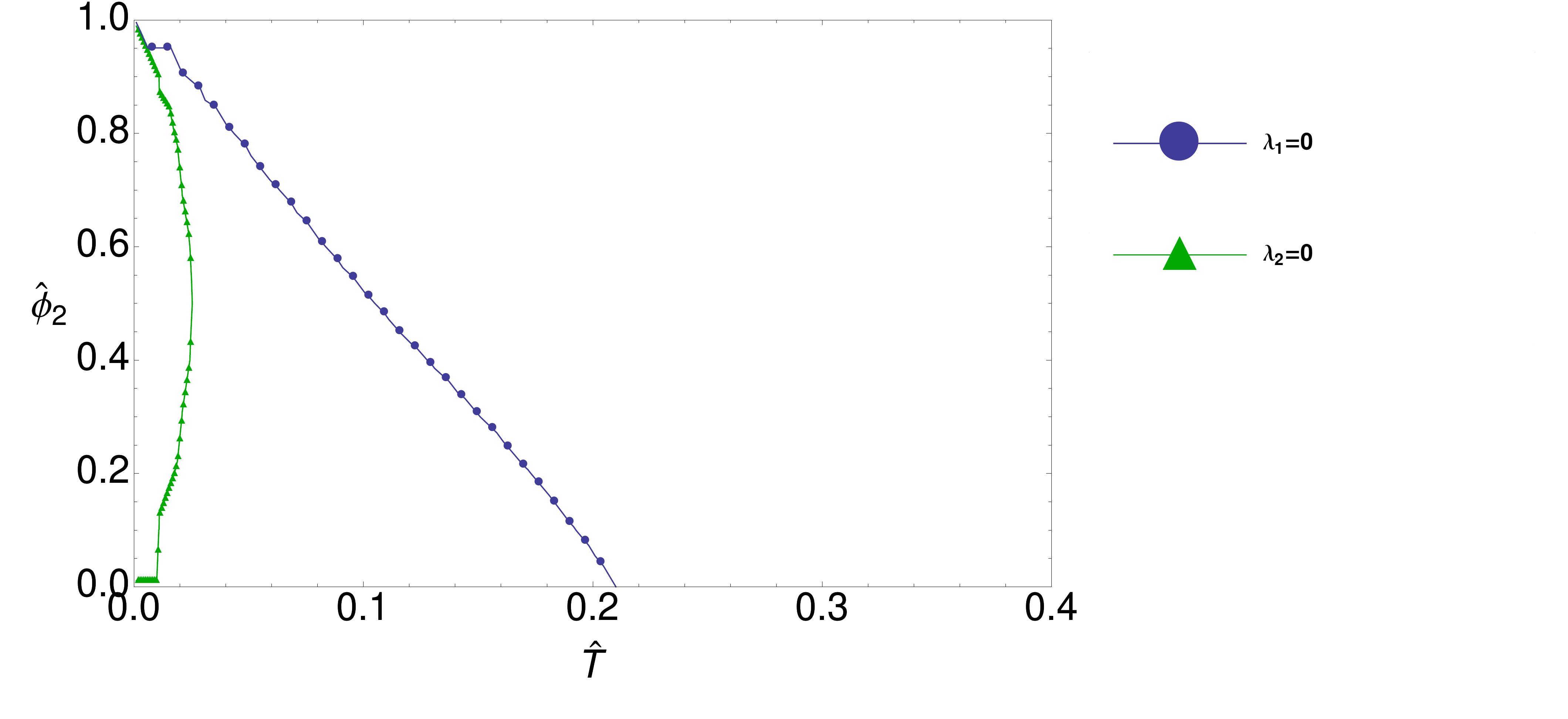}
\caption{Left: Phase diagram in the space of scaled parameters $\hat{T}=T/c, \hat{\phi}_2=\phi_2 c$ with $\phi_1=0.4$. 
Lines represent contours of $\lambda_1=0$ (circles), 
$\lambda_2=0$ (triangles) and $\lambda_3=0$ (squares). 
To the right of the line 
$\lambda_3=0$ solutions where uninfected clones are partially activated are 
stable. Lowering down the temperature and crossing the line 
$\lambda_1=0$ the $m_1$ symmetric mixtures destabilise, meaning that infected 
clones are hierarchically activated. Crossing the line $\lambda_2=0$ 
the $m_2$ symmetric mixtures destabilise, and uninfected clones are hierarchically activated. Right: Plots of $\lambda_1=0$ (circles), 
$\lambda_2=0$ (triangles) in the space $\hat{T}=T/c, \hat{\phi}_2=\phi_2 c$ for $\phi_2=1-\phi_1$.}
\label{fig:antiphase}
\end{figure}
In Fig. \ref{fig:antiphase} (left) we show the critical lines $\lambda_1=0$, $\lambda_2=0$ and $\lambda_3=0$
in the space of scaled parameters $\hat{T}=T/c, \hat{\phi_2}=\phi_2c$. As temperature is lowered, $\lambda_3$ is the 
first eigenvalue to destabilise, meaning that clonal activation will get in the form $\bm=(m_1,\ldots,m_1,m_2,\ldots,m_2)$.
Decreasing the temperature further, the system will first prioritise activation of infected clones, while keeping activation of uninfected clones symmetric,  and later, at low temperature, will activate uninfected clones in a hierarchical fashion  (see fig. \ref{fig:antiphase}, right panel).
Near the critical temperature symmetric mixtures with $\phi_2\leq 1-\phi_1$ are stable. To investigate the optimal value of $\phi_2$ we calculate the free-energy as a function of $\phi_2$ for fixed $T,\phi_1$
\bea
\hspace{-2cm}F(\phi_1,\phi_2)=-\frac{1}{\beta}\sum_{z_1,z_2}
P_{n_1}(z_1)P_{n_2}(z_2) \log(2\cosh(\beta c(m_1 + \psi)z_1+m_2 z_2)) + \frac{c}{2}(\phi_1m_1^2+\phi_2 m_2^2)
\eea
which is minimal (fig. \ref{fig:free}) at $\phi_2=1-\phi_1$, meaning that the system will keep all uninfected clones signalled. 

\section{Conclusions}
\label{sec:conclusions}
In this paper we presented a minimal model for interacting cells in the adaptive immune systems, constituting of B cells, T cells and antigens.  
Our model is able to capture important collective features of the real immune system, such as the ability of simultaneously handling multiple infections, 
the dependence of B clones' activation on the number of receptors on clonal surface and the role of idiotypic interactions in enhancing parallel 
response to multiple infections.
We analysed the dynamics of the system's order parameters quantifying the B clones' activation via linear stability analysis and Monte Carlo simulations.
We found the regions, in the parameters' space, where the system activate B clones {\em symmetrically}, {\em i.e.} with the same intensity, 
or {\em hierarchically}, whereby the system prioritises responses to particular infections. 
We showed that clones with fewer receptors are less likely to be activated (Sec. \ref{sec:q}) and 
idiotypic interactions contribute to the overall stability of the immune system, preventing unwanted activation and increasing the region where 
{\em all clones} are equally activated and ready to start an immune response upon arrival of new infections (Sec. \ref{sec:BB}). 
Furthermore, we investigated how the immune system responds to antigens, showing in particular that multiple antigens create an interference that leads to less effective response to individual 
antigens and a reduction in the {\em basal activity} of the uninfected B cells (Sec. \ref{sec:ant}). 
For higher noise level the system tends to simultaneously fight {\em all} the antigens, despite losing in terms of response strength, whereas for lower noise level 
it will prioritise some infections over others. Finally, we showed that short term memory may emerge as an hysteresis effect, without the requirement of dedicated memory cells.

\section{Acknowledgements}
It is our great pleasure to thank ACC Coolen for many useful discussions.

\end{document}